\documentclass[a4paper,reqno,10pt]{amsart}
\usepackage[all]{xy}
\usepackage[latin1]{inputenc}
\usepackage{hyperref}
\usepackage{color}
\usepackage{hhline}
\usepackage{amssymb}

\numberwithin{equation}{section}

\def\emph#1{{\bfseries\itshape{#1}}}

\newtheorem{guia}{}[section]
\newtheorem{lemma}[guia]{Lemma}
\newtheorem{definition}[guia]{Definition}
\newtheorem{proposition}[guia]{Proposition}

\newtheorem{theorem}[guia]{Theorem}
\newtheorem{assumption}[guia]{Assumption}
\newtheorem{example}[guia]{Example}
\newtheorem{remark}[guia]{Remark}

\newcommand{\lvec}[1]{\overleftarrow{#1}}
\newcommand{\rvec}[1]{\overrightarrow{#1}}

\def\R{\mathbb{R}}
\def\qquand{\qquad\text{and}\qquad}
\def\quand{\quad\text{and}\quad}
\newcommand{\sode}{\textsc{sode}}
\newcommand\map[3]{#1\colon#2\to#3}
\newcommand{\Sec}[2][]{\operatorname{Sec}_{#1}(#2)}
\newcommand{\pai}[2]{\left\langle#1,#2\right\rangle}

\newcommand{\cinfty}[1]{C^\infty(#1)}
\newcommand{\set}[2]{\{\,#1\,|\,#2\,\}}
\newcommand{\pd}[2]{\frac{\partial #1}{\partial #2}}
\newcommand{\at}[1]{\Big|_{#1}}
\newcommand{\pb}{^\star}
\newcommand{\ext}[2][]{\bigwedge\nolimits^{#1}{#2}}
\newcommand{\Ver}[1]{\operatorname{Ver}(#1)}
\newcommand{\Adm}[1]{\operatorname{Adm}(#1)}
\newcommand{\rank}[1]{\operatorname{rank}(#1)}

\DeclareMathOperator{\id}{id}

\DeclareMathOperator{\grad}{grad}
\DeclareMathOperator{\Ker}{Ker}

\renewcommand{\sup}[1]{^{\scriptscriptstyle#1}}

\newcommand{\spV}{\sup{V}}

\newcommand{\spC}{\sup{C}}

\makeatletter
\newcommand\prol{\@ifstar{\@proldf}{\@prolpf}}  
\def\@prolpf{\@ifnextchar[{\@prolpf@wrt}{\@prolpf@}}
\def\@prolpf@wrt[#1]#2{\@ifnextchar[{\@prolpf@wrt@at{#1}{#2}}{\@prolpf@wrt@{#1}{#2}}}
\def\@prolpf@wrt@at#1#2[#3]{\prolsymbol^{#1}_{#3}#2}
\def\@prolpf@wrt@#1#2{\prolsymbol^{#1}#2}
\def\@prolpf@#1{\@ifnextchar[{\@prolpf@at{#1}}{\@prolpf@@{#1}}}
\def\@prolpf@at#1[#2]{\prolsymbol_{#2}#1}
\def\@prolpf@@#1{\prolsymbol#1}
\def\@proldf{\@ifnextchar[{\@proldf@wrt}{\@proldf@}}
\def\@proldf@wrt[#1]#2{\@ifnextchar[{\@proldf@wrt@at{#1}{#2}}{\@proldf@wrt@{#1}{#2}}}
\def\@proldf@wrt@at#1#2[#3]{\prolsymbol^{*#1}_{#3}#2}
\def\@proldf@wrt@#1#2{\prolsymbol^{*#1}#2}
\def\@proldf@#1{\@ifnextchar[{\@proldf@at{#1}}{\@proldf@@{#1}}}
\def\@proldf@at#1[#2]{\prolsymbol^*_{#2}#1}
\def\@proldf@@#1{\prolsymbol^*#1}
\def\prolsymbol{\mathcal{T}}
\makeatother

\newcommand{\TEE}[1][]{\mathcal{T}^E_{#1}E}
\newcommand{\TED}[1][]{\mathcal{T}^E_{#1}D}
\newcommand{\TDD}[1][]{\mathcal{T}^D_{#1}D}
\newcommand{\X}{\mathcal{X}}
\newcommand{\V}{\mathcal{V}}

\newcommand{\Gc}{\mathcal{G}}
\newcommand{\cnabla}{\check{\nabla}}

\newcommand{\Do}[1][]{D_{#1}^\circ}
\newcommand{\tDo}[1][]{\widetilde{D_{#1}^\circ}}
\newcommand{\GLD}[1][]{G\sup{L,D}_{#1}}
\newcommand{\GLV}[1][]{G_{#1}\sup{L,\vd}}
\newcommand{\GL}[1][]{G\sup{L}_{#1}}
\newcommand{\C}{\mathcal{C}} 

\newcommand{\cm}{\mathcal{M}} 
\newcommand{\vd}[1][]{\mathcal{V}_{#1}} 
\newcommand{\cf}{\boldsymbol{\Psi}} 
\def\lcf{\lbrack\! \lbrack}
\def\rcf{\rbrack\! \rbrack}

\newcommand{\oprocendsymbol}{\hbox{$\bullet$}}
\newcommand{\oprocend}{\relax\ifmmode\else\unskip\hfill\fi\oprocendsymbol}
\def\eqoprocend{\tag*{$\bullet$}}

\begin{document}

\title{Nonholonomic Lagrangian systems on Lie algebroids}

\author[J.\ Cort\'es]{Jorge Cort\'es} \address{Jorge Cort\'es:
  Department of Mechanical and Aerospace Engineering, University of
  California at San Diego, La Jolla, California 92093, USA}
\email{cortes@ucsd.edu}

\author[M.\ de Le\'on]{Manuel de Le\'on} \address{Manuel de Le\'on:
  Instituto de Matem\'aticas y F{\'\i}sica Fundamental, Consejo
  Superior de Investigaciones Cient\'{\i}ficas, Serrano 123, 28006
  Madrid, Spain} \email{mdeleon@imaff.cfmac.csic.es}

\author[J.\ C.\ Marrero]{Juan C.\ Marrero} \address{Juan C.\ Marrero:
  Departamento de Matem\'atica Fundamental y
  Unidad Asociada ULL-CSIC Geometr\'{\i}a Diferencial y Mec\'anica
  Geom\'etrica, Facultad de   Ma\-te\-m\'a\-ti\-cas,
  Universidad de la Laguna, La Laguna, Tenerife, Canary Islands, Spain }
  \email{jcmarrer@ull.es}

\author[E.\ Mart\'{\i}nez]{Eduardo Mart\'{\i}nez} \address{Eduardo
  Mart\'{\i}nez: Departamento de Matem\'atica Aplicada, Facultad de
  Ciencias, Universidad de Zaragoza, 50009 Zaragoza, Spain}
\email{emf@unizar.es}

\thanks{This work has been partially supported by Spanish Ministry of
  Education and Culture grants MTM2004-7832, BFM2003-01319, MTM2006-03322 and
  BFM2003-02532. J. Cort\'es was partially supported by faculty
  research funds granted by the University of California, Santa Cruz.}

\keywords{Nonholonomic Mechanics, Lagrange-d'Alembert equations, Lie
  algebroids, symmetry, reduction}

\subjclass[2000]{70F25, 70H03, 70H33, 37J60, 53D17}

\begin{abstract}
  This paper presents a geometric description on Lie algebroids of
  Lagrangian systems subject to nonholonomic constraints.  The Lie
  algebroid framework provides a natural generalization of classical
  tangent bundle geometry.  We define the notion of nonholonomically
  constrained system, and characterize regularity conditions that
  guarantee that the dynamics of the system can be obtained as a
  suitable projection of the unconstrained dynamics.  The proposed
  novel formalism provides new insights into the geometry of
  nonholonomic systems, and allows us to treat in a unified way a
  variety of situations, including systems with symmetry, morphisms,
  reduction, and nonlinearly constrained systems. Various examples
  illustrate the results.
\end{abstract}

\date{\today}

\maketitle

\tableofcontents

\section{Introduction}

The category of Lie algebroids has proved useful to formulate problems
in applied mathematics, algebraic topology, and differential geometry.
In the context of Mechanics, an ambitious program was proposed
in~\cite{Weinstein} in order to develop formulations of the dynamical
behavior of Lagrangian and Hamiltonian systems on Lie algebroids and
discrete mechanics on Lie groupoids. In the last years, this program
has been actively developed by many authors, and as a result, a
powerful mathematical structure is emerging.

The main feature of the Lie algebroid framework is its inclusive
nature. Under the same umbrella, one can consider such disparate
situations as systems with symmetry, systems evolving on semidirect
products, Lagrangian and Hamiltonian systems on Lie algebras, and
field theory equations (see~\cite{CoLeMaMaMa,LeMaMa} for recent
topical reviews illustrating this).  The Lie algebroid approach to
Mechanics builds on the particular structure of the tangent bundle to
develop a geometric treatment of Lagrangian systems parallel to
Klein's formalism~\cite{Cr,Klein}.  At the same time, the attention
devoted to Lie algebroids from a purely geometrical viewpoint has led
to an spectacular development of the field, e.g.,
see~\cite{BoKoSt,CaNuSaI,Mackenzie,Sau} and references therein. The
merging of both perspectives has already provided mutual benefit, and
will undoubtedly lead to important developments in the future.

The other main theme of this paper are nonholonomic Lagrangian
systems, i.e., systems subject to constraints involving the
velocities.  This topic is a classic subject in Mathematics and
Mechanics, dating back to the early times of Lagrange; a
comprehensive list of classical references can be found
in~\cite{NF}.  At the beginning of the nineties, the
work~\cite{Ko} sparked a renewed interest in the geometric study
of nonholonomic mechanical systems, with a special emphasis on
symmetry aspects.  In the last years, several authors have
extended the ideas and techniques of the geometrical treatment of
unconstrained systems to the study of nonholonomic mechanical
systems, see the recent monographs~\cite{Bl,cortes}. These include
symplectic~\cite{CaRa,LeMa,LeMa2}, Hamiltonian~\cite{VaMa}, and
Lagrangian approaches~\cite{CuKeSnBa,KoMa1}, the study of almost
Poisson brackets~\cite{CaLeMa,IbLeMaMa,KoMa2}, and symmetry and
reduction of the
dynamics~\cite{BaSn,BlKrMaMu,CaCoLeMa,CaLeMaMa,CaLeMaMa2,CoLe,marle}.

In this paper we develop a comprehensive treatment of nonholonomic
systems on Lie algebroids. This class of systems was introduced
in~\cite{CoMa} when studying mechanical control systems (see
also~\cite{MeLa} for a recent approach to mechanical systems on
Lie algebroids subject to linear constraints).  Here, we build on
the geometry of Lie algebroids to identify suitable regularity
conditions guaranteeing that the nonholonomic system admits a
unique solution. We develop a projection procedures to obtain the
constrained dynamics as a modification of the unconstrained one,
and define an almost-Poisson nonholonomic bracket.  We show that
many of the properties that standard nonholonomic systems enjoy
have their counterpart in the proposed setup.  As important
examples, we highlight that the analysis here provides a natural
interpretation for the use of pseudo-coordinates techniques and
lends itself to the treatment of constrained systems with
symmetry, following the ideas developed in~\cite{CoMa,MaROMP}. We
carefully examine the reduction procedure for this class of
systems, paying special attention to the evolution of the momentum
map.

From a methodological point of view, the approach taken in the paper
has enormous advantages. This fact must mainly be attributed to the
inclusive nature of Lie algebroids.  Usually, the results on
nonholonomic systems available in the literature are restricted to a
particular class of nonholonomic systems, or to a specific context.
However, as illustrated in Table~\ref{tab:examples}, many different
nonholonomic systems fit under the Lie algebroid framework, and this
has the important consequence of making the results proved here widely
applicable.  With the aim of illustrating this breadth, we consider
various examples throughout the paper, including the Suslov problem,
the Chaplygin sleigh, the Veselova system, Chaplygin Gyro-type
systems, the two-wheeled planar mobile robot, and a ball rolling on a
rotating table. We envision that future developments within the
proposed framework will have a broad impact in nonholonomic mechanics.
In the course of the preparation of this manuscript, the recent
research efforts~\cite{CaNuSaII,Me} were brought to our attention.
These references, similar in spirit to the present work, deal with
nonholonomic Lagrangian systems and focus on the reduction of Lie
algebroid structures under symmetry.

{\small
\begin{table}[tbhp]
  \centering
  \begin{tabular}{|l|l|l|l|}
    \hline%
    \parbox[t]{.25\linewidth}{Nonholonomic\\ Lagrangian system}
    & Lie algebroid & Dynamics & Example\\
    \hline \hline
    Standard  & Tangent bundle & Lagrande-d'Alembert &
    Rolling disk~\cite{NF}\\
    \hline%
    On a Lie algebra & Lie algebra & Euler-Poincar\'e-Suslov &
    \parbox[t]{.15\linewidth}{Chaplygin
    sleigh~\cite{Ch}}\\
    \hline%
    \parbox[t]{.255\linewidth}{
      Nonholonomic LR\\
      systems} &\parbox[t]{.15\linewidth}{Right action Lie
      algebroid} & \parbox[t]{.25\linewidth}{Reduced Poincar\'e-Chetaev} &
    \parbox[t]{.195\linewidth}{Veselova problem \cite{VeVe}}\\
    \hline%
    \parbox[t]{.254\linewidth}{
      Nonholonomic systems with semidirect
      product symmetry} &\parbox[t]{.15\linewidth}{Left action Lie
      algebroid} & \parbox[t]{.253\linewidth}{Nonholonomic
      Euler-Poincar\'e with an advected parameter} &
    \parbox[t]{.195\linewidth}{Chaplygin's gyro \cite{Ma,MT:04}}\\
    \hline%
    Symmetry-invariant & Atiyah algebroid &
    \parbox[t]{.24\linewidth}{Nonholonomic
      Lagrange-Poincar\'e} &
    \parbox[t]{.15\linewidth}{Snakeboard~\cite{BlKrMaMu}}\\
    \hline
    \end{tabular}
  \medskip
  \caption{The Lie algebroid framework embraces different classes of
    nonholonomic systems.}
  \label{tab:examples}
\end{table}
}

The paper is organized as follows. In Section~\ref{preliminaries}
we collect some preliminary notions and geometric objects on Lie
algebroids, including differential calculus, morphisms and
prolongations. We also describe classical Lagrangian systems
within the formalism of Lie algebroids.  In Section~\ref{linear},
we introduce the class of nonholonomic Lagrangian systems subject
to linear constraints, given by a regular Lagrangian $L : E
\longrightarrow \R$ on the Lie algebroid $\tau : E \longrightarrow
M$ and a constraint subbundle $D$ of $E$.  We show that the known
results in Mechanics for these systems also hold in the context of
Lie algebroids. In particular, drawing analogies with d'Alembert
principle, we derive the Lagrange-d'Alembert equations of motion,
prove the conservation of energy and state a Noether's theorem. We
also derive local expressions for the dynamics of nonholonomic
Lagrangian systems, which are further simplified by the choice of
a convenient basis of $D$.  As an illustration, we consider the
class of nonholonomic mechanical systems. For such systems, the
Lagrangian $L$ is the polar form of a bundle metric on $E$ minus a
potential function on $M$. In Section~\ref{sec:regularity}, we
perform the analysis of the existence and uniqueness of solutions
of constrained systems on general Lie algebroids, and extend the
results in~\cite{BaSn,CaLeMaMa,CaLeMa,CoLe,LeMa} for constrained
systems evolving on tangent bundles. We obtain several
characterizations for the regularity of a nonholonomic system, and
prove that a nonholonomic system of mechanical type is always
regular.  The constrained dynamics can be obtained by projecting
the unconstrained dynamics in two different ways. Under the first
projection, we develop a distributional approach analogous to that
in~\cite{BaSn}, see also~\cite{MeLa}.  Using the second
projection, we introduce the nonholonomic bracket. The evolution
of any observable can be measured by computing its bracket with
the energy of the system. Section~\ref{sec:reduction} is devoted
to studying the reduction of the dynamics under symmetry.  Our
approach follows the ideas developed in~\cite{CeMaRa}, who defined
a minimal subcategory of the category of Lie algebroids which is
stable under Lagrangian reduction.  We study the behavior of the
different geometric objects introduced under morphisms of Lie
algebroids, and show that fiberwise surjective morphisms induce
consistent reductions of the dynamics. This result covers, but
does not reduce to, the usual case of reduction of the dynamics by
a symmetry group.  In accordance with the philosophy of the paper,
we study first the unconstrained dynamics case, and obtain later
the results for the constrained dynamics using projections.  A
(Poisson) reduction by stages procedure can also be developed
within this formalism.  It should be noticed that the reduction
under the presence of a Lie group of symmetries $G$ is performed
in two steps: first we reduce by a normal subgroup $N$ of $G$, and
then by the residual group. In Section~\ref{momentum-equation}, we
prove a general version of the momentum equation introduced
in~\cite{BlKrMaMu}. In Section~\ref{examples}, we show some
interesting examples and in Section~\ref{nonlinear}, we extend
some of the results previously obtained for linear constraints to
the case of nonlinear constraints. The paper ends with our
conclusions and a description of future research directions.


\section{Preliminaries}
\label{preliminaries}

In this section we recall some well-known facts concerning the
geometry of Lie algebroids.  We refer the reader
to~\cite{CaWe,HiMa,Mackenzie} for details about Lie groupoids, Lie
algebroids and their role in differential geometry.

\subsection{Lie algebroids}
Let $M$ be an $n$-dimensional manifold and let $\map{\tau}{E}{M}$
be a vector bundle. A vector bundle map $\map{\rho}{E}{TM}$ over
the identity is called an \emph{anchor map}. The vector bundle $E$
together with an anchor map $\rho$ is said to be an \emph{anchored
vector bundle} (see~\cite{PoPo}). A structure of \emph{Lie
algebroid} on $E$ is given by a Lie algebra structure on the
$\cinfty{M}$-module of sections of the bundle,
$(\Sec{E},[\cdot\,,\cdot])$, together with an anchor map,
satisfying the compatibility condition
\[
[\sigma,f\eta] = f[\sigma,\eta] + \bigl( \rho(\sigma)f \bigr) \eta
.
\]
Here $f$ is a smooth function on $M$, $\sigma$, $\eta$ are sections of
$E$ and $\rho(\sigma)$ denotes the vector field on $M$ given by
$\rho(\sigma)(m)=\rho(\sigma(m))$. From the compatibility condition
and the Jacobi identity, it follows that the map
$\sigma\mapsto\rho(\sigma)$ is a Lie algebra homomorphism from the set
of sections of $E$, $\Sec{E}$, to the set of vector fields on $M$,
$\mathfrak{X}(M)$.

In what concerns Mechanics, it is convenient to think of a Lie
algebroid $\map{\rho}{E}{TM}$, and more generally an anchored vector
bundle, as a substitute of the tangent bundle of $M$. In this way, one
regards an element $a$ of $E$ as a generalized velocity, and the
actual velocity $v$ is obtained when applying the anchor to $a$, i.e.,
$v=\rho(a)$. A curve $\map{a}{[t_0,t_1]}{E}$ is said to be
\emph{admissible} if $\dot{m}(t)=\rho(a(t))$, where $m(t)=\tau(a(t))$
is the base curve. We will denote by $\Adm{E}$ the space of admissible
curves on $E$.

Given local coordinates $(x^i)$ in the base manifold $M$ and a
local basis $\{e_\alpha\}$ of sections of $E$, we have local
coordinates $(x^i,y^\alpha)$ in $E$. If $a\in E$ is an element in
the fiber over $m\in M$, then we can write $a=y^\alpha
e_\alpha(m)$ and thus the coordinates of $a$ are $(m^i,y^\alpha)$,
where $m^i$ are the coordinates of the point $m$. The anchor map
is locally determined by the local functions $\rho^i_\alpha$ on
$M$ defined by $\rho(e_\alpha)=\rho^i_\alpha(\partial/\partial
x^i)$. In addition, for a Lie algebroid, the Lie bracket is
determined by the functions $C^\gamma_{\alpha\beta}$ defined by
$[e_\alpha,e_\beta]=C^\gamma_{\alpha\beta}e_\gamma$. The functions
$\rho^i_\alpha$ and $C^\gamma_{\alpha\beta}$ are called \emph{the
structure functions} of the Lie algebroid in this coordinate
system.  They satisfy the following relations
\begin{align*}
  \rho^j_\alpha\pd{\rho^i_\beta}{x^j} -
  \rho^j_\beta\pd{\rho^i_\alpha}{x^j} = \rho^i_\gamma
  C^\gamma_{\alpha\beta} \quand
  \sum_{\mathrm{cyclic}(\alpha,\beta,\gamma)} \left[\rho^i_\alpha\pd{
      C^\nu_{\beta\gamma}}{x^i} + C^\mu_{\beta\gamma}
    C^\nu_{\alpha\mu}\right]=0,
\end{align*}
which are called \emph{the structure equations} of the Lie
algebroid.

\subsection{Exterior differential}
The anchor $\rho$ allows to define the differential of a function on
the base manifold with respect to an element $a\in E$. It is given by
\[
df(a)=\rho(a)f.
\]
It follows that the differential of $f$ at the point $m\in M$ is an
element of $E_m^*$.  Moreover, a structure of Lie algebroid on $E$
allows to extend the differential to sections of the bundle
$\ext[p]{E}$, which will be called $p$-sections or just $p$-forms. If
$\omega\in\Sec{\ext[p]{E}}$, then $d\omega\in\Sec{\ext[p+1]{E}}$ is
defined by
\begin{align*}
  d\omega(\sigma_0,\sigma_1,\ldots,\sigma_p) &=
  \sum_i(-1)^i\rho(\sigma_i)(
  \omega(\sigma_0,\ldots,\widehat{\sigma_i},\ldots,\sigma_p))\\
  &\qquad{}+ \sum_{i<j}(-1)^{i+j}
  \omega([\sigma_i,\sigma_j],\sigma_0,\ldots,
  \widehat{\sigma_i},\ldots,\widehat{\sigma_j},\ldots,\sigma_p).
\end{align*}
It follows that $d$ is a cohomology operator, that is, $d^2=0$.
Locally the exterior differential is determined by
\[
dx^i=\rho^i_\alpha e^\alpha \qquand
de^\gamma=-\frac{1}{2}C^\gamma_{\alpha\beta}e^\alpha\wedge e^\beta.
\]
Throughout this paper, the symbol $d$ will refer to the exterior
differential on the Lie algebroid $E$ and not to the ordinary exterior
differential on a manifold. Of course, if $E=TM$, then both exterior
differentials coincide.

The usual Cartan calculus extends to the case of Lie algebroids
(see~\cite{Mackenzie, Nijenhuis}). For every section $\sigma$ of $E$
we have a derivation $i_\sigma$ (contraction) of degree $-1$ and a
derivation $d_\sigma=i_\sigma\circ d+d\circ i_\sigma$ (Lie derivative)
of degree $0$. Since $d^2=0$, we have that $d_\sigma\circ d=d\circ
d_\sigma$.

\subsection{Morphisms}
Let $\map{\tau}{E}{M}$ and $\map{\tau'}{E'}{M'}$ be two anchored
vector bundles, with anchor maps $\map{\rho}{E}{TM}$ and
$\map{\rho'}{E'}{TM'}$.  A vector bundle map $\map{\Phi}{E}{E'}$ over
a map $\map{\varphi}{M}{M'}$ is said to be \emph{admissible} if it
maps admissible curves onto admissible curves, or equivalently
$T\varphi\circ\rho = \rho'\circ\Phi$. If $E$ and $E'$ are Lie
algebroids, then we say that $\Phi$ is a \emph{morphism} if $\Phi\pb
d\theta=d\Phi\pb\theta$ for every $\theta\in\Sec{\ext{E'}}$. It is
easy to see that morphisms are admissible maps.

In the above expression, the pullback $\Phi\pb\beta$ of a $p$-form
$\beta$ is defined by $$
(\Phi\pb\beta)_m(a_1,a_2,\ldots,a_p)=
\beta_{\varphi(m)}\bigl(\Phi(a_1),\Phi(a_2),\ldots,\Phi(a_p)\bigr), $$
for every $a_1,\ldots,a_p\in E_m$. For a function $f\in\cinfty{M'}$
(i.e., for $p=0$), we just set $\Phi\pb f=f\circ\varphi$.

Let $(x^i)$ and $(x'{}^i)$ be local coordinate systems on $M$ and
$M'$, respectively. Let $\{e_\alpha\}$ and $\{e'_\alpha\}$ be local
basis of sections of $E$ and $E'$, respectively, and $\{e^\alpha\}$
and $\{e'{}^\alpha\}$ the corresponding dual basis.  The bundle map
$\Phi$ is determined by the relations $\Phi\pb x'{}^i = \phi^i(x)$ and
$\Phi\pb e'{}^\alpha = \phi^\alpha_\beta e^\beta$ for certain local
functions $\phi^i$ and $\phi^\alpha_\beta$ on $M$. Then, $\Phi$ is
admissible if and only if
\[
  \rho^j_\alpha\pd{\phi^i}{x^j}=\rho'{}^i_\beta\phi^\beta_\alpha.
\]
The map $\Phi$ is a morphism of Lie algebroids if and only if, in
addition to the admissibility condition above, one has
\[
\phi^\beta_\gamma C^\gamma_{\alpha\delta} =
\left(\rho^i_\alpha\pd{\phi^\beta_\delta}{x^i} -
  \rho^i_\delta\pd{\phi^\beta_\alpha}{x^i}\right) +
C'{}^\beta_{\theta\sigma}\phi^\theta_\alpha\phi^\sigma_\delta.
\]
In these expressions, $\rho^i_\alpha$, $C^\alpha_{\beta\gamma}$are the
local structure functions on $E$ and $\rho'{}^i_\alpha$,
$C'{}^\alpha_{\beta\gamma}$ are the local structure functions on $E'$.

\subsection{Prolongation of a fibered manifold with respect to a Lie
  algebroid}
Let $\map{\pi}{P}{M}$ be a fibered manifold with base manifold $M$.
Thinking of $E$ as a substitute of the tangent bundle of $M$, the
tangent bundle of $P$ is not the appropriate space to describe
dynamical systems on $P$.  This is clear if we note that the
projection to $M$ of a vector tangent to $P$ is a vector tangent to
$M$, and what one would like instead is an element of $E$, the `new'
tangent bundle of $M$.

A space which takes into account this restriction is the
\emph{$E$-tangent bundle} of $P$, also called the \emph{prolongation}
of $P$ with respect to $E$, which we denote by $\prol[E]{P}$
(see~\cite{LeMaMa,LMLA,MaMeSa,PoPo}). It is defined as the vector
bundle $\map{\tau^E_P}{\prol[E]{P}}{P}$ whose fiber at a point $p\in
P_m$ is the vector space
\[
\prol[E]{P}[p] =\set{(b,v)\in E_m\times T_pP}{\rho(b)=T_p\pi(v)}.
\]
We will frequently use the redundant notation $(p,b,v)$ to denote the
element $(b,v)\in\prol[E]{P}[p]$. In this way, the map $\tau^E_P$ is
just the projection onto the first factor. The anchor of $\prol[E]{P}$
is the projection onto the third factor, that is, the map
$\map{\rho^1}{\prol[E]{P}}{TP}$ given by $\rho^1(p,b,v)=v$. The
projection onto the second factor will be denoted by
$\map{\prol{\pi}}{\prol[E]{P}}{E}$, and it is a vector bundle map over
$\pi$. Explicitly $\prol{\pi}(p,b,v)=b$.

An element $z\in\prol[E]{P}$ is said  to be vertical if it
projects to zero, that is $\prol{\pi}(z)=0$. Therefore it is of
the form $(p,0,v)$, with $v$ a vertical vector tangent to $P$ at
$p$.

Given local coordinates $(x^i,u^A)$ on $P$ and a local basis
$\{e_\alpha\}$ of sections of $E$, we can define a local basis
$\{\X_\alpha,\V_A\}$ of sections of $\prol[E]{P}$ by
\[
\X_\alpha(p)
=\Bigl(p,e_\alpha(\pi(p)),\rho^i_\alpha\pd{}{x^i}\at{p}\Bigr) \qquand
\V_A(p) = \Bigl(p,0,\pd{}{u^A}\at{p}\Bigr).
\]
If $z=(p,b,v)$ is an element of $\prol[E]{P}$, with $b=z^\alpha
e_\alpha$, then $v$ is of the form $v=\rho^i_\alpha
z^\alpha\pd{}{x^i}+v^A\pd{}{u^A}$, and we can write
\[
z=z^\alpha\X_\alpha(p)+v^A\V_A(p).
\]
Vertical elements are linear combinations of $\{\V_A\}$.

The anchor map $\rho^1$ applied to a section $Z$  of $\prol[E]{P}$
with local expression $Z = Z^\alpha\X_\alpha+V^A\V_A$ is the
vector field on $P$ whose coordinate expression is
\[
\rho^1(Z) = \rho^i_\alpha Z^\alpha \pd{}{x^i} + V^A\pd{}{u^A}.
\]

If $E$ carries a Lie algebroid structure, then so does $\prol[E]{P}$.
The associated Lie bracket can be easily defined in terms of
projectable sections, so that $\prol{\pi}$ is a morphism of Lie
algebroids. A section $Z$ of $\prol[E]{P}$ is said to be projectable
if there exists a section $\sigma$ of $E$ such that $\prol{\pi}\circ
Z=\sigma\circ\pi$. Equivalently, a section $Z$ is projectable if and
only if it is of the form $Z(p)=(p,\sigma(\pi (p)),X(p))$, for some
section $\sigma$ of $E$ and some vector field $X$ on $E$ (which
projects to $\rho(\sigma)$). The Lie bracket of two projectable
sections $Z_1$ and $Z_2$ is then given by
\[
[Z_1,Z_2](p)=(p,[\sigma_1,\sigma_2](m),[X_1,X_2](p)), \qquad p \in
P,\,\;\;\; m=\pi(p).
\]
It is easy to see that $[Z_1,Z_2](p)$ is an element of
$\prol[E]{P}[p]$ for every $p\in P$.
Since any section of
$\prol[E]{P}$ can be locally written as a linear combination of
projectable sections, the definition of the Lie bracket for
arbitrary sections of $\prol[E]{P}$ follows.

The Lie brackets of the elements of the basis are
\[ [\X_\alpha,\X_\beta]= C^\gamma_{\alpha\beta}\:\X_\gamma, \qquad
[\X_\alpha,\V_B]=0 \qquand [\V_A,\V_B]=0,
\]
and the exterior differential is determined by
\begin{align*}
  &dx^i=\rho^i_\alpha \X^\alpha,
  &&du^A=\V^A,\\
  &d\X^\gamma=-\frac{1}{2}C^\gamma_{\alpha\beta}\X^\alpha\wedge\X^\beta,
  &&d\V^A=0,
\end{align*}
where $\{\X^\alpha,\V^A\}$ is the dual basis corresponding to
$\{\X_\alpha,\V_A\}$.

\subsection{Prolongation of a map}
Let $\map{\Psi}{P}{P'}$ be a fibered map from the fibered manifold
$\map{\pi}{P}{M}$ to the fibered manifold $\map{\pi'}{P'}{M'}$ over a
map $\map{\varphi}{M}{M'}$. Let $\map{\Phi}{E}{E'}$ be an admissible
map from $\map{\tau}{E}{M}$ to $\map{\tau'}{E'}{M'}$ over the same map
$\varphi$. The prolongation of $\Phi$ with respect to $\Psi$ is the
mapping $\map{\prol[\Phi]{\Psi}}{\prol[E]{P}}{\prol[E']{P'}}$ defined
by
\[
\prol[\Phi]{\Psi}(p,b,v) =(\Psi(p),\Phi(b),(T_p\Psi)(v)).
\]
It is clear from the definition that $\prol[\Phi]{\Psi}$ is a
vector bundle map from $\map{\tau^E_P}{\prol[E]{P}}{P}$ to
$\map{\tau^{E'}_{P'}}{\prol[E']{P'}}{P'}$ over $\Psi$. Moreover,
in ~\cite{CFTLAMF} it is proved the following result.

\begin{proposition}
  The map $\prol[\Phi]{\Psi}$ is an admissible map. Moreover,
  $\prol[\Phi]{\Psi}$ is a morphism of Lie algebroids if and only if
  $\Phi$ is a morphism of Lie algebroids.
\end{proposition}

Given local coordinate systems $(x^i)$ on $M$ and $(x'{}^i)$ on $M'$,
local adapted coordinates $(x^i,u^A)$ on $P$ and $(x'{}^i,u'{}^A)$ on
$P'$ and a local basis of sections $\{e_\alpha\}$ of $E$ and
$\{e'_\alpha\}$ of $E'$, the maps $\Phi$ and $\Psi$ are determined by
$\Phi\pb e'{}^\alpha=\Phi^\alpha_\beta e^\beta$ and
$\Psi(x,u)=(\phi^i(x),\psi^A(x,u))$. Then the action of
$\prol[\Phi]{\Psi}$ is given by
\begin{align*}
  (\prol[\Phi]{\Psi})\pb\X'{}^\alpha
  &= \Phi_\beta^\alpha\X^\beta,\\
  (\prol[\Phi]{\Psi})\pb\V'{}^A
  &=\rho^i_\alpha\pd{\psi^A}{x^i}\X^\alpha+\pd{\psi^A}{u^B}\V^B.
\end{align*}

We finally mention that the composition of prolongation maps is the
prolongation of the composition. Indeed, let $\Psi'$ be another bundle
map from $\map{\pi'}{P'}{M'}$ to another bundle
$\map{\pi''}{P''}{M''}$ and $\Phi'$ be another admissible map from
$\map{\tau'}{E'}{M'}$ to $\map{\tau''}{E''}{M''}$ both over the same
base map. Since $\Phi$ and $\Phi'$ are admissible maps then so is
$\Phi'\circ\Phi$, and thus we can define the prolongation of
$\Psi'\circ\Psi$ with respect to $\Phi'\circ\Phi$.  We have that
$\prol[\Phi'\circ\Phi]{(\Psi'\circ\Psi)}
=(\prol[\Phi']{\Psi'})\circ(\prol[\Phi]{\Psi})$.

In the particular case when the bundles $P$ and $P'$ are just $P=E$
and $P'=E'$, whenever we have an admissible map $\map{\Phi}{E}{E'}$ we
can define the prolongation of $\Phi$ along $\Phi$ itself, by
$\prol[\Phi]{\Phi}(a,b,v)=(\Phi(a),\Phi(b),T\Phi(v))$. From the result
above, we have that $\prol[\Phi]{\Phi}$ is a Lie algebroid morphism if
and only if $\Phi$ is a Lie algebroid morphism. In coordinates we
obtain
\begin{align*}
  (\prol[\Phi]{\Phi})\pb\X'{}^\alpha
  &= \Phi_\beta^\alpha\X^\beta,\\
  (\prol[\Phi]{\Phi})\pb\V'{}^\alpha
  &=\rho^i_\beta\pd{\Phi^\alpha_\gamma}{x^i}y^\gamma\X^\beta +
  \Phi^\alpha_\beta\V^\beta,
\end{align*}
where $(x^i,y^\gamma)$ are the corresponding fibred coordinates on
$E$.  From this expression it is clear that $\prol[\Phi]{\Phi}$ is
fiberwise surjective if and only if $\Phi$ is fiberwise surjective.


\subsection{Lagrangian Mechanics}
In~\cite{LMLA} (see also \cite{PoPo}) a geometric formalism for
Lagrangian Mechanics on Lie algebroids was defined. Such a
formalism is similar to Klein's formalism \cite{Klein} in standard
Lagrangian mechanics and it is developed in the prolongation
$\TEE$ of a Lie algebroid $E$ over itself. The canonical
geometrical structures defined on $\TEE$ are the following:
\begin{itemize}
\item The \emph{vertical lift} $\map{\xi\spV}{\tau^*E}{\TEE}$ given by
  $\xi\spV(a,b)=(a,0,b\spV_a)$, where $b\spV_a$ is the vector tangent
  to the curve $a+tb$ at $t=0$,
\item The \emph{vertical endomorphism} $\map{S}{\TEE}{\TEE}$ defined
  as follows:
  \[
  S(a,b,v)=\xi\spV(a,b)=(a,0,b_a\spV),
  \]
\item The \emph{Liouville section} which is the vertical section
  corresponding to the Liouville dilation vector field:
  \[
  \Delta(a)=\xi\spV(a,a)=(a,0,a_a\spV).
  \]
\end{itemize}

A section $\Gamma$ of $\TEE$ is said to be a \sode\ section if
$S\Gamma = \Delta$.

Given a Lagrangian function $L\in\cinfty{E}$ we define the
\emph{Cartan 1-form} $\theta_L$ and the \emph{Cartan 2-form}
$\omega_L$ as the forms on $\prol[E]{E}$ given by
\begin{equation}
  \label{Cartan-forms}
  \theta_L=S^*(dL)\qquand \omega_L=-d\theta_L.
\end{equation}
The real function $E_{L}$ on $E$ defined by $E_{L} = d_{\Delta}L -
L$ is the \emph{energy function} of the Lagrangian system.

By a solution of the Lagrangian system (a solution of the
\emph{Euler-Lagrange equations}) we mean a \sode\ section $\Gamma$
of $\TEE$ such that
\begin{equation}
  \label{Euler-Lagrange}
  i_\Gamma\omega_L-dE_L=0.
\end{equation}

The local expressions for the vertical endomorphism, the Liouville
section, the Cartan $2$-form and the Lagrangian energy are
\begin{equation}\label{endverlo}
  S\X_{\alpha} = \V_{\alpha}, \makebox[.3cm]{} S\V_{\alpha} = 0,
  \makebox[.3cm]{} \mbox{ for all } \alpha,
\end{equation}
\begin{equation}
  \label{Lioulo} \Delta = y^{\alpha}\V_{\alpha},
\end{equation}
\begin{equation}
  \label{omegaL}
  \omega_L
  =\pd{^2L}{y^\alpha\partial y^\beta}\X^\alpha\wedge \V^\beta
  +\frac{1}{2}\left(
    \pd{^2L}{x^i\partial y^\alpha}\rho^i_\beta-\pd{^2L}{x^i\partial
      y^\beta}\rho^i_\alpha+\pd{L}{y^\gamma}C^\gamma_{\alpha\beta}
  \right)\X^\alpha\wedge \X^\beta,
\end{equation}
\begin{equation}
  \label{EL}
  E_L=\pd{L}{y^\alpha}y^\alpha-L.
\end{equation}

Thus, a \sode\ $\Gamma$ is a section of the form
\[
\Gamma=y^\alpha\X_\alpha+f^\alpha\V_\alpha.
\]
The \sode\ $\Gamma$ is a solution of the Euler-Lagrange equations
if and only if the functions $f^\alpha$ satisfy the linear
equations
\begin{equation}
  \label{free-forces} \pd{^2L}{y^\beta\partial
    y^\alpha}f^\beta+\pd{^2L}{x^i\partial y^\alpha}\rho^i_\beta
  y^\beta +\pd{L}{y^\gamma}C^\gamma_{\alpha\beta}y^\beta
  -\rho^i_\alpha\pd{L}{x^i} =0, \mbox{ for all } \alpha.
\end{equation}
The \emph{Euler-Lagrange differential equations} are the
differential equations for the integral curves of the vector field
$\rho^1(\Gamma)$, where the section $\Gamma$ is the solution of
the Euler-Lagrange equations. Thus, these equations may be written
as
$$\dot{x}^i=\rho_\alpha^iy^\alpha,\;\;\;\; \frac{d}{dt}(\frac{\partial
  L}{\partial y^\alpha})-\rho_\alpha^i\frac{\partial L}{\partial x^i}
+ \frac{\partial L}{\partial y^\gamma}C_{\alpha\beta}^\gamma
y^\beta=0.$$

In other words, if $\delta L: \Adm{E}\to E^*$ is the
\emph{Euler-Lagrange operator}, which locally reads
\[
\delta L=(\frac{d}{dt}(\frac{\partial L}{\partial y^\alpha}) +
C_{\alpha\beta}^\gamma y^\beta \frac{\partial L}{\partial
  y^\gamma}-\rho_\alpha^i\frac{\partial L}{\partial x^i})e^\alpha,
\]
where $\{e^\alpha\}$ is the dual basis of $\{e_\alpha\}$, then the
Euler-Lagrange differential equations read
\[
\delta L=0.
\]
The function $L$ is said to be \emph{regular Lagrangian} if
$\omega_{L}$ is regular at every point as a bilinear map. In such
a case, there exists a unique section $\Gamma_{L}$ of $\TEE$ which
satisfies the equation
\[
i_{\Gamma_{L}}\omega_{L} - dE_{L} = 0.
\]
Note that from (\ref{endverlo}), (\ref{Lioulo}), (\ref{omegaL})
and (\ref{EL}), it follows that
\begin{equation}
  \label{2.4'} i_{SX} \omega_{L} = -S^*(i_{X}\omega_{L}),
  \makebox[.3cm]{} i_{\Delta}\omega_{L} = -S^*(dE_{L}),
\end{equation}
for $X \in \Sec{{\prol[E]{E}}}$. Thus, using (\ref{2.4'}), we
deduce that
\[
i_{S\Gamma_{L}}\omega_{L} = i_{\Delta}\omega_{L}
\]
which implies that $\Gamma_{L}$ is a \sode\ section. Therefore,
for a regular Lagrangian function $L$ we will say that the
dynamical equations (\ref{Euler-Lagrange}) are just the
Euler-Lagrange equations.

On the other hand, the vertical distribution is isotropic with
respect to $\omega_L$, see~\cite{LeMaMa}. This fact implies that
the contraction of $\omega_L$ with a vertical vector is a
semibasic form. This property allows us to define a symmetric
2-tensor $\GL$ along $\tau$ by
\begin{equation}
  \label{GL}
  \GL[a](b,c)=\omega_L(\tilde{b},c_a\spV),
\end{equation}
where $\tilde{b}$ is any element in $\TEE[a]$ which projects to $b$,
i.e., $\prol{\tau}(\tilde{b})=b$, and $a \in E$. In coordinates
$\GL=W_{\alpha\beta}e^\alpha\otimes e^\beta$, where the matrix
$W_{\alpha\beta}$ is given by
\begin{equation}
  \label{HessianL}
  W_{\alpha\beta}=\pd{^2L}{y^\alpha\partial y^\beta}.
\end{equation}
It is easy to see that the Lagrangian $L$ is regular if and only
the matrix $W$ is regular at every point, that is, if the tensor
$\GL$ is regular at every point. By the kernel of $\GL$ at a point
$a$ we mean the vector space
\[
\Ker\GL[a]=\set{b\in E_{\tau(a)}}{\GL[a](b,c)=0\text{ for all }c\in
  E_{\tau(a)}}.
\]
In the case of a regular Lagrangian, the Cartan 2-section $\omega_L$
is symplectic (non-degenerate and $d$-closed) and the vertical
subbundle is Lagrangian. It follows that a 1-form is semi-basic if and
only if it is the contraction of $\omega_L$ with a vertical element.

Finally, we mention that the \emph{complete lift} $\sigma\spC$ of
a section $\sigma\in\Sec{E}$ is the section of $\TEE$
characterized by the two following properties:
\begin{enumerate}
\item projects to $\sigma$, i.e.,
  $\prol{\tau}\circ\sigma\spC=\sigma\circ\tau$,
\item $d_{\sigma\spC}\hat{\mu}=\widehat{d_\sigma\mu}$,
\end{enumerate}
where by $\hat{\alpha}\in\cinfty{E}$ we denote the linear function
associated to a 1-section $\alpha\in\Sec{E^*}$. Note that
\begin{equation}\label{SODEcompl}
\Gamma \mbox{ {\sc sode} section, } \sigma \in \Sec{E} \Rightarrow
S[\sigma^{c}, \Gamma ] = 0,
\end{equation}

\vspace{-.5cm}

\begin{equation}\label{Vertcompl}
S \gamma = 0,\;\;\; \sigma \in \Sec{E} \Rightarrow S[\sigma^{c},
\gamma ] = 0.
\end{equation}


\section{Linearly constrained Lagrangian systems}
\label{linear}

Nonholonomic systems on Lie algebroids were introduced in~\cite{CoMa}.
This class of systems includes, as particular cases, standard
nonholonomic systems defined on the tangent bundle of a manifold and
systems obtained by the reduction of the action of a symmetry group.
The situation is similar to the non-constrained case, where the
general equation $\delta L=0$ comprises as particular cases the
standard Lagrangian Mechanics, Lagrangian Mechanics with symmetry,
Lagrangian systems with holonomic constraints, systems on semi-direct
products and systems evolving on Lie algebras, see e.g.,~\cite{LMLA}.

We start with a free Lagrangian system on a Lie algebroid $E$. As
mentioned above, these two objects can describe a wide class of
systems. Now, we plug in some nonholonomic linear constraints
described by a subbundle $D$ of the bundle $E$ of admissible
directions. If we impose to the solution curves $a(t)$ the condition
to stay on the manifold $D$, we arrive at the equations $\delta
L_{a(t)}=\lambda(t)$ and $a(t)\in D$, where the constraint force
$\lambda(t)\in E^*_{\tau(a(t))}$ is to be determined.  In the tangent
bundle geometry case ($E=TM$), the d'Alembert principle establishes
that the mechanical work done by the constraint forces vanishes, which
implies that $\lambda$ takes values in the annihilator of the
constraint manifold $D$. Therefore, in the case of a general Lie
algebroid, the natural equations one should pose are (see~\cite{CoMa})
\[
\delta L_{a(t)}\in\Do[\tau(a(t))]\qquand a(t)\in D.
\]
In more explicit terms, we look for curves $a(t)$ on $E$ such that
\begin{itemize}
\item[--] they are admissible, $\rho(a(t))=\dot{m}(t)$, where
  $m=\tau\circ a$,
\item[--] they stay in $D$, $a(t)\in D_{m(t)}$,
\item[--] there exists $\lambda(t)\in \Do[m(t)]$ such that $\delta
  L_{{a}(t)}=\lambda(t)$.
\end{itemize}
If $a(t)$ is one of such curves, then $(a(t),\dot{a}(t))$ is a
curve in $\TEE$. Moreover, since $a(t)$ is in $D$, we have
$\dot{a}(t)$ is tangent to $D$, that is,
$(a(t),\dot{a}(t))\in\TDD$. Under some regularity conditions (to
be made precise later on), we may assume that the above curves are
integral curves of a section $\Gamma$, which as a consequence will
be a \sode\ section taking values in $\TDD$. Based on these
arguments, we may reformulate geometrically our problem as the
search for a \sode~$\Gamma$ (defined at least on a neighborhood of
$D$) satisfying $(i_\Gamma\omega_L-dE_L)_a\in\tDo[\tau(a)]$ and
$\Gamma(a)\in {\mathcal T}_{a}^{D}D$, at every point $a\in D$. In
the above expression $\tDo$ is the pullback of $\Do$ to $\TEE$,
that is, $\alpha\in\tDo[\tau(a)]$ if and only if there exists
$\lambda\in\Do[\tau(a)]$ such that
$\alpha=\lambda\circ\prol{\tau}[a]$.
\begin{definition}
  A nonholonomically \emph{constrained Lagrangian system} on a Lie
  algebroid $E$ is a pair $(L,D)$, where $L$ is a smooth function on
  $E$, \emph{the Lagrangian}, and $i\colon D\hookrightarrow E$ is a
  smooth subbundle of $E$, known as the \emph{constraint subbundle}.
  By a solution of the nonholonomically constrained Lagrangian system
  $(L,D)$ we mean a \sode\ section $\Gamma\in\TEE$ which satisfies the
  \emph{Lagrange-d'Alembert equations}
  \begin{equation}
    \label{Lagrange-d'Alembert}
    \begin{aligned}
      &(i_\Gamma\omega_L-dE_L)|_D\in \Sec{\tDo},\\
      &\Gamma|_D\in \Sec{\TDD}.
    \end{aligned}
  \end{equation}
\end{definition}
With a slight abuse of language, we will interchangeably refer to
a solution of the constrained Lagrangian system as a section or
the collection of its corresponding integral curves.  The
restriction of the projection $\map{\tau}{E}{M}$ to $D$ will be
denoted by $\pi$, that is, $\map{\pi=\tau|_D}{D}{M}$.
\begin{remark}[Domain of definition of solutions of the
  Lagrange-d'Alembert equations]
  {\rm We want to stress that a solution of the Lagrange-d'Alembert
    equations needs to be defined only over $D$, but for practical
    purposes we consider it extended to $E$ (or just to a neighborhood
    of $D$ in $E$). We will not make any notational distinction
    between a solution on $D$ and any of its extensions. Solutions
    which coincide on $D$ will be considered as equal.
    See~\cite{GrMe,LeMa} for a more in-depth discussion.  In
    accordance with this convention, by a \sode\ on $D$ we mean a
    section of $\TDD$ which is the restriction to $D$ of some \sode\
    defined in a neighborhood of $D$. Alternatively, a \sode\ on $D$
    is a section $\Gamma$ of $\TDD$ such that
    $\prol{\tau}(\Gamma(a))=a$ for every $a\in D$.} \oprocend
\end{remark}
\begin{remark}[Holonomic constraints]
  {\rm A nonholonomically constrained Lagrangian system $(L, D)$ on a
    Lie algebroid $E$ is said to be \emph{holonomic} if $D$ is a Lie
    subalgebroid of $E$. This means that $[X, Y] \in \Sec{D}$, for $X,
    Y \in \Sec{D}$. Thus, the real function $L_{D} = L_{|D}: D \to \R$
    defines an unconstrained (free) Lagrangian system on the Lie
    algebroid $D$. Moreover, it is easy to prove that $\mathcal{I}
    \circ \Delta_{D} = \Delta \circ i$ and $\mathcal{I} \circ S_{D} =
    S \circ \mathcal{I}$, where $\mathcal{I} = {\prol[i]{i}}: \TDD \to
    \TEE$ is the prolongation of the Lie algebroid morphism $i\colon
    D\hookrightarrow E$ and $\Delta_{D}$ (respectively, $S_{D}$) is
    the Liouville section (respectively, the vertical endomorphism) of
    the Lie algebroid $\TDD$. Therefore, since $L \circ i = L_D$, we
    deduce that
    \[
    \mathcal{I}^*(\theta_{L}) = \theta_{L_{D}}, \makebox[.4cm]{}
    \mathcal{I}^*(\omega_{L}) = \omega_{L_{D}}, \makebox[.4cm]{}
    \mathcal{I}^*(dE_{L}) = dE_{L_{D}}.
    \]
    Consequently, if $\Gamma$ is a \sode\ section of $\TEE$, $a, b \in
    D$, $(b, X) \in \prol[E]{E}[a]$ and $(b, Y) \in \prol[E]{D}[a]$
    then
    \[
    (i_{\Gamma}\omega_{L} - dE_{L})(a)(b, X) = (i_{\Gamma_{|D}}
    \omega_{L_{D}} - dE_{L_{D}})(a)(b, Y) + (i_{\Gamma}\omega_{L} -
    dE_{L})(a)(0, Z),
    \]
    $(0, Z)$ being a vertical element of $\prol[E]{E}[a]$.
    
    Now, using (\ref{2.4'}), we have that $(i_{\Gamma}\omega_{L} -
    dE_{L})(a)(0, Z) = 0$ which implies that
    \[
    (i_{\Gamma}\omega_{L} - dE_{L})(a)(b, X) = (i_{\Gamma_{|D}}
    \omega_{L_{D}} - dE_{L_{D}})(a)(b, Y).
    \]
    The above facts prove that a \sode\ section $\Gamma$ of $\TEE$ is
    a solution of the holonomic Lagrangian system $(L, D)$ on $E$ if
    and only if $\Gamma_{|D}$ is a solution of the Euler-Lagrange
    equations for the (unconstrained) Lagrangian function $L_D$ on the
    Lie algebroid $D$. In other words, the holonomic Lagrangian system
    $(L, D)$ on $E$ may be considered as an unconstrained (free)
    Lagrangian system on the Lie algebroid $D$.} \oprocend
\end{remark}

Next, suppose that $(L, D)$ is a nonholonomically constrained
Lagrangian system on the Lie algebroid $E$. Then, the different spaces
we will consider are shown in the following commutative diagram
\[
\xymatrix{ &&TM\ar@{=}[r]&TM&\\
  &&D\ar[u]_{\rho_D}\ar[r]^i&E\ar[u]^\rho\\
  &TD\ar[ruu]^{T\pi}\ar[rd]_{\tau_D}&
  \TDD\ar[u]_{\prol{\pi}}\ar[r]^{\mathcal{I}}\ar[d]^{\pi^D_D}\ar[l]_{\rho^1}&
  \TEE\ar[u]^{\prol{\tau}}\ar[d]_{\tau^E_E}\ar[r]^{\rho^1} &
  TE\ar[luu]_{T\tau}\ar[dl]^{\tau_E}\\
  &&D\ar[r]^i\ar[d]_{\pi}&E\ar[d]^{\tau}&\\
  &&M\ar@{=}[r]&M }
\]

As an intermediate space in our analysis of the regularity of the
constrained systems, we will also consider $\TED$, the $E$-tangent to
$D$. The main difference between $\TED$ and $\TDD$ is that the former
has a natural Lie algebroid structure while the later does not.

The following two results are immediate consequences of the above form
of the Lagrange-d'Alembert equations.

\begin{theorem}[Conservation of energy]
  If $(L,D)$ is a constrained Lagrangian system and $\Gamma$ is a
  solution of the dynamics, then $d_\Gamma E_L=0$ (on $D$).
\end{theorem}
\begin{proof}
  Indeed, for every $a\in D$, we have $\Gamma(a)\in\TDD[a]$, so that
  $\prol{\tau}(\Gamma(a))\in D$.  Therefore $i_\Gamma\tDo=0$ and
  contracting $0=i_\Gamma(i_\Gamma\omega_L-dE_L)=-d_\Gamma E_L$ at
  every point in $D$.
\end{proof}

\begin{theorem}[Noether's theorem]
  Let $(L,D)$ be a constrained Lagrangian system which admits a unique
  \sode\ $\Gamma$ solution of the dynamics. If $\sigma$ is a section
  of $D$ such that there exists a function $f\in\cinfty{M}$ satisfying
  \[
  d_{\sigma\spC}L=\dot{f},
  \]
  then the function $F=\pai{\theta_L}{\sigma\spC}-f$ is a constant of
  the motion, that is, $d_\Gamma F=0$ (on $D$).
\end{theorem}
\begin{proof}
  Using that $\theta_L(\Gamma)=d_\Delta(L)$, we obtain
  $i_{\sigma\spC}(i_\Gamma\omega_L-dE_L) = i_{\sigma\spC}(-d_\Gamma
  \theta_L + dL)= d_{\sigma\spC}L-d_\Gamma\pai{\theta_L}{\sigma\spC} +
  \theta_L [\Gamma,\sigma\spC]$ and, since $[\Gamma,\sigma\spC]$ is
  vertical, we deduce \[ i_{\sigma^c}(i_\Gamma
  \omega_L-dE_L)=d_{\sigma^c}L-d_\Gamma\pai{\theta_L}{\sigma\spC}.\]
  Thus, taking into account that $i_{\sigma\spC}\tDo=0$, we get
  $0=d_\Gamma(\pai{\theta_L}{\sigma\spC}-f)=-d_\Gamma F$.
\end{proof}

\begin{example}[Mechanical systems with nonholonomic constraints] {\rm
    Let $\Gc:E\times_M E\to \R$ be a bundle metric on $E$. The \emph{
      Levi-Civita connection} $\nabla^\Gc$ is determined by the
    formula
  \begin{align*}
    2\Gc(\nabla^{\Gc}_\sigma\eta,\zeta)
    &=\rho(\sigma)(\Gc(\eta,\zeta)) +
    \rho(\eta)(\Gc(\sigma,\zeta))-\rho(\zeta)(\Gc(\eta,\sigma))\\
    & \qquad + \Gc(\sigma,[\zeta,\eta]) +
    \Gc(\eta,[\zeta,\sigma])-\Gc(\zeta, [\eta,\sigma]) ,
      \end{align*}
  for $\sigma,\eta,\zeta\in \Sec{E}$. The coefficients of the
  connection $\nabla^{\Gc}$ are given by
  \[
  \Gamma_{\beta\gamma}^\alpha =
  \frac{1}{2}\Gc^{\alpha\nu}([\nu,\beta;\gamma]+[\nu,\gamma; \beta] +
  [\beta,\gamma; \nu]),
  \]
  where $\Gc_{\alpha\nu}$ are the coefficients of the metric $\Gc$,
  $(\Gc^{\alpha\nu})$ is the inverse matrix of $(\Gc_{\alpha\nu})$
  and
  \[
  [\alpha,\beta;\gamma]=\frac{\partial \Gc_{\alpha\beta}}{\partial
    x^i}\rho_\gamma^i + C_{\alpha\beta}^\mu\Gc_{\mu\gamma}.
  \]

  Using the covariant derivative induced by $\nabla^{\Gc}$, one may
  introduce the notion of a geodesic of $\nabla^{\Gc}$ as follows.  An
  admissible curve $a:I\to E$ is said to be a \emph{geodesic} if
  $\nabla_{a(t)}^{\Gc}a(t)=0$, for all $t\in I$. In local coordinates,
  the conditions for being a geodesic read
  \[
  \frac{da^\gamma}{dt} + \frac{1}{2}(\Gamma_{\alpha\beta}^\gamma +
  \Gamma_{\beta\alpha}^\gamma)a^\alpha a^\beta=0,\;\;\; \mbox{for all
  }\gamma.
  \]
  The geodesics are the integral curves of a \sode\ section
  $\Gamma_{\nabla^{\Gc}}$ of $\prol[E]{E}$, which is locally given
  by
  \[
  \Gamma_{\nabla^{\Gc}}=y^\gamma{\mathcal
    X}_\gamma-\frac{1}{2}(\Gamma_{\alpha\beta}^\gamma +
  \Gamma_{\beta\alpha}^\gamma)y^\alpha y^\beta\V_\gamma.
  \]
  $\Gamma_{\nabla^{\Gc}}$ is called the \emph{geodesic flow} (for more
  details, see \cite{CoMa}).

  The class of systems that were considered in detail in~\cite{CoMa}
  is that of \emph{mechanical systems with nonholonomic
    constraints}\footnote{In fact, in~\cite{CoMa}, we considered
    controlled mechanical systems with nonholonomic constraints, that
    is, mechanical systems evolving on Lie algebroids and subject to
    some external control forces.}. The Lagrangian function $L$ is of
  mechanical type, i.e., it is of the form
  \[
  L(a)=\frac{1}{2} \Gc(a,a) - V(\tau(a)),
  \quad
  a\in E,
  \]
  with $V$ a function on $M$.

  The Euler-Lagrange section for the unconstrained system can be
  written as
  $$\Gamma_L=\Gamma_{\nabla^\Gc} - (\grad_\Gc V) \spV.$$
  In this expression, by $\grad_\Gc{V}$ we mean the section of $E$
  such that $\pai{dV(m)}{a}=\Gc(\grad_\Gc V(m),a)$, for all $m\in M$
  and all $a\in E_m$, and where we remind that $d$ is the differential
  in the Lie algebroid. The Euler-Lagrange differential equations can
  be written as
  \begin{equation}\label{Mechanical:eqs-motion}\begin{array}{ll}
      \dot{x}^i & =\rho^i_\alpha y^\alpha , \\[5pt]
      \dot{y}^\alpha &= -\displaystyle\frac{1}{2} \left(
        \Gamma^\alpha_{\beta \gamma} +\Gamma^\alpha_{\gamma\beta}\right)
      y^\beta y^\gamma - \Gc^{\alpha \beta} \rho^i_{\beta} \pd{V}{x^i}.
    \end{array}
  \end{equation}
  Alternatively, one can describe the dynamical behavior of the
  mechanical control system by means of an equation on $E$ via the
  covariant derivative. An admissible curve $a: t \mapsto a(t)$ with
  base curve $t\mapsto m(t)$ is a solution of the
  system~\eqref{Mechanical:eqs-motion} if and only if
  \begin{align}
    \label{Mechanical:eqs-motion-connection}
    \nabla^{\Gc}_{a(t)}a(t) + \grad_\Gc V(m(t)) = 0.
  \end{align}

  Note that
  \[
  \Gc (m(t))(\nabla_{a(t)}^\Gc a(t) + \grad_\Gc V(m(t)),b)=\delta
  L(a(t))(b),\;\;\;\; \mbox{ for } b\in E_{m(t)}.
  \]
  If this mechanical control system is subject to the constraints
  determined by a subbundle $D$ of $E$, we can do the following.
  Consider the orthogonal decomposition $E=D\oplus D^\perp$, and the
  associated orthogonal projectors $\map{P}{E}{D}$,
  $\map{Q}{E}{D^\perp}$. Using the fact that $\Gc(P\cdot,\cdot) =
  \Gc(\cdot,P\cdot)$, one can write the Lagrange-d'Alembert equations
  in the form
  \[
  P(\nabla^\Gc_{a(t)}a(t)) + P(\grad_\Gc V(m(t))) = 0, \qquad Q(a)=0.
  \]
  A specially well-suited form of these equations makes use of the
  \emph{constrained connection} $\cnabla$ defined by
  $\cnabla_\sigma\eta
  =P(\nabla^\Gc_\sigma\eta)+\nabla^\Gc_\sigma(Q\eta)$. In terms of
  $\cnabla$, we can rewrite this equation as $\cnabla_{a(t)}a(t) +
  P(\grad_\Gc V(m(t))) = 0$, $Q(a)=0$, where we have used the fact
  that the connection $\cnabla$ restricts to the subbundle $D$.

  Moreover, following the ideas in~\cite{Lewis}, we proved
  in~\cite{CoMa} that the subbundle $D$ is geodesically invariant for
  the connection $\cnabla$, that is, any integral curve of the spray
  $\Gamma_{\cnabla}$ associated with $\cnabla$ starting from a point
  in $D$ is entirely contained in $D$. Since the terms coming from the
  potential $V$ also belongs to $D$, we have that the constrained
  equations of motion can be simply stated as
  \begin{align}
    \label{eq:eqs-motion-connection-nh}
    \cnabla_{a(t)}a(t) +  P(\grad_\Gc V(m(t))) = 0,
    \qquad
    a(0)\in D.
  \end{align}
  Note that one can write the constrained equations of the motion as
  follows
  \[
  \dot{a}(t)=\rho^1(\Gamma_{\cnabla}(a(t))-{P}(\grad_{\Gc}V)^v(a(t)))
  \]
  and that the restriction to $D$ of the vector field
  $\rho^1(\Gamma_{\cnabla}-{P}(\grad_{\Gc}V)^v)$ is tangent to ${
    D}.$

  The coordinate expression of equations
  (\ref{eq:eqs-motion-connection-nh}) is greatly simplified if we take
  a basis $\{e_\alpha\}=\{e_a,e_A\}$ of $E$ adapted to the orthogonal
  decomposition $E=D\oplus D^\perp$, i.e., $D
  =\operatorname{span}\{e_a\}$, $D^\perp =
  \operatorname{span}\{e_A\}$. Denoting by $(y^\alpha)=(y^a,y^A)$ the
  induced coordinates, the constraint equations $Q(a)=0$ just read
  $y^A=0$. The differential equations of the motion are then
  \begin{align*}
    \dot{x}^i & = \rho^i_a y^a , \\
    \dot{y}^a & = - \frac{1}{2}
    \left(\check{\Gamma}^a_{bc}+\check{\Gamma}^a_{cb}\right)
    y^by^c-\Gc^{ab}\rho^i_b\pd{V}{x^i},  \\
    y^A &= 0,
  \end{align*}
  where $\check{\Gamma}^\alpha_{\beta \gamma}$ are the connection
  coefficients of the constrained connection $\check{\nabla}$.  } \oprocend
\end{example}

In the above example the dynamics exists and is completely determined
whatever the (linear) constraints are. As we will see in
Section~\ref{sec:regularity}, this property is lost in the general
case.

\subsection{Lagrange-d'Alembert equations in local coordinates}

Let us analyze the form of the Lagrange-d'Alembert equations in local
coordinates.  Following the example above, let us choose a special
coordinate system adapted to the structure of the problem as follows.
We consider local coordinates $(x^i)$ on an open set $\mathcal{U}$ of
$M$ and we take a basis $\{e_a\}$ of local sections of $D$ and
complete it to a basis $\{e_a,e_A\}$ of local sections of $E$ (both
defined on the open $\mathcal{U}$). In this way, we have coordinates
$(x^i,y^a,y^A)$ on $E$. In this set of coordinates, the constraints
imposed by the submanifold $D\subset E$ are simply $y^A=0$. If
$\{e^a,e^A\}$ is the dual basis of $\{e_a,e_A\}$, then a basis for the
annihilator $\Do$ of $D$ is $\{e^A\}$ and a basis for $\tDo$ is
$\X^A$.

An element $z$ of $\TED$ is of the form $z= u^\alpha\X_\alpha+z^a\V_a
= u^a\X_a+u^A\X_A+z^a\V_a$, that is, the component $\V_A$ vanishes
since $\rho^1(z)$ is a vector tangent to the manifold $D$ with
equations $y^A=0$. The projection of $z$ to $E$ is $\prol{\tau}(z)=u^a
e_a+u^A e_A$, so that the element $z$ is in $\TDD$ if and only if
$u^A=0$. In other words, an element in $\TDD$ is of the form
$z=u^a\X_a+z^a\V_a$.

Let us find the local expression of the Lagrange-d'Alembert equations
in these coordinates. We consider a section $\Gamma$ such that
$\Gamma_{|D} \in \Sec{\TDD}$, which is therefore of the form
$\Gamma=g^a\X_a+f^a\V_a$.  From the local expression~\eqref{omegaL} of
the Cartan 2-form and the local expression~\eqref{EL} of the energy
function, we get
\[
0=\pai{i_\Gamma\omega_L-dE_L}{\V_\alpha}=
-y^B\pd{^2L}{y^\alpha\partial y^B} -(y^b-g^b)\pd{^2L}{y^\alpha\partial
  y^b}.
\]
If we assume that the Lagrangian $L$ is regular, when we evaluate at
$y^A=0$, we have that $g^a=y^a$ and thus $\Gamma$ is a \sode.
Moreover, contracting with $\X_a$, after a few calculations we get
\[
0=\pai{i_\Gamma\omega_L-dE_L}{\X_a} =-\left\{
  d_\Gamma\left(\pd{L}{y^a}\right)
  +\pd{L}{y^\gamma}C^\gamma_{a\beta}y^\beta -\rho^i_a\pd{L}{x^i}
\right\},
\]
so that (again after evaluation at $y^A=0$), the functions $f^a$ are
solution of the linear equations
\begin{equation}
  \label{LD-explicit} \pd{^2L}{y^b\partial
    y^a}f^b+\pd{^2L}{x^i\partial y^a}\rho^i_by^b
  +\pd{L}{y^\gamma}C^\gamma_{ab}y^b -\rho^i_a\pd{L}{x^i} =0,
\end{equation}
where all the partial derivatives of the Lagrangian are to be
evaluated on $y^A=0$.

As a consequence, we get that there exists a unique solution of the
Lagrange-d'Alembert equations if and only if the matrix
\begin{equation}
  \label{gld-local}
  \C_{ab}(x^i,y^c)=\pd{^2L}{y^a\partial y^b}(x^i,y^c,0)
\end{equation}
is regular. Notice that $\C_{ab}$ is a submatrix of $W_{\alpha\beta}$,
evaluated at $y^A=0$ and that, as we know, if $L$ is of mechanical
type then the Lagrange-d'Alembert equations have a unique solution.
The differential equations for the integral curves of the vector field
$\rho^1(\Gamma)$ are the Lagrange-d'Alembert differential equations,
which read
\begin{equation}
  \label{LD-edo}
  \begin{aligned}
    &\dot{x}^i=\rho^i_ay^a,\\
    &\frac{d}{dt}\left(\pd{L}{y^a}\right) +
    \pd{L}{y^\gamma}C^\gamma_{ab}y^b -\rho^i_a\pd{L}{x^i}=0,\\
    &y^A=0.
  \end{aligned}
\end{equation}
Finally, notice that the contraction with $\X_A$ just gives the
components $\lambda_A=\pai{i_\Gamma\omega_L-dE_L}{\X_A}|_{y^A=0}$ of
the constraint forces $\lambda=\lambda_Ae^A$.

\begin{remark}[Equations in terms of the constrained Lagrangian]
  {\rm In some occasions, it is useful to write the equations in the
    form
    \begin{equation}\label{LD-edo2}
      \begin{aligned}
        &\dot{x}^i=\rho^i_ay^a,\\
        &\frac{d}{dt}\left(\pd{L}{y^a}\right) + \pd{L}{y^c}C^c_{ab}y^b
        -\rho^i_a\pd{L}{x^i}=-\pd{L}{y^A}C^A_{ab}y^b,\\
        &y^A=0,
      \end{aligned}
    \end{equation}
    where, on the left-hand side of the second equation, all the
    derivatives can be calculated from the value of the Lagrangian on
    the constraint submanifold $D$. In other words, we can substitute
    $L$ by the constrained Lagrangian $L_c$ defined by
    $L_c(x^i,y^a)=L(x^i,y^a,0)$.} \oprocend
\end{remark}

\begin{remark}[Lagrange-d'Alembert equations in quasicoordinates]
  {\rm A particular case of this construction is given by constrained
    systems defined in the standard Lie algebroid
    $\map{\tau_M}{TM}{M}$. In this case, the equations~\eqref{LD-edo}
    are the Lagrange-d'Alembert equations written in quasicoordinates,
    where $C^{\alpha}_{\beta\gamma}$ are the so-called Hamel's
    transpositional symbols, which obviously are nothing but the
    structure coefficients (in the Cartan's sense) of the moving frame
    $\{e_\alpha\}$, see e.g.,~\cite{EhKoMoRi,Ha}.} \oprocend
\end{remark}


\subsection{Solution of Lagrange-d'Alembert equations}
\label{sec:regularity}

\begin{assumption}
  In what follows, we will assume that the Lagrangian $L$ is regular
  at least in a neighborhood of $D$.
\end{assumption}

Let us now perform a precise global analysis of the existence and
uniqueness of the solution of Lagrange-d'Alembert equations.

\begin{definition}
  A constrained Lagrangian system $(L,D)$ is said to be
  \emph{regular} if the Lagrange-d'Alembert equations have a unique
  solution.
\end{definition}

In order to characterize geometrically those nonholonomic systems
which are regular, we define the tensor $\GLD$ as the restriction of
$\GL$ to $D$, that is, $\GLD[a](b,c)=\GL[a](b,c)$ for every $a\in D$
and every $b,c\in D_{\tau(a)}$. In coordinates adapted to $D$, we have
that the local expression of $\GLD$ is $\GLD=\C_{ab}e^a\otimes e^b$
where the matrix $\C_{ab}$ is given by equation~\eqref{gld-local}.

A second important geometric object is the subbundle
$F\subset\TEE|_D\to D$ whose fiber at the point $a\in D$ is
$F_a=\omega_L^{-1}(\tDo[\tau(a)])$. More explicitly,
\[
F_a = \set{z\in\TEE[a]}{\text{exists $\zeta\in\Do[\tau(a)]$ s.t.
    $\omega_L(z,u)=\pai{\zeta}{\prol{\tau}(u)}$ for all
    $u\in\TEE[a]$}}.
\]
From the definition, it is clear that the rank of $F$ is $\rank
{F}=\rank{\Do}=\rank{E}-\rank{D}$.

Finally, we also consider the subbundle $(\TDD)^\perp\subset\TEE|_D\to
D$, the orthogonal to $\TDD$ with respect to the symplectic form
$\omega_L$. The rank of $(\TDD)^\perp $ is
$\rank{\TDD}^\perp=\rank{\TEE} - \rank{\TDD} = 2(\rank{E} - \rank{D})
= 2 \rank{\Do}$.

The relation among these three objects is described by the following
result.
\begin{lemma}
  \label{F-TDD} The following properties are satisfied:
  \begin{enumerate}
  \item The elements in $F$ are vertical.  An element $\xi\spV(a,b)\in
    F_a$ if and only if $\GL[a](b,c)=0$ for all $c\in D_{\tau(a)}$.
  \item $(\TDD)^\perp\cap\Ver{\TEE}=F$.
  \end{enumerate}
\end{lemma}
\begin{proof}
  (1) The elements in $F$ are vertical because the elements in $\tDo$
  are semi-basic. If $\xi\spV(a,b)\in F_a$ then there exists $\zeta\in
  \Do[\tau (a)]$ such that
  $\omega_L(\xi\spV(a,b),u)=\pai{\zeta}{\prol{\tau}(u)}$ for all
  $u\in\TEE[a]$. In terms of $\GL$ and writing $c=\prol{\tau}(u)$, the
  above equation reads $-\GL[a](b,c)=\pai{\zeta}{c}$. By taking
  $u\in\prol{\tau}^{-1}(D)$, then $c$ is in $D$ and therefore
  $\GL[a](b,c)=0$ for all $c\in D_{\tau(a)}$. Conversely, if
  $\GL[a](b,c)=0$ for all $c\in D_{\tau(a)}$, then the 1-form
  $\zeta=-\GL[a](b,\ )$ is in $\Do[\tau(a)]$. Therefore
  $\omega_L(\xi\spV(a,b),u) = - \GL[a](b,\prol{\tau}(u)) =
  \pai{\zeta}{\prol{\tau}(u)})$, which is the condition for
  $\xi\spV(a,b)\in F_a$.

  \noindent (2) The condition for a vertical element $\xi\spV(a,b)$ to
  be in $(\TDD)^\perp$ is $\omega_L(\xi\spV(a,b),w)=0$ for all
  $w\in\TDD[a]$, or equivalently, $\GL[a](b,\prol{\tau}(w))=0$. The
  vector $c=\prol{\tau}(w)$ is an arbitrary element of $D_{\tau(a)}$,
  so that the above condition reads $\GL[a](b,c)=0$, for all $c\in
  D_{\tau(a)}$, which is precisely the condition for $\xi\spV(a,b)$ to
  be in $F_a$.
\end{proof}

\begin{theorem}
  \label{regularity} The following properties are equivalent:
  \begin{enumerate}
  \item The constrained Lagrangian system $(L,D)$ is regular,
  \item $\Ker\GLD=\{0\}$,
  \item $\TED \cap F=\{0\}$,
  \item $\TDD\cap(\TDD)^\perp=\{0\}$.
  \end{enumerate}
\end{theorem}
\begin{proof}{[(1)$\Leftrightarrow$(2)]}
  The equivalence between the first two conditions is clear from the
  local form of the Lagrange-d'Alembert equations~\eqref{LD-explicit},
  since the coefficients of the unknowns $f^a$ are precisely the
  components~\eqref{gld-local} of $\GLD$.

  [(2)$\Leftrightarrow$(3)] ($\Rightarrow$) Let $a\in D$ and consider
  an element $z\in \TED[a]\cap F_a$. Since the elements of $F$ are
  vertical, we have $z=\xi\spV(a,b)$ for some $b\in E_{\tau(a)}$.
  Moreover, $z\in\TED[a]$ implies that $b$ is an element in
  $D_{\tau(a)}$. On the other hand, if $z=\xi\spV(a,b)$ is in $F_a$,
  then Lemma~\ref{F-TDD} implies that $\GL[a](b,c)=0$ for all $c\in
  D_{\tau(a)}$. Thus $\GLD[a](b,c)=0$ for all $c\in D_{\tau(a)}$, from
  where $b=0$, and hence $z=0$.

  ($\Leftarrow$) Conversely, if for some $a\in D$, there exists
  $b\in\Ker\GLD[a]$ with $b\neq0$ then, using Lemma \ref{F-TDD}, we
  deduce that $z=\xi\spV(a,b)\in\TED[a]\cap F_a$ and $z\neq0$.

  [(2)$\Leftrightarrow$(4)] ($\Rightarrow$) Let $a\in D$ and consider
  an element $v\in \TDD[a]\cap(\TDD[a])^\perp$, that is,
  $\omega_L(v,w)=0$ for all $w\in\TDD[a]$. If we take $w=\xi\spV(a,b)$
  for $b\in D_{\tau(a)}$ arbitrary, then we have
  $\omega_L(v,\xi\spV(a,b)) = \GLD[a](\prol{\tau}(v),b)=0$ for all
  $b\in D_{\tau(a)}$, from where it follows that $\prol{\tau}(v)=0$.
  Thus $v$ is vertical, $v=\xi\spV(a,c)$, for some $c\in D$ and then
  $\omega_L(\xi\spV(a,c),w) = -\GLD[a](c,\prol{\tau}(w))=0$ for all
  $w\in\TDD[a]$. Therefore $c=0$ and hence $v=0$.

  ($\Leftarrow$) Conversely, if for some $a\in D$, there exists
  $b\in\Ker\GLD[a]$ with $b\neq0$, then $0\neq\xi\spV(a,b) \in \TDD[a]
  \cap(\TDD[a])^\perp$, because $\omega_L(\xi\spV(a,b),w) =
  \GLD(b,\prol{\tau}(w))=0$ for all $w\in\TDD[a]$.
\end{proof}

In the case of a constrained mechanical system, the tensor $\GL$ is
given by $\GL[a](b,c) = \Gc_{\tau(a)}(b,c)$, so that it is positive
definite at every point. Thus the restriction to any subbundle $D$ is
also positive definite and hence regular. Thus, nonholonomic
mechanical systems are always regular.

\begin{proposition}
  Conditions (3) and (4) in Theorem~\ref{regularity} are equivalent,
  respectively, to
  \begin{itemize}
  \item[(3')] $\TEE|_D=\TED\oplus F$,
  \item[(4')] $\TEE|_D=\TDD\oplus(\TDD)^\perp$.
  \end{itemize}
\end{proposition}
\begin{proof}
  The equivalence between (4) and (4') is obvious, since we are
  assuming that the free Lagrangian is regular, i.e., $\omega_L$ is
  symplectic. The equivalence of (3) and (3') follows by computing the
  dimension of the corresponding spaces. The ranks of $\TEE$, $\TED$
  and $F$ are
  \begin{align*}
    &\rank {\TEE} = 2 \rank{E} , \\
    &\rank{\TED} = \rank{E} +\rank{D} , \\
    &\rank{F} = \rank{\Do} = \rank{E} - \rank{D}.
  \end{align*}
  Thus $\rank{\TEE}=\rank{\TED} + \rank{F}$, and the result follows.
\end{proof}

\subsection{Projectors}\label{Projectors}
We can express the constrained dynamical section in terms of the free
dynamical section by projecting to the adequate space, either $\TED$
or $\TDD$, according to each of the above decompositions of $\TEE|_D$.
Of course, both procedures give the same result.

\subsubsection*{Projection to $\TED$}
Assuming that the constrained system is regular, we have a direct sum
decomposition
\[
\TEE[a]=\TED[a]\oplus F_a,
\]
for every $a\in D$, where we recall that the subbundle $F\subset\TED$
is defined by $F=\omega_L^{-1}(\tDo)$, or equivalently
$\tDo=\omega_L(F)$.

Let us denote by $P$ and $Q$ the complementary projectors defined by
this decomposition, that is,
\[
\map{P_a}{\TEE[a]}{\TED[a]} \quand \map{Q_a}{\TEE[a]}{F_a}, \mbox{
  for all }a\in D.
\]
Then we have,
\begin{theorem}\label{projection-external}
  Let $(L,D)$ be a regular constrained Lagrangian system and let
  $\Gamma_L$ be the solution of the free dynamics, i.e.,
  $i_{\Gamma_L}\omega_L=dE_L$. Then the solution of the constrained
  dynamics is the \sode\ $\Gamma_{(L, D)}$ obtained by projection
  $\Gamma_{(L, D)}=P(\Gamma_L|_D)$.
\end{theorem}
\begin{proof}
  Indeed, if we write $\Gamma_{(L, D)}(a) = \Gamma_L(a)-Q(\Gamma_L(a))$ for
  $a\in D$, then we have
  \[
  i_{\Gamma_{(L, D)}(a)}\omega_L-dE_L(a) =
  i_{\Gamma_L(a)}\omega_L-i_{Q(\Gamma_L(a))}\omega_L-dE_L(a) =
  -i_{Q(\Gamma_L(a))}\omega_L\in\tDo[\tau(a)] ,
  \]
  which is an element of $\tDo[\tau(a)]$ because $Q(\Gamma_L(a))$ is
  in $F_a$. Moreover, since $\Gamma_L$ is a \sode\ and $Q(\Gamma_L)$
  is vertical (since it is in $F$), we have that $\Gamma_{(L, D)}$ is also a
  \sode.
\end{proof}

We consider adapted local coordinates $(x^i,y^a,y^A)$ corresponding to
the choice of an adapted basis of sections $\{e_a,e_A\},$ where
$\{e_a\}$ generate $D$. The annihilator $\Do$ of $D$ is generated by
$\{e^A\}$, and thus $\tDo$ is generated by $\{\X^A\}$. A simple
calculation shows that a basis $\{Z_A\}$ of local sections of $F$ is
given by
\begin{equation}\label{zetas}
  Z_A=\V_A-Q^a_A\V_a,
\end{equation}
where $Q^a_A = W_{Ab}\C^{ab}$ and $\C^{ab}$ are the components of the
inverse of the matrix $\C_{ab}$ given by equation~\eqref{gld-local}.
The local expression of the projector over $F$ is then
\[
Q = Z_A\otimes \V^A.
\]

If the expression of the free dynamical section $\Gamma_L$ in this
local coordinates is
\[
\Gamma_L=y^\alpha\X_\alpha+f^\alpha\V_\alpha,
\]
(where $f^\alpha$ are given by equation~\eqref{free-forces}),
then the expression of the constrained dynamical section is
\[
\Gamma_{(L, D)} =y^a\X_a+(f^a+f^AQ^a_A)\V_a,
\]
where all the functions $f^\alpha$ are evaluated at $y^A=0$.

\subsubsection*{Projection to $\TDD$}
We have seen that the regularity condition for the constrained system
$(L,D)$ can be equivalently expressed by requiring that the subbundle
$\TDD$ is a symplectic subbundle of $(\TEE,\omega_L)$.  It follows
that, for every $a\in D$, we have a direct sum decomposition
\[
\TEE[a]=\TDD[a]\oplus(\TDD[a])^\perp.
\]
Let us denote by $\bar{P}$ and $\bar{Q}$ the complementary projectors
defined by this decomposition, that is,
\[
\map{\bar{P}_a}{\TEE[a]}{\TDD[a]} \qquand
\map{\bar{Q}_a}{\TEE[a]}{(\TDD[a])^\perp},\;\;\mbox{ for all }
a\in D.
\]
Then, we have the following result:
\begin{theorem}\label{projection-internal}
  Let $(L,D)$ be a regular constrained Lagrangian system and let
  $\Gamma_L$ be the solution of the free dynamics, i.e.,
  $i_{\Gamma_L}\omega_L=dE_L$. Then the solution of the constrained
  dynamics is the \sode\ $\Gamma_{(L, D)}$ obtained by projection
  $\Gamma_{(L, D)}=\bar{P}(\Gamma_L|_D)$.
\end{theorem}
\begin{proof}
  From Theorem~\ref{projection-external} we have that the solution
  $\Gamma_{(L, D)}$ of the constrained dynamics is related to the free
  dynamics by $\Gamma_L|_D=\Gamma+Q(\Gamma_L|_D)$. Let us prove that
  $Q(\Gamma_L|_D)$ takes values in $(\TDD)^\perp$. Indeed,
  $Q(\Gamma_L|_D)$ takes values in $F=(\TDD)^\perp\cap\Ver{\TEE}$, so
  that, in particular, it takes values in $(\TDD)^\perp$.  Thus, since
  $\Gamma$ is a section of $\TDD$, it follows that
  $\Gamma_L|_D=\Gamma_{(L, D)}+Q(\Gamma_L|_D)$ is a decomposition of
  $\Gamma_L|_D$ according to $\TEE|_D=\TDD\oplus(\TDD)^\perp$, which
  implies $\Gamma_{(L, D)}=\bar{P}(\Gamma_L|_D)$.
\end{proof}

In adapted coordinates, a local basis of sections of
$(\TDD)^\perp$ is $\{Y_A,Z_A\},$ where the sections $Z_A$ are
given by~\eqref{zetas} and the sections $Y_A$ are
\[
Y_A = \X_A-Q^a_A\X_a+\C^{bc}(M_{Ab}-M_{ab}Q^a_A)\V_c ,
\]
with $M_{\alpha\beta}=\omega_L(\X_\alpha,\X_\beta)$. Therefore the
expression of the projector onto $(\TDD)^\perp$ is
\[
\bar{Q}=Z_A\otimes\V^A+Y_A\otimes\X^A.
\]
Note that $S(Y_A) = Z_A$.

\subsection{The distributional approach}
The equations for the Lagrange-d'Alembert section $\Gamma$ can be
entirely written in terms of objects in the manifold $\TDD$. Recall
that $\TDD$ is not a Lie algebroid. In order to do this, define the
2-section $\omega\sup{L,D}$ as the restriction of $\omega_L$ to
$\TDD$. If $(L,D)$ is regular, then $\TDD$ is a symplectic subbundle
of $(\TEE,\omega_L)$.  From this, it follows that $\omega\sup{L,D}$ is
a symplectic section on that bundle.  We also define
$\varepsilon\sup{L,D}$ to be the restriction of $dE_L$ to $\TDD$.
Then, taking the restriction of Lagrange-d'Alembert equations to
$\TDD$, we get the following equation
\begin{equation}\label{LDBS}
i_\Gamma\omega\sup{L,D}=\varepsilon\sup{L,D},
\end{equation}
which uniquely determines the section $\Gamma$.  Indeed, the unique
solution $\Gamma$ of the above equations is the solution of
Lagrange-d'Alembert equations: if we denote by $\lambda$ the
constraint force, we have for every $u\in\TDD[a]$ that
\[
\omega_L(\Gamma(a),u)-\pai{dE_L(a)}{u} =
\pai{\lambda(a)}{\prol{\tau}(u)}=0 ,
\]
where we have taken into account that $\prol{\tau}(u)\in D$ and
$\lambda(a)\in\Do$.

This approach, the so called \emph{distributional approach}, was
initiated by Bo\-cha\-rov and Vinogradov (see \cite{ViKu}) and
further developed by \'Sniatycki and
coworkers~\cite{BaSn,CuSn,Sn98}. Similar equations, within the
framework of Lie algebroids, are the base of the theory proposed
in~\cite{MeLa}.

\begin{remark}[Alternative description with $\TED$] {\rm One can also
    consider the restriction to $\TED$, which is a Lie algebroid, but
    no further simplification is achieved by this.  If $\bar{\omega}$
    is the restriction of $\omega_L$ to $\TED$ and $\bar{\varepsilon}$
    is the restriction of $dE_L$ to $\TED$, then the
    Lagrange-d'Alembert equations can be written in the form
    $i_\Gamma\bar{\omega}-\bar{\varepsilon}=\bar{\lambda}$, where
    $\bar{\lambda}$ is the restriction of the constraint force to
    $\TED$, which, in general, does not vanish. Also notice that the
    2-form $\bar{\omega}$ is closed but, in general, degenerated.}
  \oprocend
\end{remark}

\subsection{The nonholonomic bracket}\label{NHBracket}
Let $f, g$ be two smooth functions on $D$ and take arbitrary
extensions to $E$ denoted by the same letters (if there is no
possibility of confusion). Suppose that $X_f$ and $X_g$ are the
Hamiltonian sections on $\TEE$ given respectively by
\[
i_{X_f} \, \omega_L = df
\qquand
i_{X_g} \, \omega_L = dg.
\]
We define the \emph{nonholonomic bracket} of $f$ and $g$ as follows:
\begin{equation}
\label{nhb}
\{f, g\}_{nh} = \omega_L(\bar{P}(X_f), \bar{P}(X_g)).
\end{equation}
Note that if $f'$ is another extension of $f$, then
$(X_f-X_{f'})_{|D}$ is a section of $(\TDD)^{\perp}$ and, thus, we
deduce that~\eqref{nhb} does not depend on the chosen extensions. The
nonholonomic bracket is an almost-Poisson bracket, i.e., it is
skew-symmetric, a derivation in each argument with respect to the
usual product of functions and does not satisfy the Jacobi identity.

In addition, one can prove the following formula
\begin{equation}
  \label{evolution}
  \dot{f} = \{f, E_L\}_{nh}.
\end{equation}
Indeed, we have
\begin{align*}
  \dot{f} & = d_{\Gamma_{(L, D)}} f = i_{\Gamma_{(L, D)}} df =
  i_{\Gamma_{(L, D)}} i_{X_f} \omega_L \\
  &= \omega_L(X_f, \Gamma_{(L, D)}) =
  \omega_L(X_f, \bar{P}(\Gamma_L)) \\
  &= \omega_L(\bar{P}(X_f), \bar{P}(\Gamma_L)) = \{f, E_L\}_{nh}.
\end{align*}
Equation~\eqref{evolution} implies once more the conservation of the
energy (by the skew-symmetric character of the nonholonomic bracket).

Alternatively, since $\TDD$ is an anchored vector bundle, one can take
the function $f \in \cinfty{D}$ and its differential
$\bar{d}f\in\Sec{(\TDD)^*}$. Since $\omega\sup{L,D}$ is regular, we
have a unique section $\bar{X}_f \in \Sec{\TDD}$ defined by
$i_{\bar{X}_f}\omega\sup{L,D} = \bar{d}f$. Then the nonholonomic
bracket of two functions $f$ and $g$ is $\{f,g\}_{nh} =
\omega\sup{L,D}(\bar{X}_f,\bar{X}_g)$. Note that if $\tilde{f}\in
C^\infty(E)$ (resp. $\tilde{g}\in C^\infty(E)$) is an extension to $E$
of $f$ (resp., $g$), then $\bar{X}_f = \bar{P}(X_{\tilde{f}})_{|D}$
(resp., $\bar{X}_g = \bar{P}(X_{\tilde{g}})_{|D}$).


\section{Morphisms and reduction}
\label{sec:reduction}

One important advantage of dealing with Lagrangian systems evolving on
Lie algebroids is that the reduction procedure can be naturally
handled by considering morphisms of Lie algebroids, as it was already
observed by Weinstein~\cite{Weinstein}. We study in this section the
transformation laws of the different geometric objects in our theory
and we apply these results to the study of the reduction theory.

\begin{proposition}\label{transformation-S}
  Let $\map{\Phi}{E}{E'}$ be a morphism of Lie algebroids, and
  consider the $\Phi$-tangent prolongation of $\Phi$, i.e
  $\map{\prol[\Phi]{\Phi}}{\TEE}{{\mathcal T}^{E'}E'}$.  Let $\xi\spV$
  and $\xi'{}\spV$, $S$ and $S'$, and $\Delta$ and $\Delta'$, be the
  vertical liftings, the vertical endomorphisms, and the Liouville
  sections on $E$ and $E'$, respectively. Then,
  \begin{enumerate}
  \item $\prol[\Phi]{\Phi}(\xi\spV(a,b)) =
    \xi'{}\spV(\Phi(a),\Phi(b))$, for all $(a,b)\in E\times_M E$,
  \item $\prol[\Phi]{\Phi}\circ\Delta = \Delta'\circ\Phi$,
  \item $\prol[\Phi]{\Phi}\circ S = S'\circ \prol[\Phi]{\Phi}$.
  \end{enumerate}
\end{proposition}
\begin{proof}
  For the first property, we notice that both terms are vertical, so
  that we just have to show that their action on functions coincide.
  For every function $f'\in\cinfty{E'}$, we deduce that
  \[
  \begin{aligned}
    \rho'^{1}(\prol[\Phi]{\Phi}(\xi\spV(a,b)))f'
    &=T\Phi(\rho^{1}(\xi\spV(a,b)))f' =T\Phi(b\spV_a)f'
    =b\spV_a(f'\circ\Phi)\\
    &=\frac{d}{dt}f'(\Phi(a+tb))\at{t=0}
    =\frac{d}{dt}f'(\Phi(a)+t\Phi(b))\at{t=0}\\
    &=\Phi(b)\spV_{\Phi(a)}(f')
    =\rho'^{1}(\xi'{}\spV(\Phi(a),\Phi(b)))f'.
  \end{aligned}
  \]
  For the second property, we have $\Delta(a)=\xi\spV(a,a)$ so that
  applying the first property it follows that
  \[
  \prol[\Phi]{\Phi}(\Delta(a))
  =\prol[\Phi]{\Phi}(\xi\spV(a,a))
  =\xi'{}\spV(\Phi(a),\Phi(a))
  =\Delta'(\Phi(a)).
  \]
  Finally, for any $z=(a,b,V)\in\TEE$, we obtain that
  \[
  \begin{aligned}
    \prol[\Phi]{\Phi}(S(z)) &=\prol[\Phi]{\Phi}(\xi\spV(a,b))
    =\xi'{}\spV(\Phi(a),\Phi(b))\\
    &=S'(\Phi(a),\Phi(b),T\Phi(V)) =S'(\prol[\Phi]{\Phi}(z)),
  \end{aligned}
  \]
  which concludes the proof.
\end{proof}

\begin{proposition}\label{transformation-omegaL}
  Let $L\in\cinfty{E}$ be a Lagrangian function, $\theta_L$ the Cartan
  form and $\omega_L=-d\theta_L$.  Let $\map{\Phi}{E}{E'}$ be a Lie
  algebroid morphism and suppose that $L=L'\circ \Phi$, with $L'\in
  \cinfty{E'}$ a Lagrangian function.  Then, we have
  \begin{enumerate}
  \item $(\prol[\Phi]{\Phi})\pb\theta_{L'}=\theta_L$,
  \item $(\prol[\Phi]{\Phi})\pb\omega_{L'}=\omega_L$,
  \item $(\prol[\Phi]{\Phi})\pb E_{L'}=E_L$,
  \item $G\sup{L'}_{\Phi(a)}(\Phi(b),\Phi(c))=\GL[a](b,c)$, for every
    $a\in E$ and every $b,c\in E_{\tau(a)}$.
  \end{enumerate}
\end{proposition}
\begin{proof}
  Indeed, for every $Z\in\TEE$ we have
  \[
  \begin{aligned}
    \pai{(\prol[\Phi]{\Phi})\pb\theta_{L'}}{Z}
    &=\pai{\theta_{L'}}{\prol[\Phi]{\Phi}(Z)}
    =\pai{dL'}{S'(\prol[\Phi]{\Phi}(Z))}
    =\pai{dL'}{\prol[\Phi]{\Phi}(S(Z))}\\
    &=\pai{(\prol[\Phi]{\Phi})\pb dL'}{S(Z)}
    =\pai{d(\prol[\Phi]{\Phi})\pb L'}{S(Z)}
    =\pai{d(L'\circ\Phi)}{S(Z)}\\
    &=\pai{dL}{S(Z)} =\pai{\theta_L}{Z},
  \end{aligned}
  \]
  where we have used the transformation rule for the vertical
  endomorphism. The second property follows from the fact that
  $\prol[\Phi]{\Phi}$ is a morphism, so that $(\prol[\Phi]{\Phi})\pb
  d=d(\prol[\Phi]{\Phi})\pb$. The third one follows similarly and the
  fourth is a consequence of the second property and the definitions
  of the tensors $\GL$ and $G\sup{L'}$.
\end{proof}

Let $\Gamma$ be a \sode\ and $L\in\cinfty{E}$ be a Lagrangian. For
convenience, we define the 1-form $\delta_\Gamma L\in\Sec{(\TEE)^*}$
by
\[
\pai{\delta_\Gamma L}{Z}=\pai{dE_L-i_\Gamma\omega_L}{Z}
=\pai{dE_L}{Z}-\omega_L(\Gamma,Z),
\]
for every section $Z$ of $\TEE$. We notice that  $\Gamma$ is the
solution of the free dynamics if and only if $\delta_\Gamma L=0$.
On the other hand, notice that the 1-form $\delta_\Gamma L$ is
semibasic, because $\Gamma$ is a \sode.

\begin{proposition}\label{transformation-deltaL}
  Let $\Gamma$ be a \sode\ in $E$ and $\Gamma'$ a \sode\ in $E'$. Let
  $L\in\cinfty{E}$ and $L'\in\cinfty{E'}$ be Lagrangian functions
  defined on $E$ and $E'$, respectively, such that $L=L'\circ\Phi$.
  Then,
  \begin{equation}
    \label{Gamma-relation}
    \pai{\delta_\Gamma L-(\prol[\Phi]{\Phi})\pb\delta_{\Gamma'} L'}{Z}
    =\omega_{L'}(\Gamma'\circ\Phi-\prol[\Phi]{\Phi} \circ
    \Gamma,\prol[\Phi]{\Phi}(Z)),
  \end{equation}
  for every section $Z$ of $\TEE$.
\end{proposition}
\begin{proof}
  Indeed, from $(\prol[\Phi]{\Phi})\pb dE_{L'}=d E_L$, we have that
  \[
  \begin{aligned}
    \pai{\delta_\Gamma L-(\prol[\Phi]{\Phi})\pb\delta_{\Gamma'} L'}{Z}
    &=\pai{(\prol[\Phi]{\Phi})\pb i_{\Gamma'}\omega_{L'}-i_\Gamma\omega_L} {Z}\\
    &=\pai{(\prol[\Phi]{\Phi})\pb i_{\Gamma'}\omega_{L'} -
      i_\Gamma(\prol[\Phi]{\Phi})\pb\omega_{L'}}{Z}\\
    & = \omega_{L'}(\Gamma'\circ\Phi -
    \prol[\Phi]{\Phi}\circ\Gamma,\prol[\Phi]{\Phi}(Z)),
  \end{aligned}
\]
which concludes the proof.
\end{proof}

\subsection{Reduction of the free dynamics}
Here, we build on Propositions~\ref{transformation-omegaL}
and~\ref{transformation-deltaL} to identify conditions under which the
dynamics can be reduced under a morphism of Lie algebroids.  We first
notice that, from Proposition~\ref{transformation-omegaL}, if $\Phi$
is fiberwise surjective morphism and $L$ is a regular Lagrangian on
$E$, then $L'$ is a regular Lagrangian on $E'$ (note that
$\prol[\Phi]{\Phi}:{\mathcal T}^EE\to {\mathcal T}^{E'}{E'}$ is a
fiberwise surjective morphism).  Thus, the dynamics of both systems is
uniquely defined.

\begin{theorem}[Reduction of the free
  dynamics]\label{reduction-free-dynamics} Suppose that the
  Lagrangian functions $L$ and $L'$ are $\Phi$-related, that is,
  $L = L' \circ \Phi$. If $\Phi$ is a fiberwise surjective
  morphism and $L$ is a regular Lagrangian then $L'$ is also a
  regular Lagrangian. Moreover, if $\Gamma_{L}$ and $\Gamma_{L'}$
  are the solutions of the free dynamics defined by $L$ and $L'$
  then
  \[
  \prol[\Phi]{\Phi}\circ\Gamma_L=\Gamma_{L'}\circ\Phi.
  \]
  Therefore, if $a(t)$ is a solution of the free dynamics defined
  by~$L$, then $\Phi(a(t))$ is a solution of the free dynamics defined
  by~$L'$.
\end{theorem}
\begin{proof}
  If $\Gamma_L$ and $\Gamma_{L'}$ are the solutions of the dynamics,
  then $\delta_{\Gamma_L} L=0$ and $\delta_{\Gamma_{L'}}L'=0$ so that
  the left-hand side in equation~\eqref{Gamma-relation} vanishes. Thus
  \[
  \omega_{L'}(\Gamma_{L'}\circ\Phi-\prol[\Phi]{\Phi} \circ
  \Gamma_L,\prol[\Phi]{\Phi}(Z)) = 0,
  \]
  for every $Z\in\Sec{\TEE}$. Therefore, using that $L'$ is regular
  and the fact that $\prol[\Phi]{\Phi}$ is a fiberwise surjective
  morphism, we conclude the result.
\end{proof}

We will say that the unconstrained dynamics $\Gamma_{L'}$ is the
\emph{reduction of the unconstrained dynamics} $\Gamma_L$ by the
morphism $\Phi$.

\subsection{Reduction of the constrained dynamics}
The above results about reduction of unconstrained Lagrangian systems
can be easily generalized to nonholonomic constrained Lagrangian
systems whenever the constraints of one system are mapped by the
morphism to the constraints of the second system. Let us elaborate on
this.

Let $(L,D)$ be a constrained Lagrangian system on a Lie algebroid $E$
and let $(L',D')$ be another constrained Lagrangian system on a second
Lie algebroid $E'$. Along this section, we assume that there is a
fiberwise surjective morphism of Lie algebroids $\map{\Phi}{E}{E'}$
such that $L=L'\circ\Phi$ and $\Phi(D)=D'$.  The latter condition
implies that the base map is also surjective, so that we will assume
that $\Phi$ is an epimorphism (i.e., in addition to being fiberwise
surjective, the base map $\varphi$ is a submersion).

As a first consequence, we have $G\sup{L',D'}_{\Phi(a)}
(\Phi(b),\Phi(c)) = \GLD[a](b,c)$, for every $a\in D$ and every
$b,c\in D_{\pi(a)}$, and therefore, if $(L,D)$ is regular, then so is
$(L',D')$.

\begin{lemma}\label{l5.5}
  With respect to the decompositions $\TEE|_D=\TED\oplus F$ and
  $\prol[E']{E'}|_{D'}=\prol[E']{D'}\oplus F'$, we have the following
  properties:
  \begin{enumerate}
  \item $\prol[\Phi]{\Phi}(\TED)=\prol[E']{D'}$,
  \item $\prol[\Phi]{\Phi}(F)=F'$,
  \item If $P,Q$ and $P',Q'$ are the projectors associated with
    $(L,D)$ and $(L',D'),$ respectively, then
    $P'\circ\prol[\Phi]{\Phi}=\prol[\Phi]{\Phi}\circ P$ and
    $Q'\circ\prol[\Phi]{\Phi}=\prol[\Phi]{\Phi}\circ Q$.
  \end{enumerate}
  With respect to the decompositions $\TEE|_D=\TDD\oplus(\TDD)^\perp$
  and $\prol[E']{E'}|_{D'}=\prol[D']{D'}\oplus(\prol[D']{D'})^\perp$
  we have the following properties:
  \begin{enumerate}\setcounter{enumi}{3}
  \item $\prol[\Phi]{\Phi}(\TDD)=\prol[D']{D'}$,
  \item $\prol[\Phi]{\Phi}\bigl((\TDD)^\perp\bigr) =
    (\prol[D']{D'})^\perp$,
  \item If $\bar{P},\bar{Q}$ and $\bar{P}',\bar{Q}'$ are the
    projectors associated with $(L,D)$ and $(L',D'),$ respectively,
    then
    $\bar{P}'\circ\prol[\Phi]{\Phi}=\prol[\Phi]{\Phi}\circ\bar{P}$ and
    $\bar{Q}'\circ\prol[\Phi]{\Phi}=\prol[\Phi]{\Phi}\circ\bar{Q}$.
  \end{enumerate}
\end{lemma}
\begin{proof}
  From the definition of $\prol[\Phi]{\Phi}$, it follows that
  \[
  (\prol[\Phi]{\Phi})({\mathcal T}^ED)\subseteq {\mathcal
    T}^{E'}D',\;\;\;\;(\prol[\Phi]{\Phi})({\mathcal T}^DD)\subseteq
  {\mathcal T}^{D'}D'.
  \]
  Thus, one may consider the vector bundle morphisms
  \[
  \prol[\Phi]{\Phi}:{\mathcal T}^ED\to {\mathcal T}^{E'}D',\;\;\;
  \prol[\Phi]{\Phi}:{\mathcal T}^DD\to {\mathcal T}^{D'}D'.
  \]
  Moreover, using that $\Phi$ is fiberwise surjective and that
  $\varphi$ is a submersion, we deduce that the rank of the above
  morphisms is maximum. This proves (1) and (4).

  The proof of (5) is as follows. For every $a'\in D'$, one can choose
  $a\in D$ such that $\Phi(a)=a'$, and one can write any element
  $w'\in\prol[D']{D'}[a']$ as $w'=\prol[\Phi]{\Phi}(w)$ for some
  $w\in\prol[D]{D}[a]$. Thus, if $z\in(\TDD[a])^\perp$, for every
  $w'\in\prol[D']{D'}[a']$ we have
  \[
  \omega_{L'}(\prol[\Phi]{\Phi}(z),w')
  =\omega_{L'}(\prol[\Phi]{\Phi}(z),\prol[\Phi]{\Phi}(w))
  =\omega_L(z,w)=0 ,
  \]
  from where it follows that
  $\prol[\Phi]{\Phi}(z)\in(\prol[D']{D'})^\perp$. In a similar way,
  using that ${\mathcal T}^\Phi\Phi:(\prol[E]{E})_{|D}\to
  (\prol[E']{E'})_{|D'}$ is fiberwise surjective, (2) in
  Proposition~\ref{transformation-omegaL} and (4), we obtain that
  $(\prol[D']{D'})^\perp\subseteq
  (\prol[\Phi]{\Phi})((\prol[D]{D})^\perp)$.

  For the proof of (2) we have that
  \[
  \prol[\Phi]{\Phi}(F) =\prol[\Phi]{\Phi}((\TDD)^\perp\cap\Ver{\TEE})
  \subseteq (\prol[D']{D'})^\perp\cap\Ver{\prol[E']{E'}} =F'.
  \]
  Thus, using that $\prol[\Phi]{\Phi}: (\prol[E]{E})_{|D}\to
  (\prol[E']{E'})_{|D'}$ is fiberwise surjective, the fact that
  $(\prol[E]{E})_{|D}=\prol[E]{D}\oplus F$ and (1), it follows that
  \[
  (\prol[E']{E'})_{|D'}=\prol[E']{D'} \oplus (\prol[\Phi]{\Phi})(F).
  \]
  Therefore, since $(\prol[E']{E'})_{|D'}=\prol[E']{D'}\oplus F'$,
  we conclude that (2) holds.

  Finally, (3) is an immediate consequence of (1) and (2), and
  similarly, (6) is an immediate consequence of (4) and (5).
\end{proof}

From the properties above, we get the following result.

\begin{theorem}[Reduction of the constrained dynamics]\label{t5.6}
  Let $(L,D)$ be a regular constrained Lagrangian system on a Lie
  algebroid $E$ and let $(L',D')$ be a constrained Lagrangian system
  on a second Lie algebroid $E'$. Assume that a fiberwise surjective
  morphism of Lie algebroids $\map{\Phi}{E}{E'}$ exists such that
  $L=L'\circ\Phi$ and $\Phi(D)=D'$. If $\Gamma_{(L, D)}$ is the
  constrained dynamics for $L$ and $\Gamma_{(L', D')}$ is the
  constrained dynamics for $L'$, then
  $\prol[\Phi]{\Phi}\circ\Gamma_{(L, D)}=\Gamma_{(L', D')} \circ\Phi$.
  If $a(t)$ is a solution of Lagrange-d'Alembert differential
  equations for $L$, then $\Phi(a(t))$ is a solution of
  Lagrange-d'Alembert differential equations for~$L'$.
\end{theorem}
\begin{proof}
  For the free dynamics, we have that
  $\prol[\Phi]{\Phi}\circ\Gamma_L=\Gamma_{L'}\circ\Phi$. Moreover,
  from property (3) in Lemma~\ref{l5.5}, for every $a\in D$, we have
  that
  \[
  \begin{array}{l}
  \prol[\Phi]{\Phi}(\Gamma_{(L, D)}(a)) =\prol[\Phi]{\Phi}(P(\Gamma_L(a)))
  =P'(\prol[\Phi]{\Phi}(\Gamma_L(a))) \\[5pt] =P'(\Gamma_{L'}(\Phi(a)))
  =\Gamma_{(L', D')}(\Phi(a)),
  \end{array}
  \]
  which concludes the proof.
\end{proof}

We will say that the constrained dynamics $\Gamma_{(L', D')}$ is
the \emph{reduction of the constrained dynamics} $\Gamma_{(L, D)}$
by the morphism $\Phi$.

\begin{theorem}\label{t5.7}
  Under the same hypotheses as in Theorem \ref{t5.6}, we have that
  \[
  \{f'\circ \Phi,g'\circ \Phi\}_{nh}=\{f',g'\}'_{nh}\circ \Phi,
  \]
  for $f',g'\in C^\infty(D'),$ where $\{\cdot,\cdot\}_{nh}$
  (respectively, $\{\cdot,\cdot\}_{nh}')$ is the nonholonomic bracket
  for the constrained system $(L,D)$ (respectively, $(L',D')$). In
  other words, $\Phi:D\to D'$ is an almost-Poisson morphism.
\end{theorem}

\begin{proof}
  Using (2) in Proposition~\ref{transformation-omegaL} and
  the fact that $\Phi$ is a Lie algebroid morphism, we deduce that
  \[
  (i_{X_{f'\circ
      \Phi}}(\prol[\Phi]{\Phi})^*\omega_{L'})=i_{X_{f'}}\omega_{L'}\circ
  \Phi.
  \]
  Thus, since $\prol[\Phi]{\Phi}$ is fiberwise surjective, we obtain
  that
  \[
  \prol^\Phi\Phi\circ X_{f'\circ \Phi}=X_{f'}\circ \Phi.
  \]
  Now, from (\ref{nhb}) and Lemma~\ref{l5.5}, we conclude that
  \[
  \{f'\circ \Phi,g'\circ \Phi\}_{nh}=\{f',g'\}_{nh}'\circ \Phi.
  \]
\end{proof}

One of the most important cases in the theory of reduction is the case
of reduction by a symmetry group. In this respect, we have the
following result.

\begin{theorem}[\cite{LeMaMa,Mackenzie}]
  \label{quotient-Lie-algebroid}
  Let $\map{q^Q_G}{Q}{M}$ be a principal $G$-bundle, let
  $\map{\tau}{E}{Q}$ be a Lie algebroid, and assume that we have an
  action of $G$ on $E$ such that the quotient vector bundle $E/G$ is
  well-defined. If the set $\Sec{E}^G$ of equivariant sections of $E$
  is a Lie subalgebra of $\Sec{E}$, then the quotient $E'=E/G$ has a
  canonical Lie algebroid structure over $M$ such that the canonical
  projection $\map{q^E_G}{E}{E/G}$, given by $a\mapsto[a]_G$, is a
  (fiberwise bijective) Lie algebroid morphism over $q^Q_G$.
\end{theorem}

As a concrete example of application of the above theorem, we have the
well-known case of the Atiyah or Gauge algebroid. In this case, the
Lie algebroid $E$ is the standard Lie algebroid $TQ\to Q$, the action
is by tangent maps $gv\equiv T\psi_g(v)$, the reduction is the Atiyah
Lie algebroid $TQ/G\to Q/G$ and the quotient map
$\map{q^{TQ}_G}{TQ}{TQ/G}$ is a Lie algebroid epimorphism. It follows
that if $L$ is a $G$-invariant regular Lagrangian on $TQ$ then the
unconstrained dynamics for $L$ projects to the unconstrained dynamics
for the reduced Lagrangian $L'$. Moreover, if the constraints $D$ are
also $G$-invariant, then the constrained dynamics for $(L,D)$ reduces
to the constrained dynamics for $(L',D/G)$.

On a final note, we mention that the pullback of the distributional
equation $i_{\Gamma'}\omega\sup{L',D'}-\varepsilon\sup{L',D'}=0$ by
$\prol[\Phi]{\Phi}$ is precisely
$(i_\Gamma\omega\sup{L,D}-\varepsilon\sup{L,D})\circ\prol[\Phi]{\Phi}=0$.

\subsection{Reduction by stages}
As a direct consequence of the results exposed above, one can obtain a
theory of reduction by stages. In Poisson geometry, reduction by
stages is a straightforward procedure. Given the fact that the
Lagrangian counterpart of Poisson reduction is Lagrangian reduction,
it is not strange that reduction by stages in our framework becomes
also straightforward.

The Lagrangian theory of reduction by stages is a consequence of the
following basic observation:
\begin{quote}
  Let $\map{\Phi_1}{E_0}{E_1}$ and $\map{\Phi_2}{E_1}{E_2}$ be a
  fiberwise surjective morphisms of Lie algebroids and let
  $\map{\Phi}{E_0}{E_2}$ be the composition $\Phi=\Phi_2\circ\Phi_1$.
  The reduction of a Lagrangian system in $E_0$ by $\Phi$ can be
  obtained by first reducing by $\Phi_1$ and then reducing the
  resulting Lagrangian system by $\Phi_2$.
\end{quote}

This result follows using that $\prol[\Phi]{\Phi} =
\prol[\Phi_2]{\Phi_2} \circ \prol[\Phi_1]{\Phi_1}$. Based on this
fact, one can analyze one of the most interesting cases of
reduction: the reduction by the action of a symmetry group. We
consider a group $G$ acting on a manifold $Q$ and a closed normal
subgroup $N$ of $G$.  The process of reduction by stages is
illustrated in the following diagram
\[
\xymatrix{ \ar@/_{15pt}/[dd]^{\ \cdot/G}_{\text{Total reduction }}
  &\map{\tau_Q}{E_0=TQ}{M_0=Q}\ar[d]^{\ \cdot/N}_{\text{First
      reduction }}\\
  &\map{\tau_1}{E_1=TQ/N}{M_1=Q/N}\ar[d]^{\ \cdot/(G/N)}_{\text{Second
      reduction }}\\
  &\map{\tau_2}{E_2=(TQ/N)/(G/N)}{M_2=(Q/N)/(G/N)} }
\]

In order to prove our results about reduction by stages, we have to
prove that $E_0, E_1$ and $E_2$ are Lie algebroids, that the quotient
maps $\map{\Phi_1}{E_0}{E_1}$, $\map{\Phi_2}{E_1}{E_2}$ and
$\map{\Phi}{E_0}{E_2}$ are Lie algebroids morphisms and that the
composition $\Phi_1\circ\Phi_2$ equals to $\Phi$.  Our proof is based
on the following well-known result (see~\cite{CeMaRa}), which contains
most of the ingredients in the theory of reduction by stages.

\begin{theorem}(\cite{CeMaRa})\label{principal-reduction}
  Let $\map{q^Q_G}{Q}{M}$ be a principal $G$-bundle and $N$ a closed
  normal subgroup of $G$. Then,
  \begin{enumerate}
  \item $\map{q^Q_N}{Q}{Q/N}$ is a principal $N$-bundle,
  \item $G/N$ acts on $Q/N$ by the rule $[g]_N[q]_N=[gq]_N$,
  \item $\map{q^{Q/N}_{G/N}}{Q/N}{(Q/N)/(G/N)}$ is a principal
    $(G/N)$-bundle.
  \item The map $\map{i}{Q/G}{(Q/N)/(G/N)}$ defined by
    $[q]_G\mapsto[[q]_N]_{G/N}$ is a diffeomorphism.
  \end{enumerate}
\end{theorem}

Building on the previous results, one can deduce the following
theorem, which states that the reduction of a Lie algebroid can be
done by stages.
\begin{theorem}
  Let $\map{q^Q_G}{Q}{M}$ be a principal $G$-bundle and $N$ be a
  closed normal subgroup of $G$. Then,
  \begin{enumerate}
  \item $\map{\tau_{TQ/G}}{TQ/G}{Q/G}$ is a Lie algebroid and
    $\map{q^{TQ}_G}{TQ}{TQ/G}$ is a Lie algebroid epimorphism,
  \item $\map{\tau_{TQ/N}}{TQ/N}{Q/N}$ is a Lie algebroid and
    $\map{q^{TQ}_N}{TQ}{TQ/N}$ is a Lie algebroid epimorphism,
  \item $G/N$ acts on $TQ/N$ by the rule $[g]_N[v]_N=[gv]_N$,
  \item $\map{\tau_{(TQ/N)/(G/N)}}{(TQ/N)/(G/N)}{(Q/N)/(G/N)}$ is a
    Lie algebroid and $\map{q^{TQ/N}_{G/N}}{TQ/N}{(TQ/N)/(G/N)}$ is a
    Lie algebroid epimorphism,
  \item The map $\map{I}{TQ/G}{(TQ/N)/(G/N)}$ defined by
    $[v]_G\mapsto[[v]_N]_{G/N}$ is an isomorphism of Lie algebroids
    over the map $i$.
  \end{enumerate}
\end{theorem}

\begin{proof}
  The vector bundle $\tau_{TQ/G}:TQ/G\to Q/G$ (respectively,
  $\tau_{TQ/N}:TQ/N\to Q/N)$ is the Atiyah algebroid for the principal
  $G$-bundle $q_G^Q:Q\to Q/G$ (respectively, $q_N^Q:Q\to Q/N$), so
  that (1) and (2) are obvious. Condition (3) is just condition (2) of
  Theorem~\ref{principal-reduction} applied to the principal
  $N$-bundle $TQ\to TQ/N$. To prove condition (4), we notice that the
  action of $G/N$ on the Lie algebroid $TQ/N$ is free and satisfies
  the conditions of Theorem~\ref{quotient-Lie-algebroid}. Finally, the
  Lie algebroid morphism $\map{j}{TQ}{TQ/N}$ is equivariant with
  respect to the $G$-action on $TQ$ and the $(G/N)$-action on $TQ/N$.
  Thus, it induces a morphism of Lie algebroids in the quotient. It is
  an isomorphism since it is a diffeomorphism by
  Theorem~\ref{principal-reduction}.
\end{proof}

The following diagram illustrates the above situation:
\[
\xymatrix{%
  TQ\ar@/^{20pt}/[rr]^{\Phi} \ar[d]_{\tau_Q}\ar[r]_{\kern-10pt\Phi_1}
  &
  TQ/N\ar[d]^{\tau_1}\ar[r]_{\kern-20pt \Phi_2} & (TQ/N)/(G/N)\ar[d]^{\tau_2}\\
  Q\ar@/_{20pt}/[rr]_{q^Q_G}\ar[r]^{q_N^Q} & Q/N\ar[r]^{\kern-15pt
    q_{G/N}^{Q/N}}&(Q/N)/(G/N) }
\]

In particular, for the unconstrained case one has the following
result.
\begin{theorem}[Reduction by stages of the free dynamics]
  Let $\map{q^Q_G}{Q}{Q/G}$ be a principal $G$-bundle, and $N$ a
  closed normal subgroup of $G$. Let $L$ be a Lagrangian function on
  $Q$ which is $G$-invariant. Then the reduction by the symmetry group
  $G$ can be performed in two stages:
  \begin{itemize}
  \item[1.] reduce by the normal subgroup $N$,
  \item[2.] reduce the resulting dynamics from 1. by the residual
    symmetry group $G/N$.
  \end{itemize}
\end{theorem}

Since the dynamics of a constrained system is obtained by projection
of the free dynamics, we also the following result.
\begin{theorem}[Reduction by stages of the constrained dynamics]
  Let $\map{q^Q_G}{Q}{Q/G}$ be a principal $G$-bundle and $N$ a closed
  normal subgroup of $G$. Let $(L,D)$ be a $G$-invariant constrained
  Lagrangian system. Then the reduction by the symmetry group $G$ can
  be performed in two stages:
  \begin{itemize}
  \item[1.] reduce by the normal subgroup $N$,
  \item[2.] reduce the resulting dynamics from 1. by the residual
  symmetry group $G/N$.
  \end{itemize}
\end{theorem}


\section{The momentum equation}
\label{momentum-equation}

In this section, we introduce the momentum map for a constrained
system on a Lie algebroid, and examine its evolution along the
dynamics. This gives rise to the so-called momentum equation.

\subsection{Unconstrained case}

Let us start by discussing the unconstrained case.  Let $\tau_E:E\to
M$ be a Lie algebroid over a manifold $M$ and $L:E\to \R$ be a regular
Lagrangian function.  Suppose that $\tau_K:K\to M$ is a vector bundle
over $M$ and that $\Psi:K\to E$ is a vector bundle morphism (over the
identity of $M$) between $K$ and $E$. Then, we can define \emph{the
  unconstrained momentum map} $J_{(L,\Psi)}:E\to K^*$ \emph{associated
  with $L$ and $\Psi$} as follows
\[
J_{(L,\Psi)}(a)\in K_x^*, \mbox{ for } a\in E_x,
\]
and
\[
(J_{(L,\Psi)}(a))(k)=\frac{d}{dt}_{|t=0}L(a+t\Psi(k))=\Psi(k)_a\spV(L),\mbox{
  for }k\in K_x.
\]
If $\sigma:M\to K$ is a section of $\tau_K:K\to M$ then, using the
momentum map $J_{(L,\Psi)}$, we may introduce the real function
$J_{(L,\Psi)}^\sigma:E\to \R$ given by
\begin{equation}\label{Momentum1}
  J_{(L,\Psi)}^\sigma(a) =
  J_{(L,\Psi)}(a)(\sigma(x))=\Psi(\sigma(x))_a\spV(L),\mbox{
    for }a\in E_x.
\end{equation}
\begin{theorem}[The unconstrained momentum equation]\label{theorem6.1}
  Let $\Gamma_L$ be the Euler-Lagrange section associated with the
  regular Lagrangian function $L:E\to \R$. If $\sigma:M\to K$ is a
  section of $\tau_K:K\to M$ and $(\Psi\circ \sigma)^c\in
  Sec(\prol[E]{E})$ is the complete lift of $(\Psi\circ \sigma)\in
  Sec(E)$, we have that
  \begin{equation}\label{EqMomen1}
    <d^{\prol[E]{E}}J_{(L,\Psi)}^\sigma,\Gamma_L>=<d^{\prol[E]{E}}L,
    (\Psi\circ \sigma)^c>,
  \end{equation}
  where $d^{{\mathcal T}^EE}$ is the differential of Lie algebroid
  ${\prol[E]{E}}\to E.$ In particular, if \linebreak
  $<d^{\prol[E]{E}}L,(\Psi\circ \sigma)^c>=0$, then the real function
  $J_{(L,\Psi)}^\sigma$ is a constant of the motion for the Lagrangian
  dynamics associated with the Lagrangian function $L.$
\end{theorem}
\begin{proof}
  Let $S:{\prol[E]{E}}\to {\prol[E]{E}}$ be the vertical endomorphism.
  If $(\Psi\circ \sigma)^v\in Sec(\prol[E]{E})$ is the vertical lift of
  $(\Psi\circ \sigma)\in Sec(E)$ then, using (\ref{Momentum1}) and the
  fact that $S(\Psi\circ \sigma)^c=(\Psi\circ \sigma)^v,$ it follows
  that
  \begin{equation}\label{Momen2}
    J_{(L,\Psi)}^\sigma=\theta_L((\Psi\circ \sigma)^c),
  \end{equation}
  where $\theta_L$ is the Cartan $1$-form associated with $L$.

  Thus, from (\ref{Momen2}), we deduce that
  \[
  d^{\prol[E]{E}}J_{(L,\Psi)}^\sigma={\mathcal L}_{(\Psi\circ
    \sigma)^c}^{\prol[E]{E}}\theta_L + i(\Psi\circ
  \sigma)^c(\omega_L),\] $\omega_L$ being the Cartan $2$-form
  associated with $L$.

  Therefore, if $E_L:E\to \R$ is the Lagrangian energy, we obtain
  that
  \begin{equation}\label{Formula}
    \begin{array}{rcl}
      <d^{\prol[E]{E}}J^\sigma_{(L,\Psi)},\Gamma_L> & = &
      <d^{\prol[E]{E}}(\theta_L(\Gamma_L)),
      (\Psi\circ \sigma)^c>-<d^{\prol[E]{E}}E_L,(\Psi\circ \sigma)^c>\\[8pt]
      &&-<\theta_L,[(\Psi\circ
      \sigma)^c,\Gamma_L]>.
    \end{array}
  \end{equation}
  Now, from (\ref{SODEcompl}) and since $\Gamma_L$ is a \sode\
  section, it follows that
  \[
  \Theta_L(\Gamma_L)=<d^{\prol[E]{E}}L,\Delta>,\;\;\;\;
  <\theta_L, [(\Psi\circ \sigma)^c,\Gamma_L]> =0,
  \]
  where $\Delta\in Sec({\prol[E]{E}})$ is the Liouville section.
  Consequently, using (\ref{Formula}) we deduce that
  (\ref{EqMomen1}) holds.
\end{proof}

\begin{remark}[Conservation of momentum on $TM$]\label{remark6.2}
  {\rm Let $L:TM\to \R$ be an standard regular Lagrangian function on
    $TM.$ Suppose that $G$ is a Lie group with Lie algebra ${\frak g}$
    and that $\psi:G\times M\to M$ is a (left) action of $G$ on $M.$
    Then, we may consider the trivial vector bundle over $M$
    \[
    K=M\times {\frak g}\to M
    \]
    and the vector bundle morphism $\Psi:K\to TM$ (over the identity
    of $M$) defined by
    \begin{equation}\label{Psi}
      \Psi(x,\xi)=\xi_M(x),
    \end{equation}
    where $\xi_M\in {\frak X}(M)$ is the infinitesimal generator of the
    action $\psi$ associated with $\xi\in {\frak g}$.
    
    A direct computation proves that the (unconstrained) momentum map
    $J_{L,\Psi)}:E=TM\to K^*=M\times {\frak g}^*$ associated with $L$
    and $\Psi$ is given by $$J_{(L,\Psi)}(v_x)=(x,J(v_x)),\mbox{ for }
    v_x\in T_xM,$$ where $J:TM\to {\frak g}^*$ is the standard
    momentum map associated with $L$ and the action $\psi$ defined by
    $$J(v_x)(\xi)=\frac{d}{dt}_{|t=0}L(v_x+t\xi_M(x)), \mbox{ for }
    v_x\in T_xM \mbox{ and } \xi\in {\frak g}$$ (see, for instance,
    \cite{AM}).
    
    Now, each $\xi\in {\frak g}$ defines a (constant) section $\sigma$
    of the vector bundle $K=M\times {\frak g}\to M$ and the real
    function $J_{(L,\Psi)}^\sigma$ is just the momentum $J_\xi:TM\to
    \R$ in the direction of $\xi$.

    On the other hand, if $\eta\in {\frak g}$, then the infinitesimal
    generator $\eta_{TM}$ of the tangent action $T\psi:G\times TM\to
    TM$ associated with $\eta$ is the (standard) complete lift
    $\eta_M^c\in {\frak X}(TM)$ of $\eta_M.$ Therefore, using Theorem
    \ref{theorem6.1}, we deduce a well-known result~\cite{AM}: ``If
    the Lagrangian function $L:TM\to \R$ is invariant under the
    tangent action $T\psi$ of $G$ on $TM$ then, for every $\xi\in
    {\frak g}$, the momentum $J_\xi:TM\to \R$ in the direction of
    $\xi$ is a constant of the motion of the Lagrangian dynamics.''}
  \oprocend
\end{remark}

\subsection{Constrained case}

Next, let us discuss the constrained case.  Suppose that $L:E\to \R$
is a regular Lagrangian function on a Lie algebroid $\tau_E:E\to M$,
that $\tau_K:K\to M$ is a vector bundle over $M$ and that $\Psi:K\to
E$ is a vector bundle morphism (over the identity of $M$) between $K$
and $E$.

In addition, let $\tau_D:D\to M$ be a vector subbundle of $\tau_E:E\to
M$ such that the nonholonomic Lagrangian system $(L,D)$ is regular.

If $x$ is point of $M$ we consider the vector subspace $K_x^D$ of
$K_x$ given by
\[
K_x^D=\{k\in K_x/\Psi(k)\in D_x\}.
\]
We will denote by $i_x:K_x^D\to K_x$ the canonical inclusion, by
$i_x^*:K_x^*\to (K_x^D)^*$ the canonical projection and by $K^D$ and
$(K^D)^*$ the sets
\[
K^D=\bigcup_{x\in M}K_x^D,\;\;\;\; (K^D)^*=\bigcup_{x\in
M}(K_x^D)^*.
\]
Then, we define the \emph{nonholonomic momentum map
  $J_{(L,D,\Psi)}:E\to (K^D)^*$ associated with the system $(L,D)$ and
  the morphism $\Psi$} as follows
\[
(J_{(L,D,\Psi)})_{|E_x}=i_x^*\circ (J_{(L,\Psi)})_{|E_x},\;\;\;
\mbox{ for } x\in M.
\]
Now, if $\sigma:M\to K$ is a section of $\tau_{K}:K\to M$ such
that $\sigma(x)\in K_x^D$, for all $x\in M,$ we may introduce the
real function $J^\sigma_{(L,D,\Psi)}:E\to \R$ given by
\[
J^\sigma_{(L,D,\Psi)}(a)=J_{(L,D,\Psi)}(a)(\sigma(x)),\mbox{ for }
a\in E_x,
\]
that is, $J_{(L,D,\Psi)}^\sigma=J_{(L,\Psi)}^\sigma.$

\begin{theorem}[The nonholonomic momentum equation]\label{theorem6.3}
  Let $\Gamma_{(L,D)}$ be the solution of the constrained dynamics for
  the nonholonomic Lagrangian system $(L,D)$. If $\sigma:M\to K$ is a
  section of $\tau_K:K\to M$ such that $\sigma(x)\in K_x^D$, for all
  $x\in M$, and $(\Psi\circ \sigma)^c\in Sec({\prol[E]{E}})$ is the
  complete lift of $(\Psi\circ \sigma)\in Sec(E)$ then we have that
  \begin{equation}\label{EqMomen2}
    <d^{\prol[E]{D}}((J_{(L,D,\Psi)})_{|D}),\Gamma_{(L,D)}> =
    <d^{\prol[E]{E}}L,(\Psi\circ \sigma)^c>_{|D},
  \end{equation}
  where $d^{\prol[E]{D}}$ (respectively, $d^{\prol[E]{E}}$) is the
  differential of Lie algebroid ${\prol[E]{D}}\to D$ (respectively,
  ${\prol[E]{E}}\to E$). In particular, if
  $<d^{\prol[E]{E}}L,(\Psi\circ \sigma)^c>_{|D}=0$, then the real
  function $J_{(L,D,\Psi)}^\sigma$ is a constant of the motion for the
  constrained dynamics associated with the nonholonomic Lagrangian
  system $(L,D)$.
\end{theorem}
\begin{proof}
  Denote by $j:D\to E$ and by ${\mathcal J}:{\prol[E]{D}}\to {\prol[E]{E}}$
  the canonical inclusions and by $Q:{{\mathcal T}_D^EE}\to F$ the
  corresponding projector, where $F=\omega_L^{-1}(\widetilde{D}^0)$
  (see Section \ref{Projectors}). Then, as we know,
  \[
  \Gamma_{(L,D)}=(\Gamma_L-Q\Gamma_L)_{|D}.
  \]
  Moreover, the pair $({\mathcal J},j)$ is a Lie algebroid monomorphism
  which implies that
  \[
  d^{\prol[E]{D}}((J_{(L,D,\Psi)}^\sigma)_{|D})=({\mathcal
    J},j)^*(d^{\prol[E]{E}}J^\sigma_{(L,D,\Psi)}).
  \]
  Thus, using that $J_{(L,D,\Psi)}^\sigma=J^\sigma_{(L,\Psi)}$ and
  proceedings as in the proof of Theorem \ref{theorem6.1}, we deduce
  that
  \begin{equation}\label{step1}
    \begin{array}{rcl}
      <d^{\prol[E]{D}}((J_{(L,D,\Psi)}^\sigma)_{|D}),\Gamma_{(L,D)}> &
      = &<d^{\prol[E]{E}}L,(\Psi\circ
      \sigma)^c>_{|D}\\&&\kern-80pt-\{({\mathcal L}_{(\Psi\circ
        \sigma)^c}^{\prol[E]{E}}\theta_L)(Q\Gamma_L) + (i(\Psi\circ
      \sigma)^c(\omega_L))(Q\Gamma_L)\}_{|D}.
    \end{array}
  \end{equation}
  Now, since $S(Q\Gamma_L)=0$, then $S[(\Psi\circ
  \sigma)^c,Q\Gamma_L]=0$ (see (\ref{Vertcompl})) and it follows
  that
  \[
  \theta_L(Q\Gamma_L)=0,\;\;\;\; \theta_L[(\Psi\circ
  \sigma)^c,Q\Gamma_L]=0.
  \]
  Therefore,
  \begin{equation}\label{Lider}
    ({\mathcal L}_{(\Psi\circ
      \sigma)^c}^{\prol[E]{E}}\theta_L)(Q\Gamma_L)=0.
  \end{equation}
  On the other hand, we have that
  \[
  (i(Q\Gamma_L)\omega_L)_{|D}=S^*(\alpha_{(L,D)}),\mbox{ with }
  \alpha_{(L,D)}\in Sec(({\prol[E]{D}})^0).
  \]
  Consequently,
  $$\{(i(\psi\circ
  \sigma)^c\omega_L)(Q\Gamma_L)\}_{|D}=-\alpha_{(L,D)}((\Psi\circ
  \sigma)^v_{|D}).$$

  But, since $\Psi\circ \sigma$ is a section of $\tau_D:D\to M$, it
  follows that $(\Psi\circ \sigma)^v_{|D}$ is a section of
  ${\prol[E]{D}}\to D.$ This implies that
  \begin{equation}\label{QGamma}
    \{(i(\Psi\circ \sigma)^c\omega_L)(Q\Gamma_L)\}_{|D}=0.
  \end{equation}
  Finally, using (\ref{step1}), (\ref{Lider}) and (\ref{QGamma}), we
  conclude that (\ref{EqMomen2}) holds.
\end{proof}

\begin{remark}[Nonholonomic momentum equation on $TM$ and horizontal
  symmetries]\label{Remark6.4}
  {\rm Suppose that $L:TM\to \R$ is an standard regular Lagrangian
    function on $E=TM$ and that $\psi:G\times M\to M$ is a (left)
    action of a Lie group $G$ on $M$. Then, we consider the trivial
    vector bundle $\tau_K:K=M\times {\frak g}\to M$ and the vector
    bundle morphism $\Psi:K\to TM$ (over the identity of $M$) defined
    by (\ref{Psi}).

    Now, let $D$ be a vector subbundle (over $M$) of the vector bundle
    $\tau_{M}:TM\to M$, that is, $D$ is a distribution on $M$, and
    assume that the nonholonomic Lagrangian system $(L,D)$ is regular.
    If $x$ is a point of $M,$ we have that $K_x^D=\{x\}\times {\frak
      g}^x,$ where ${\frak g}^x$ is the vector subspace of ${\frak g}$
    given by
  \[
  {\frak g}^x=\{\xi\in {\frak g}/\xi_M(x)\in D_x\}.
  \]
  We also remark that the sets $K^D$ and $(K^D)^*$ may be identified
  with the sets
  \[
  {\frak g}^D=\bigcup_{x\in M}{\frak g}^x,\;\;\;\; ({\frak
    g}^D)^*=\bigcup_{x\in M}({\frak g}^x)^*.
  \]
  Under this identification, the nonholonomic momentum map
  $J_{(L,D,\Psi)}:E\to (K^D)^*$ associated with the system $(L,D)$ and
  the morphism $\Psi$ is just the standard nonholonomic momentum map
  $J^{nh}:TM\to ({\frak g}^D)^*$ associated with the system $(L,D)$
  and the action $\psi$ (see \cite{BlKrMaMu,CaLeMaMa,CaLeMaMa2}).

  Now, if $\widetilde{\xi}:M\to {\frak g}$ is an smooth map the
  $\widetilde{\xi}$ defines, in a natural way, a section
  $\sigma_{\widetilde{\xi}}:M\to K=M\times {\frak g}$ of the vector
  bundle $\tau_K:K=M\times {\frak g}\to M.$ We denote by
  $J^{nh}_{\widetilde{\xi}}:TM\to \R$ the real function
  $J_{(L,D,\Psi)}^{\sigma_{\widetilde{\xi}}}:E\to \R$ and by
  $\Xi_{\widetilde{\xi}}$ the vector field $\Psi\circ
  \sigma_{\widetilde{\xi}}$ on $M.$ Then, using Theorem
  \ref{theorem6.3}, we deduce a well-known result (see
  \cite{BlKrMaMu,CaLeMaMa,CaLeMaMa2}): ``If $\Gamma_{(L,D)}$ is the
  solution of the constrained dynamics for the nonholonomic system
  $(L,D)$, we have that
  \[
  \Gamma_{L,D}((J_{\widetilde{\xi}}^{nh})_{|D})=(\Xi_{\widetilde{\xi}})^c_{|D}(L)."
  \]
  The above equality is an intrinsic expression of the \emph{standard
    nonholonomic momentum equation}. In addition, using again Theorem
  \ref{theorem6.3} we also deduce another well-known result (see
  \cite{BlKrMaMu,CaLeMaMa,CaLeMaMa2}): ``If the Lagrangian function
  $L:TM\to \R$ is invariant under the tangent action $T\psi$ of $G$ on
  $TM$ and $\xi\in {\frak g}$ is \emph{a horizontal symmetry} (that
  is, $\xi\in {\frak g}^x$, for all $x\in M$) then the real function
  $(J^{nh}_{\widetilde{\xi}})_{|D}$ is a constant of the motion for
  the constrained Lagrangian dynamics, where $\widetilde{\xi}:M\to
  {\frak g}$ is the constant map
  \[
  \widetilde{\xi}(x)=\xi, \mbox{ for all } x\in M." \eqoprocend
  \]
} 
\end{remark}


\section{Examples}\label{examples}

\newcommand{\g}{\mathfrak{g}} \renewcommand{\d}{\mathfrak{d}}
\newcommand{\h}{\mathfrak{h}}
\newcommand{\pe}[2]{\langle\langle#1,#2\rangle\rangle}

As in the unconstrained case, constrained Lagrangian systems on
Lie algebroids appear frequently.  We show some examples next.

\subsection{Nonholonomic Lagrangian systems on Lie algebras}
Let ${\frak g}$ be a real algebra of finite dimension. Then, it is
clear that ${\frak g}$ is a Lie algebroid over a single point.  Now,
suppose that $(l,{\frak d})$ is a nonholonomic Lagrangian system on
${\frak g}$, that is, $l:{\frak g}\to \R$ is a Lagrangian function and
${\frak d}$ is a vector subspace of ${\frak g}$.  If $w:I\to {\frak
  g}$ is a curve on ${\frak g}$ then
\[
dl(\omega(t))\in T_{\omega(t)}^*{\frak g}\cong {\frak
  g}^*,\;\;\;\; \forall t\in I,
\]
and thus, the map $dl\circ \omega$ may be considered as a curve on
${\frak g}^*$
\[
dl\circ \omega:I\to {\frak g}^*.
\]
Therefore,
\[
(dl\circ \omega)'(t)\in T_{dl(\omega(t))}{\frak g}^*\cong {\frak
  g}^*,\;\;\;\forall t\in I.
\]
Moreover, from (\ref{LD-edo}), it follows that $\omega$ is a
solution of the Lagrange-d'Alembert equations for the system
$(l,{\frak d})$ if and only if
\begin{equation}\label{EPSeq}
  (dl\circ \omega)'(t)-ad^*_{\omega(t)}(dl(\omega(t)))\in {\frak
    d}^\circ,\;\;\; \omega(t)\in {\frak d}, \; \; \; \; \forall t
\end{equation}
where $ad^*:{\frak g}\times {\frak g}^*\to {\frak g}^*$ is the
infinitesimal coadjoint action.

The above equations are just the so-called \emph{Euler-Poincar{\'e}
  -Suslov equations} for the system $(l,{\frak d})$ (see \cite{FeZe}).
We remark that in the particular case when the system is
unconstrained, that is, ${\frak d}={\frak g}$, then one recovers the
\emph{the standard Euler-Poincar{\'e} equations }for the Lagrangian
function $l:{\frak g}\to \R.$

If $G$ is a Lie group with Lie algebra ${\frak g}$ then nonholonomic
Lagrangian systems on ${\frak g}$ may be obtained (by reduction) from
nonholonomic LL mechanical systems with configuration space the Lie
group $G.$

In fact, let $e$ be the identity element of $G$ and $\mathbb{I}:{\frak
  g} \to {\frak g}^*$ be a symmetric positive definite inertia
operator. Denote by $g_e:{\frak g}\times {\frak g}\to \R$ the
corresponding scalar product on ${\frak g}$ given by
\[
g_e(\omega,\omega')=<\mathbb{I}(\omega),\omega'>, \mbox{ for
}\omega,\omega'\in {\frak g}\cong T_eG.
\]
$g_e$ induces a left-invariant Riemannian metric $g$ on $G$. Thus, we
way consider the Lagrangian function $L:TG\to \R$ defined by
\[
L(v_h)=\frac{1}{2}g_h(v_h,v_h),\;\;\; \mbox{ for }v_h\in T_hG.
\]
In other words, $L$ is the kinetic energy associated with the
Riemannian metric $g$.

Now, let $D$ be a left-invariant distribution on $G$. Then, since $L$
is a left-invariant function, the pair $(L,D)$ is an standard
nonholonomic LL system in the terminology of \cite{FeZe}.

On the other hand, the Lagrangian momentum map $\Phi:TG\to {\frak g}$
given by
\[
\Phi(v_h)=(T_h l_{h^{-1}})(v_h),\mbox{ for } v_h\in T_hG
\]
is a fiberwise bijective morphism of Lie algebroids. Moreover, if
$l=L_{|{\frak g}}$ and ${\frak d}=D_e$ then the pair $(l,{\frak d} )$
is a nonholonomic Lagrangian system on ${\frak g}$ and
\[
l\circ \Phi=L \mbox{ and } \Phi(D)={\frak d}. \] Thus, the system
$(l,{\frak d})$ is regular. In addition, if $v:I\to TG$ is a
solution of the Lagrange-d'Alembert equations for the system
$(L,D)$ then, using Theorem \ref{t5.6}, we deduce that the curve
$\Phi\circ v:I\to {\frak g}$ is a solution of the
Lagrange-d'Alembert equations for the system $(l,{\frak d}).$

We remark that
\[
l(\omega)=\frac{1}{2}g_e(\omega,\omega) =
\frac{1}{2}<\mathbb{I}(\omega),\omega>,\mbox{ for }\omega\in {\frak
  g}.
\]
Therefore, if $\omega:I\to {\frak g}$ is a curve on ${\frak g},$ we
have that
\[
(dl\circ \omega)(t)=\mathbb{I}(\omega(t)),\mbox{ for all } t
\]
and, using (\ref{EPSeq}), it follows that $\omega$ is a solution of
the Lagrange-d'Alembert equations for the system $(l,{\frak d})$ if
and only if
\[
\dot{\omega}-\mathbb{I}^{-1}(ad_{\omega(t)}^*\mathbb{I}(\omega(t)))\in
{\frak d}^{\perp},\;\;\; \omega(t)\in {\frak d}, \mbox{ for all }
t,
\]
where ${\frak d}^\perp$ is the orthogonal complement of the subspace
${\frak d}$, that is,
\[
{\frak d}^\perp=\{\omega'\in {\frak
  g}/<\mathbb{I}(\omega'),\omega>=0,\forall \omega\in {\frak d}\}.
\]
Two simple examples of the above general situation are the following
ones.

\subsection*{The Suslov  system}
The most natural example of LL system is the \emph{nonholonomic Suslov
  problem}, which describes the motion of a rigid body about a fixed
point under the action of the following nonholonomic constraint: the
body angular velocity vector is orthogonal to a some fixed direction
in the body frame.

The configuration space of the problem is the group $G=SO(3)$.  Thus,
in this case, the Lie algebra ${\frak g}$ may be identified with
$\R^3$ and, under this identification, the Lie bracket on ${\frak g}$
is just the cross product $\times$ on $\R^3.$

Moreover, if $\mathbb{I}:\R^3\to (\R^3)^*\cong \R^3$ is the inertia
tensor of the body then a curve $\omega:I\to \R^3$ on $\R^3$ is a
solution of the Euler-Poincar{\'e}-Suslov equations for the system if
and only if
\begin{equation}\label{Suseq}
  \dot{\omega}=\mathbb{I}^{-1}((\mathbb{I}\omega) \times \omega)) +
  \lambda \mathbb{I}^{-1}(\Gamma),\;\;\;\; <\omega,\Gamma>=0,
\end{equation}
where $\lambda$ is the Lagrange multiplier, $\Gamma$ is a fixed unit
vector in $\R^3$ and $<\cdot,\cdot>$ is the standard scalar product in
$\R^3$. Since the nonholonomic system is regular, the Lagrange
multiplier $\lambda$ is uniquely determined. In fact, differentiating
the equation $<\omega,\Gamma>=0$, we find
\[
\lambda=-\frac{<\mathbb{I}\omega\times \omega,
  \mathbb{I}^{-1}\Gamma>}{<\Gamma, \mathbb{I}^{-1}\Gamma>}
\]
and, consequently, Eqs. (\ref{Suseq}) are equivalent to
\[
\dot{\omega}=\mathbb{I}^{-1}(<\mathbb{I}\omega,\Gamma>\omega\times
\mathbb{I}^{-1}\Gamma),\;\;\;\; <\omega,\Gamma>=0.
\]
Multidimensional generalizations of the Suslov problem have been
discussed by several authors (see~\cite{FeKo,Jo,ZeBl1}).

\subsection*{The Chaplygin sleigh}
The Chaplygin sleigh is a rigid body sliding on a horizontal plane.
The body is supported at three points, two of which slide freely
without friction while the third is a knife edge, a constraint that
allows no motion orthogonal to this edge. This mechanical system was
introduced and studied in 1911 by Chaplygin \cite{Ch} (see also
\cite{NF}).

The configuration space of this system is the group $SE(2)$ of
Euclidean motions of the two-dimensional plane $\R^2.$ As we know, we
may choose local coordinates $(\theta,x,y)$ on $SE(2)$.  $\theta$ and
$(x,y)$ are the angular orientation of the blade and position of the
contact point of the blade on the plane, respectively.

Now, we introduce a coordinate system called the body frame by placing
the origin at the contact point and choosing the first coordinate axis
in the direction of the knife edge. Denote the angular velocity of the
body by $\omega=\dot{\theta}$ and the components of the linear
velocity of the contact point relative to the body frame by $v_1,v_2$.
The set $(\omega,v_1,v_2)$ is regarded as an element of the Lie
algebra ${\frak{se}}(2)$. Note that
\[
v_1=\dot{x}\cos\theta + \dot{y} \sin \theta,\;\;\;\;
v_2=\dot{y}\cos\theta - \dot{x}\sin\theta.
\]
The position of the center of mass is specified by the coordinates
$(a,b)$ relative to the body frame. Let $m$ and $J$ denote the mass
and moment of inertia of the sleigh relative to the contact point.
Then, the corresponding symmetric positive definite inertia operator
$\mathbb{I}:{\frak se}(2)\to {\frak se}(2)^*$ and the reduced
nonholonomic Lagrangian system $(l,{\frak d})$ on ${\frak se}(2)$ are
given by
\[
\begin{array}{rcl}
  \mathbb{I} (\omega,v_1,v_2)&=&\left(
    \begin{array}{ccc}
      J + m(a^2+b^2)&-bm&am\\
      -bm&m&0\\
      am&0&m \end{array}
  \right)\left(\begin{array}{c}\omega\\v_1\\v_2\end{array}\right),
  \\[20pt]
  l(\omega,v_1,v_2)&=&\frac{1}{2}[(J+m(a^2+b^2))\omega^2 + m(v_1^2 +
  v_2^2)\\[5pt]&&-2mb\omega v_1 + 2am\omega v_2],\\[8pt]
  {\frak d}&=&\{(\omega, v_1,v_2)\in {\frak
    se}(2)/v_2=0\},\end{array}
\]
(see \cite{FeZe}). Thus, the Lagrange-d'Alembert equations for the
system $(l,{\frak d})$ are $$\begin{array}{rcl}
  \dot{\omega}&=&\displaystyle\frac{am\omega}{J+ma^2}(b\omega-v_1),\\[8pt]
  \dot{v_1} & = &
  \displaystyle\frac{a\omega}{J+ma^2}((J+m(a^2+b^2))\omega-mbv_1),\\[8pt]
  v_2&=&0.
\end{array}$$
Multidimensional generalizations of the Chaplygin sleigh were
discussed in \cite{FeZe} (see also \cite{NF} and \cite{ZeBl2}).

\subsection{Nonholonomic LR systems and right action Lie algebroids}
Here, we show how the reduction of a nonholonomic LR system produces a
nonholonomic Lagrangian system on a right action Lie algebroid.

Let us start by recalling the definition of a right action Lie
algebroid (see \cite{HiMa}).  Let $(F, [\cdot, \cdot]_{F}, \rho_{F})$
be a Lie algebroid over a manifold $N$ and $\pi: M \to N$ be a smooth
map. \emph{A right action of $F$ on $\pi: M \to N$} is a $\R$-linear
map
\[
\Psi: Sec (F) \to {\frak X}(M), \; \; \; X\in Sec (F) \to
\Psi(X) \in {\frak X}(M)
\]
such that
\[
\begin{array}{l}
  \Psi (f X) = (f \circ \pi)\Psi (X), \; \; \Psi([X, Y]_{F}) =
  [\Psi(X), \Psi(Y)], \\[5pt] (T_{m}\pi)(\Psi(X)(m)) =
  \rho_{F}(X(\pi(m))),
\end{array}
\]
for $f \in C^{\infty}(N)$, $X, Y \in Sec (F)$ and $m \in M$. If
$\Psi: Sec (E) \to {\frak X}(M)$ is a right action of $F$ on
$\pi: M \to N$ and $\tau_{F}: F \to N$ is the vector bundle
projection then the pullback vector bundle of $F$ over $\pi$,
\[
E = \pi^{*}F = \{(m, f) \in M \times F / \tau_{F}(f) = \pi (m)\}
\]
is a Lie algebroid over $M$ with Lie algebroid structure
$([\cdot, \cdot]_{E}, \rho_{E})$ which is characterized by
\[
[X, Y]_{E} = [X, Y]_{F} \circ \pi, \; \; \; \rho_{E}(X)(m) =
\Psi(X)(m),
\]
for $X, Y \in Sec(E)$ and $m \in M$. The triple $(E, [\cdot,
\cdot]_{E}, \rho_{E})$ is called the \emph{right action Lie
  algebroid of $F$ over $\pi$} and it is denoted by $\pi_{\Psi}F$
(see \cite{HiMa}).

Note that if the Lie algebroid $F$ is a real Lie algebra ${\frak g}$
of finite dimension and $\pi: M \to \{\mbox{a point}\}$ is the
constant map then a right action of ${\frak g}$ on $\pi$ is just a
right infinitesimal action $\Psi: {\frak g} \to {\frak X}(M)$ of
${\frak g}$ on the manifold $M$. In such a case, the corresponding
right action Lie algebroid is the trivial vector bundle $pr_{1}: M
\times {\frak g} \to M$.

Next we recall the definition of a nonholonomic LR system
following~\cite{FeJo,Jo2}.  Let $G$ be a compact connected Lie group
with Lie algebra ${\frak g}$ and $<\cdot,\cdot>:{\frak g}\times {\frak
  g}\to \R$ be an $Ad_G$-invariant scalar product on ${\frak g}$.
Now, suppose that ${\mathcal I}:{\frak g} \to {\frak g}$ is a inertia
operator which is symmetric and definite positive with respect to the
scalar product $<\cdot,\cdot>$. Denote by $g$ the left-invariant
Riemannian metric given by
\begin{equation}\label{metrica}
  g_h(v_h,v_{h'})=<{\mathcal
    I}(T_hl_{h^{-1}}(v_h)),(T_hl_{h^{-1}})(v_h')>
\end{equation}
for $h\in G$ and $v_h,v_h'\in T_hG.$

Then, the Lagrangian function $L:TG\to \R$ of the system is
\begin{equation}\label{L}
  L(v_h)=\frac{1}{2}g_h(v_h,v_h)-V(h), \mbox{ for }v_h\in T_hG,
\end{equation}
$V:G\to \R$ being the potential energy.  The constraint distribution
$D$ is a right-invariant distribution on $G$. Thus, if $e$ is the
identity element of $G$ and ${\frak d} =D_e$, we have that
\begin{equation}\label{eq:D}
  D_h=(T_er_h)({\frak d} )=(T_el_h)(Ad_{h^{-1}}({\frak d})),\mbox{
    for }h\in G
\end{equation}
where $Ad:G\times {\frak g}\to {\frak g}$ is the adjoint action.

The nonholonomic Lagrangian system $(L,D)$ on $TG$ is called a
\emph{nonholonomic LR system} in the terminology of \cite{FeJo,Jo2}.
Note that, since $L$ is a Lagrangian function of mechanical type, the
system $(L,D)$ is regular.  Now, assume that
\[
{\frak s}={\frak d}^{\perp}=\{\omega' \in {\frak g} / <
\omega,\omega'>=0,\forall \omega\in {\frak d}\}
\]
is a Lie subalgebra of ${\frak g}$, that $S$ is a closed Lie subgroup
of $G$ with Lie algebra ${\frak s}$ and that the potential energy $V$
is $S$-invariant.

Next, let us show that the nonholonomic LR system $(L,D)$ may be
reduced to a nonholonomic Lagrangian system on a right action Lie
algebroid.  In fact, consider the Riemannian homogeneous space
$M=S\setminus G$ and the standard transitive right action $\psi$ of
$G$ on $M=S\setminus G$. Denote by $\Psi:{\frak g}\to {\frak
  X}(S\setminus G)$ the corresponding right infinitesimal action of
${\frak g}$ on $S\setminus G$. Then, $\Psi$ induces a Lie algebroid
structure on the trivial vector bundle $pr_1:S\setminus G\times {\frak
  g}\to S\setminus G$.

On the other hand, using that the potential energy $V$ is $S$
invariant, we deduce that $V$ induces a real function
$\widetilde{V}:S\setminus G\to \R$ on $S\setminus G$ such that
\begin{equation}\label{vtil}
  \widetilde{V}\circ \pi=V,
\end{equation}
where $\pi:G\to S\setminus G$ is the canonical projection.  Thus, we
can introduce the Lagrangian function $\tilde{L}: S\setminus G\times
{\frak g}\to \R$ on the action Lie algebroid $pr_1:S\setminus G\times
{\frak g}\to S\setminus G$ defined by
\begin{equation}\label{Ltil}
  \widetilde{L}(\widetilde{h},\omega)=\frac{1}{2}<{\mathcal
    I}(\omega),\omega>-\widetilde{V}(\widetilde{h}),\;\;\; \mbox{ for
  }\widetilde{h}\in S\setminus G \mbox{ and }\omega\in  {\frak g}.
\end{equation}

Now, for every $h\in G$, we consider the subspace ${\frak d} (h)$ of
${\frak g}$ given by
\begin{equation}\label{Deltah}
  {\frak d}(h)=Ad_{h^{-1}}({\frak d}).
\end{equation}
The dimension of ${\frak d}(h)$ is equal to the dimension of
${\frak d}$. Moreover, since $<\cdot,\cdot>$ is $Ad_G$-invariant,
it follows that
\[
{\frak d}(h)=(Ad_{h^{-1}}({\frak s}))^\perp=\{\omega'\in {\frak
  g}/<\omega',Ad_{h^{-1}}(\omega)>=0,\;\;\forall \omega\in {\frak
  s}\}.
\]
In particular, we have that
\[
{\frak d}(s)={\frak s}^\perp={\frak d}, \;\;\forall s\in S
\]
which implies that ${\frak d} (sh)={\frak d} (h)$, for all $h\in
G.$

Therefore, we can define a vector subbundle $D$ of the Lie algebroid
$pr_1:S\setminus G\times {\frak g}\to S\setminus G$ as follows
\begin{equation}\label{Dtil}
  \widetilde{D}_{\widetilde{h}}=\{\widetilde{h}\}\times {\frak d}
  (h),\mbox{ for } \widetilde{h}\in S\setminus G
\end{equation}
with $h\in G$ and $\pi(h)=\widetilde{h}.$ Consequently, the pair
$(\widetilde{L},\widetilde{D})$ is a nonholonomic Lagrangian system
on the action Lie algebroid $pr_1:S\setminus G\times {\frak g}\to
S\setminus G.$

In addition, we may prove the following result
\begin{proposition}\label{propositon7.1}
  \begin{enumerate}
  \item If $\widetilde{\Phi}:TG\to S\setminus G\times {\frak g}$ is
    the map given by
    \begin{equation}\label{Fitil}
      \widetilde{\Phi}(v_h)=(\pi(h),(T_hl_{h^{-1}}(v_h)),\mbox{ for all
      } v_h\in T_hG
    \end{equation}
    then $\widetilde{\Phi}$ is a fiberwise bijective Lie algebroid
    morphism over $\pi$.
  \item The nonholonomic Lagrangian systems
    $(L,D)$ and $(\tilde{L},\widetilde{D})$ on $TG$ and $S/G\times
    {\frak g}$ are $\widetilde{\Phi}$-related, that is,
    \[
    \tilde{L} \circ \widetilde{\Phi}=L,\;\;\;
    \widetilde{\Phi}(D)=\widetilde{D}.
    \]
  \item The system $(\widetilde{L},\widetilde{D})$ is regular and if
    $\gamma:I\to TG$ is a solution of the Lagrange-d'Alembert
    equations for the system $(L,D)$ then $\widetilde{\Phi}\circ
    \gamma: I \to S\setminus G\times {\frak g}$ is a solution of the
    Lagrange-d'Alembert equations for the system
    $(\widetilde{L},\widetilde{D}).$
  \end{enumerate}
\end{proposition}
\begin{proof}
  $(1)$ Consider the standard (right) action $r$ of $G$ on itself
  \[
  r:G\times G\to G, \; \; \; (h,h')\in G\times G\to r_{h'}(h)=hh'\in
  G.
  \]
  As we know, the infinitesimal generator of $r$ associated with an
  element $\omega$ of ${\frak g}$ is
  \[
  \omega_G=\lvec{\omega},
  \]
  where $\lvec{\omega}$ is the left-invariant vector field on $G$
  such that $\lvec{\omega}(e)=\omega$.

  On the other hand, it is clear that the projection $\pi:G\to
  S\setminus G$ is equivariant with respect to the actions $r$ and
  $\psi$. Thus,
  \[
  (T_h\pi)(\lvec{\omega}(h))=\Psi(\omega)(\pi(h)), \mbox{ for } h\in
  G.
  \]
  Therefore, if $\rho:S/G\times {\frak g}\to T(S\setminus G)$ is the
  anchor map of the Lie algebroid $pr_{1}: S\setminus G\times {\frak
    g} \to S\setminus G$,  it follows that
  \[
  \rho(\widetilde{\Phi}(\lvec{\omega}(h))) =
  \rho(\pi(h),\omega)=(T_h\pi)(\rvec{\omega}(h)), \; \; \; \mbox{ for
  } h\in G.
  \]
  Furthermore, since
  \[
  [\lvec{\omega},\lvec{\omega}']=\lvec{[\omega,\omega']_{\frak g}},
  \; \; \; \; \mbox{ for } \omega,\omega'\in {\frak g},
  \]
  we conclude that $\widetilde{\Phi}$ is a Lie algebroid morphism
  over $\pi.$

  In addition, it is obvious that if $h\in G$ then
  \[
  \widetilde{\Phi}_{|T_hG}:T_hG\to \{\pi(h)\}\times {\frak g}
  \]
  is a linear isomorphism.

  \medskip

  \noindent $(2)$ From (\ref{metrica}), (\ref{L}), (\ref{vtil}),
  (\ref{Ltil}) and $(\ref{Fitil})$, we deduce that
  \[
  \widetilde{L}\circ \widetilde{\Phi}=L.
  \]
  Moreover, using (\ref{eq:D}), (\ref{Deltah}), (\ref{Dtil}) and
  (\ref{Fitil}), we obtain that
  \[
  \widetilde{\Phi}(D)=\widetilde{D}.
  \]
  $(3)$ It follows from $(1),$ $(2)$ and using the results of
  Section \ref{sec:reduction} (see Theorem \ref{t5.6}).
\end{proof}

Next, we obtain the necessary and sufficient conditions for a curve
$(\widetilde{h},\omega):I\to S\setminus G\times {\frak g}$ to be a
solution of the Lagrange-d'Alembert equations for the system
$(\widetilde{L},\widetilde{D})$. Let $\flat_{<\cdot,\cdot>}:{\frak
  g}\to {\frak g}^*$ be the linear isomorphism induced by the scalar
product $<\cdot,\cdot>:{\frak g}\times {\frak g}\to \R$ and
$\mathbb{I}:{\frak g}\to {\frak g}^*$ be the inertia operator given by
\begin{equation}\label{Inertia2}
  \mathbb{I}(w_1)(w_2)=<{\mathcal I}(\omega_1),\omega_2>, \mbox{ for
  }\omega_1,\omega_2\in {\frak g}.
\end{equation}
On the other hand, if $\widetilde{h}'\in S\setminus G$ we will denote
by $\Psi_{\tilde{h}'}:{\frak g}\to T_{\widetilde{h}'}(S\setminus G)$
the linear epimorphism defined by
\[
\Psi_{\widetilde{h}'}(\omega')=\Psi({\omega'})(\widetilde{h}'),
\mbox{ for } \omega'\in {\frak g}.
\]

In addition, if $\pi(h')=\widetilde{h}'$, we identify the vector space
$\widetilde{D}_{\widetilde{h}'}$ with the vector subspace ${\frak
  d}(h')$ of ${\frak g}$. Then, using (\ref{LD-edo}), (\ref{Ltil}) and
(\ref{Inertia2}), we deduce that the curve $(\widetilde{h},\omega)$ is
a solution of the Lagrange-d'Alembert equations for the system
$(\widetilde{L},\widetilde{D})$ if and only if
\[
\begin{array}{l}
  \dot{\widetilde{h}}(t)=\Psi_{\widetilde{h}(t)}(\omega(t))\\
  \{\dot{\omega}(t)-\mathbb{I}^{-1}(ad_{\omega(t)}^{*}\mathbb{I}
  (\omega(t)))-\mathbb{I}^{-1}(\Psi^{*}_{\widetilde{h}(t)}
  (d\widetilde{V}(\widetilde{h}(t))))\}\in
  \widetilde{D}_{\widetilde{h}(t)}^\perp,\\
  \omega(t)\in \widetilde{D}_{\widetilde{h}(t)},
\end{array}
\]
for all $t$, where $\widetilde{D}_{\widetilde{h}(t)}^\perp$ is the
orthogonal complement of the vector subspace
$\widetilde{D}_{\widetilde{h}(t)}\subseteq {\frak g}$ with respect
to the scalar product $<\cdot,\cdot>$. These equations will be
called \emph{the reduced Poincar\'e-Chetaev equations}.

We treat next a simple example of the above general situation.

\subsection*{The Veselova system}
The most descriptive illustration of an LR system is the Veselova
problem on the motion of a rigid body about a fixed point under the
action of the nonholonomic constraint
$$<\omega,\gamma>=0.$$
Here, $\omega$ is the vector of the angular velocity in the body
frame, $\gamma$ is a unit vector which is fixed in an space frame and
$<\cdot,\cdot>$ denotes the standard scalar product in $\R^3$
(see~\cite{VeVe}).

The Veselova system is an LR system on the Lie group $G=SO(3)$ which
is the configuration space of the rigid body motion. Thus, in this
case, the Lie algebra ${\frak g}$ may be identified with $\R^3$ and,
under this identification, the Lie bracket $[\cdot,\cdot]_{\frak g}$
is the cross product $\times$ on $\R^3$.  Moreover, the adjoint action
of $G=SO(3)$ on ${\frak g}\cong \R^3$ is the standard action of
$SO(3)$ on $\R^3$. This implies that $<\cdot,\cdot>$ is an
$Ad_{SO(3)}$-invariant scalar product on ${\frak g}\cong \R^3.$

The vector subspace ${\frak d}$ of $\R^3$ is just the orthogonal
complement (with respect to $<\cdot,\cdot>$) of a vector subspace
$<\gamma_0>$ of dimension $1$, with $\gamma_0$ a unit vector in
$\R^3$, that is,
\[
{\frak d}=\{\omega\in \R^3/<\omega,\gamma_0>=0\}.
\]
Therefore,
\[
{\frak s}={\frak d}^\perp = <\gamma_0>
\]
is a Lie subalgebra of ${\frak g}\cong \R^3$. Furthermore, the
isotropy group $S$ of $\gamma_0$ with respect to the adjoint
action of $G=SO(3),$
\[
S=\{s\in SO(3)/s\gamma_0^T=\gamma_0^T\},
\]
is a closed Lie subgroup with Lie algebra ${\frak s}$. We remark
that $S$ is isomorphic to the circle $S^1.$

Consequently, the corresponding homogeneous space $M=S\setminus SO(3)$
is the orbit of the adjoint action of $SO(3)$ on $\R^3$ over the point
$\gamma_0$ and, it is well-known that, such an orbit may be identified
with the unit sphere $S^2$. In fact, the map
\[
S\setminus SO(3)\to S^2, \;\;\; [h]\to \gamma_0
h=(h^{-1}\gamma_0^T)^T
\]
is a diffeomorphism (see, for instance, \cite{MaRa}).

Under the above identification the (right) action of SO(3) on
$M=S\setminus SO(3)$ is just the standard (right) action of $SO(3)$ on
$S^2$. Thus, our action Lie algebroid is the trivial vector bundle
$pr_1:S^2\times \R^3\to S^2$ and the Lie algebroid structure on it is
induced by the standard infinitesimal (right) action $\Psi:\R^3\to
{\frak X}(S^2)$ of the Lie algebra $(\R^3,\times)$ on $S^2$ defined by
\[
\Psi(\omega)(\gamma)=\gamma\times \omega,\mbox{ for } \omega\in \R^3
\mbox{ and } \gamma\in S^2.
\]
In the presence of a potential $\tilde{V}: \gamma\to
\widetilde{V}(\gamma)$ the nonholonomic Lagrangian system
$(\widetilde{L},\widetilde{D})$ on the Lie algebroid
$pr_1:S^2\times \R^3\to \R^3$ is given by
\[
\widetilde{L}(\gamma,\omega)=\frac{1}{2}
\mathbb{I}(\omega)(\omega)-\widetilde{V}(\gamma), \; \; \;
\widetilde{D}(\gamma)=\{\gamma\}\times
\{\omega\in\R^3/<\omega,\gamma>=0\},
\]
$\mathbb{I}:\R^3\to\R^3$ being the inertia tensor of the rigid
body.

The Lagrange-d'Alembert equations for
$(\widetilde{L},\widetilde{D})$ are
\begin{equation}\label{Vese1}
  \dot{\gamma}=\gamma\times \omega,\;\;\;
  \dot{\omega}=\mathbb{I}^{-1}\{(\mathbb{I}\omega\times\omega) +
  \omega\times \frac{\partial \widetilde{V}}{\partial \gamma} +
  \lambda\gamma\},\;\;\; <\omega,\gamma>=0
\end{equation}
where $\lambda$ is the Lagrange multiplier. Since the system
$(\widetilde{L}, \widetilde{D})$ is regular, $\lambda$ is uniquely
determined. In fact,
\begin{equation}\label{Vese2}
  \lambda=-\frac{<\mathbb{I}\omega\times \omega + \gamma\times
    \frac{\partial \widetilde{V}}{\partial \gamma},
    \mathbb{I}^{-1}\gamma>}{<\mathbb{I}^{-1}\gamma,\gamma>}.
\end{equation}
Eqs (\ref{Vese1}) and (\ref{Vese2}) are just the classical dynamical
equations for the Veselova system (see \cite{VeVe}; see also
\cite{FeJo,Jo2}).

\subsection{Semidirect product symmetry and left action Lie
  algebroids}

Here, we show how the reduction of some nonholonomic mechanical
systems with semidirect product symmetry produces nonholonomic
Lagrangian systems on left action Lie algebroids.

Let us start by recalling the definition of a left action Lie
algebroid (see \cite{HiMa}).  Let $(F, [\cdot, \cdot]_{F}, \rho_{F})$
be a Lie algebroid over a manifold $N$ and $\pi: M \to N$ be a smooth
map. \emph{A left action of $F$ on $\pi: M \to N$} is a $\R$-linear
map
\[
\Psi: Sec (F) \to {\frak X}(M), \; \; \; X\in Sec (F) \to
\Psi(X) \in {\frak X}(M)
\]
such that
\[
\begin{array}{l}
  \Psi (f X) = (f \circ \pi)\Psi (X), \; \; \Psi([X, Y]_{F}) =
  -[\Psi(X), \Psi(Y)], \\[5pt] (T_{m}\pi)(\Psi(X)(m)) =
  -\rho_{F}(X(\pi(m))),
\end{array}
\]
for $f \in C^{\infty}(N)$, $X, Y \in Sec (F)$ and $m \in M$. If
$\Psi: Sec (E) \to {\frak X}(M)$ is a left action of $F$ on
$\pi: M \to N$ and $\tau_{F}: F \to N$ is the vector bundle
projection then the pullback vector bundle of $F$ over $\pi$,
\[
E = F^{*}\pi = \{(f, m) \in F \times M / \tau_{F}(f) = \pi (m)\}
\]
is a Lie algebroid over $M$ with Lie algebroid structure
$([\cdot, \cdot]_{E}, \rho_{E})$ which is characterized by
\[
[X, Y]_{E} = [X, Y]_{F} \circ \pi, \; \; \; \rho_{E}(X)(m) =
-\Psi(X)(m),
\]
for $X, Y \in Sec(E)$ and $m \in M$. The triple $(E, [\cdot,
\cdot]_{E}, \rho_{E})$ is called \emph{the left action Lie algebroid
  of $F$ over $\pi$} and it is denoted by $F_{\Psi}\pi$ (see
\cite{HiMa}).

Next, we consider a particular class of nonholonomic Lagrangian
systems on left action Lie algebroids.  Let $V$ be a real vector space
of finite dimension and $\cdot: G \times V \to V$ be a left
representation of a Lie group $G$ on $V$. We also denote by $\cdot:
{\frak g} \times V \to V$ the left infinitesimal representation of the
Lie algebra ${\frak g}$ of $G$ on $V$.  Then, we can consider the
semidirect Lie group $S= G \circledS V$ with the multiplication
\[
(g, v) (g', v') = (gg', v + g \cdot v').
\]
The Lie algebra ${\frak s}$ of $S$ is the semidirect product
${\frak s} = {\frak g} \circledS V$ with the Lie bracket $[\cdot,
\cdot]_{\frak s}: {\frak s} \times {\frak s} \to {\frak s}$ given
by
\[
[(\omega, \dot{v}), (\omega', \dot{v}')]_{\frak s} = ([\omega,
\omega']_{\frak g}, \omega \cdot \dot{v}' - \omega' \cdot \dot{v})
\]
for $\omega, \omega' \in {\frak g}$ and $\dot{v}, \dot{v}' \in V$.
Here, $[\cdot, \cdot]_{\frak g}$ is the Lie bracket on ${\frak
g}$.

Moreover, we use the following notation. If $v \in V$ then $\rho_{v}:
{\frak g} \to V$ is the linear map defined by
\[
\rho_{v}(\omega) = \omega \cdot v, \; \; \; \mbox{ for } \omega
\in {\frak g},
\]
and $\rho_{v}^{*}: V^{*} \to {\frak g}^{*}$ is the dual map of
$\rho_{v}: {\frak g} \to V$.

Now, let $N$ be a smooth manifold. Then, it is clear that the
product manifold $F = {\frak s} \times TN$ is the total space of a
vector bundle over $N$. Moreover, if $(\omega, \dot{v}) \in {\frak
s}$ and $X$ is a vector field on $N$ then the pair $((\omega,
\dot{v}), X)$ defines a section of the vector bundle $\tau_{F}: F
= {\frak s} \times TN \to N$. In fact, if $\{\omega_{i}\}$ is a
basis of ${\frak g}$, $\{\dot{v}_{j}\}$ is a basis of $V$ and
$\{X_{k}\}$ is a local basis of ${\frak X}(N)$ then
$\{((\omega_{i}, 0), 0), ((0, \dot{v}_{j}), 0), ((0, 0), X_{k})\}$
is a local basis of $Sec (F)$.

The vector bundle $\tau_{F}: F \to N$ admits a Lie algebroid structure
$([\cdot, \cdot]_{F}, \rho_{F})$, which is characterized by the
following relations
\begin{equation}\label{LiealgF}
  \begin{array}{rcl}
    [((\omega, \dot{v}), X), ((\omega', \dot{v}'), X')]_{F} & = &
    ([(\omega, \dot{v}), (\omega', \dot{v}')]_{\frak s}, [X, X']) \\
    & = &([\omega, \omega']_{\frak g}, \omega \cdot \dot{v}' - \omega'
    \cdot \dot{v},
    [X, X']), \\[5pt]
    \rho_{F}((\omega, \dot{v}), X) & = & X,
  \end{array}
\end{equation}
for $((\omega, \dot{v}), X), ((\omega', \dot{v}'), X') \in {\frak
s} \times {\frak X}(N)$.

Next, suppose that $v_{0}$ is a point of $V$ and that ${\mathcal
  O}_{v_{0}}$ is the orbit of the action of $G$ on $V$ by $v_{0}$,
that is,
\[
{\mathcal O}_{v_{0}} = \{g \cdot v_{0} \in V / g \in G \}.
\]
Denote by $\pi: M = N \times {\mathcal O}_{v_{0}} \to N$ the
canonical projection on the first factor and by $\Psi: Sec (F) \to
{\frak X}(M)$ the left action of $F$ on $\pi$, which is
characterized by the following relation
\[
\Psi ((\omega, \dot{u}), X)(n, v) = (-X(n), \omega \cdot v)
\]
for $((\omega, \dot{u}), X) \in {\frak s} \times {\frak X}(N)$ and
$(n, v) \in N \times {\mathcal O}_{v_{0}} = M$.

Then, we have the corresponding left action Lie algebroid
$\tau_{E}: E = ({\frak s} \times TN)_{\Psi} \pi \to M = N \times
{\mathcal O}_{v_{0}}$. Note that $E = ({\frak s} \times
TN)_{\Psi}\pi = ({\frak s} \times TN) \times {\mathcal O}_{v_{0}}$
and that the anchor map $\rho_{E}: E = ({\frak s} \times TN)
\times {\mathcal O}_{v_{0}} \to TM = TN \times T{\mathcal
O}_{v_{0}}$ of $\tau_{E}: E \to M$ is given by
\begin{equation}\label{AnclaE}
  \rho_{E}((\omega, \dot{u}), X_{n}, v) = (X_{n}, -\omega \cdot v)
\end{equation}
for $((\omega, \dot{u}), X_{n}, v) \in {\frak s} \times T_{n}N
\times {\mathcal O}_{v_{0}}$.

Now, let $L: ({\frak s} \times TN) \times {\mathcal O}_{v_{0}} \to
\R$ be a Lagrangian function and $D$ be the vector subbundle of
$\tau_{E}: E \to M$ whose fiber $D_{(n, v)}$ over the point $(n,
v) \in N \times {\mathcal O}_{v_{0}} = M$ is defined by
\begin{equation}\label{D}
  \begin{array}{l}
    D_{(n, v)} = \{(((\omega, \omega \cdot v), X_{n}), v) / \omega \in
    {\frak g}, X_{n} \in T_{n}N \} \\ [5pt] \; \; \; \; \subseteq
    E_{(n, v)} = ({\frak s} \times T_{n}N) \times \{v\}.
  \end{array}
\end{equation}
Next, we obtain the Lagrange-d'Alembert equations for the system $(L,
D)$. For this purpose, we choose a basis $\{\omega_{\alpha}\}$ of
${\frak g}$, a basis $\{u_{A}\}$ of $V$, a system of local fibred
coordinates $(x^{i}, \dot{x}^{i})$ on $TN$ and a system of local
coordinates $(v^{i})$ on ${\mathcal O}_{v_{0}}$. Denote by
$\omega^{\alpha}$ (respectively, $u^{A}$) the global coordinates on
${\frak g}$ (respectively, $V$) induced by the basis
$\{\omega_{\alpha}\}$ (respectively, $\{u_{A}\}$).

Suppose that
\[
[\omega_{\alpha}, \omega_{\beta}]_{\frak g} = c_{\alpha
\beta}^{\gamma} \omega_{\gamma}, \; \; \; \omega_{\alpha} \cdot
u_{A} = a_{\alpha A}^{B} u_{B}.
\]
Then, we have that
\[
c_{\alpha \beta}^{\gamma} a_{\gamma A}^{B} = a_{\beta
  A}^{C}a_{\alpha C}^{B} - a_{\alpha A}^{C}a_{\beta C}^{B}.
\]
Next, we consider the local basis of sections $\{e_{i},
e_{\alpha}, e_{A}\}$ of $E$ given by
\[
\begin{array}{rcl}
  e_{i}(n, v) & = & ((0, 0, \frac{\partial}{\partial x^{i}}_{|n}),
  v),  \; \; \; e_{\alpha}(n, v) \; = \; ((\omega_{\alpha},
  \omega_{\alpha} \cdot v, 0_{n}), v) \\ [5pt] e_{A}(n, v) & = &
  ((0, u_{A}, 0_{n}), v)
\end{array}
\]
for $(n, v) \in N \times {\mathcal O}_{v_{0}} = M$. Note that
$\{e_{i}, e_{\alpha}\}$ is a local basis of sections of the constraint
subbundle $D$.  In addition, if $(x^{i}, v^{j}; y^{i}, y^{\alpha},
y^{A})$ are the local coordinates on $E$ induced by the basis
$\{e_{i}, e_{\alpha}, e_{A}\}$, it follows that
\begin{equation}\label{Changecoor}
  y^{i} = \dot{x}^{i}, \; \; \; y^{\alpha} = \omega^{\alpha}, \; \;
  \; y^{A} = u^{A} - a_{\alpha B}^{A}u^{B}_{0} \omega^{\alpha},
\end{equation}
where $u^{B}_{0}$ is the local function on $M = N \times {\mathcal
  O}_{v_{0}}$ defined by $u^{B}_{0} = (u^{B})_{|{\mathcal
    O}_{v_{0}}}.$ Moreover,
\[
\begin{array}{l}
  \rho_{E}(e_{i}) = -\frac{\partial}{\partial
    x^{i}}, \; \; \; \rho_{E}(e_{\alpha}) =
  \rho^{i}_{\alpha}\frac{\partial}{\partial v^{i}}, \; \; \;
  \rho_{E}(e_{A}) = 0, \\[5pt]
  [e_{\alpha}, e_{\beta}]_{E} = c_{\alpha \beta}^{\gamma}
  (e_{\gamma} + a_{\gamma A}^{B}u^{A}_{0} e_{B}), \; \; \;
  [e_{\alpha}, e_{A}]_{E} = -[e_{A}, e_{\alpha}]_{E} = a_{\alpha
    A}^{B} e_{B},
\end{array}
\]
and the rest of the fundamental Lie brackets are zero.  Thus, a curve
\[
t \to (x^{i}(t), v^{j}(t); y^{i}(t), y^{\alpha}(t), y^{A}(t))
\]
is a solution of the Lagrange-d'Alembert equations for the system
$(L, D)$ if and only if
\[
\begin{array}{l}
  \dot{x}^{i} = y^{i}, \; \; \; \dot{v}^{j} =
  \rho_{\alpha}^{j}y^{\alpha}, \mbox{ for all } i \mbox{ and } j,
  \\[5pt]
  \displaystyle \frac{d}{dt}(\displaystyle \frac{\partial
    L}{\partial y^{i}}) - \displaystyle \frac{\partial L}{\partial
    x^{i}} = 0, \mbox{ for all } i,
  \\[8pt]
  \displaystyle \frac{d}{dt}(\displaystyle \frac{\partial
    L}{\partial y^{\alpha}}) + \displaystyle (\frac{\partial
    L}{\partial y^{\gamma}} + \displaystyle \frac{\partial L}{\partial
    y^{B}} a_{\gamma A}^{B}u^{A}_{0})c_{\alpha
    \beta}^{\gamma}y^{\beta} - \rho_{\alpha}^{i}\displaystyle
  \frac{\partial L}{\partial v^{i}} = 0, \mbox{ for all } \alpha,
  \\[8pt]
  y^{A} = 0, \mbox{ for all } A.
\end{array}
\]
If we consider the local expression of the curve in the
coordinates $(x^{i}, v^{j}; \dot{x}^{i}, \omega^{\alpha}, u^{A})$
then, from (\ref{Changecoor}), we deduce that the above equations
are equivalent to
\[
\begin{array}{l}
  \dot{x}^{i} = y^{i}, \; \; \; \dot{v}^{j} =
  \rho_{\alpha}^{j}\omega^{\alpha}, \mbox{ for all } i \mbox{ and }
  j,
  \\[5pt]
  \displaystyle \frac{d}{dt}(\displaystyle \frac{\partial
    L}{\partial \dot{x}^{i}}) - \displaystyle \frac{\partial
    L}{\partial x^{i}} = 0, \mbox{ for all } i,
  \\[8pt]
  \displaystyle \frac{d}{dt}(\displaystyle \frac{\partial
    L}{\partial \omega^{\alpha}}) + \displaystyle \frac{\partial
    L}{\partial \omega^{\gamma}} c_{\alpha \beta}^{\gamma}
  \omega^{\beta} + \displaystyle \frac{d}{dt}(a_{\alpha
    B}^{A}u^{B}_{0} \displaystyle \frac{\partial L}{\partial u^{A}})
  \\[8pt]
  \hspace{1cm} + 2a_{\gamma B}^{A}u^{B}_{0} \displaystyle
  \frac{\partial L}{\partial u^{A}} c_{\alpha
    \beta}^{\gamma}\omega^{\beta} - \rho_{\alpha}^{i} \displaystyle
  \frac{\partial L}{\partial v^{i}} = 0, \mbox{ for all } \alpha,
  \\[5pt]
  u^{A} = a_{\alpha B}^{A}u^{B}_{0} \omega^{\alpha}, \mbox{ for all
  } A,
\end{array}
\]
or, in vector notation,
\[
\begin{array}{l}
  \dot{v} = - \omega \cdot v,
  \\[5pt]
  \displaystyle \frac{d}{dt}(\displaystyle \frac{\partial
    L}{\partial \dot{x}}) - \displaystyle \frac{\partial L}{\partial
    x} = 0,
  \\[8pt]
  \displaystyle \frac{d}{dt}(\displaystyle \frac{\partial
    L}{\partial \omega}) + (ad^{*}_{\omega} \displaystyle
  \frac{\partial L}{\partial \omega}) = -\displaystyle \frac{d}{dt}
  (\rho^{*}_{v} \displaystyle \frac{\partial L}{\partial u}) -2
  ad^{*}_{\omega}(\rho^{*}_{v} \displaystyle \frac{\partial
    L}{\partial u}) - \rho^{*}_{v} \displaystyle \frac{\partial
    L}{\partial v},
  \\[5pt]
  u = \rho_{v}\omega.
\end{array}
\]
Nonholonomic Lagrangian systems, of the above type, on the left
action Lie algebroid $\tau_{E}: E = ({\frak s} \times TN) \times
{\mathcal O}_{v_{0}} \to M = N \times {\mathcal O}_{v_{0}}$ may be
obtained (by reduction) from an standard nonholonomic Lagrangian
system with semidirect product symmetry.

In fact, let $Q$ be the product manifold $S \times N$ and suppose
that we have a Lagrangian function $\tilde{L}: TQ \to \R$ and a
distribution $\tilde{D}$ on $Q$ whose characteristic space
$\tilde{D}_{((g, v), n)} \subseteq T_{g}G \times T_{v}V \times
T_{n}N \simeq T_{g}G \times V \times T_{n}N$ at the point $((g,
v), n) \in S \times N$ is
\begin{equation}\label{Dtilv0}
  \tilde{D}_{((g, v), n)} = \{((\dot{g}, \dot{v}), \dot{n}) \in
  T_{g}G \times V \times T_{n}N / \dot{v} =
  (T_{g}r_{g^{-1}})(\dot{g}) \cdot v_{0} \},
\end{equation}
where $v_{0}$ is a fixed point of $V$.

We can consider the natural left action of the Lie group $S$ on
$Q$ and, thus, the left action $A$ of the Lie subgroup $H_{v_{0}}
= G_{v_{0}} \circledS V$ of $S$ on $Q$, where $G_{v_{0}}$ is the
isotropy group of $v_{0}$ with respect to the action of $G$ on
$V$. The tangent lift $TA$ of $A$ is given by
\begin{equation}\label{TA}
  TA ((\tilde{g}, \tilde{u}), (v_{g}, (v, \dot{v}), X_{n})) =
  ((T_{g}l_{\tilde{g}})(v_{g}), (\tilde{u} + \tilde{g} \cdot v,
  \tilde{g} \cdot \dot{v}), X_{n})
\end{equation}
for $(\tilde{g}, \tilde{u}) \in H_{v_{0}}$ and $(v_{g}, (v,
\dot{v}), X_{n}) \in T_{((g, v), n)}Q \simeq T_{g}G \times V
\times T_{n}N$.

Using (\ref{TA}), it follows that the distribution $\tilde{D}$ is
invariant under the action $TA$ of $H_{v_{0}}$ on $TQ$. Moreover,
we will assume that the Lagrangian function is also
$H_{v_{0}}$-invariant. Therefore, we have a nonholonomic
Lagrangian system $(\tilde{L}, \tilde{D})$ on the standard Lie
algebroid $TQ \to Q$ which is $H_{v_{0}}$-invariant. This type of
systems were considered in \cite{MT:04}.

Since the function $\tilde{L}$ is $H_{v_{0}}$-invariant, we deduce
that there exists a real function $L: ({\frak s} \times TN) \times
{\mathcal O}_{v_{0}} \to \R$ on the left action Lie algebroid
$\tau_{E}: E = ({\frak s} \times TN) \times {\mathcal O}_{v_{0}}
\to M = N \times {\mathcal O}_{v_{0}}$ which is defined by
\begin{equation}\label{Ele}
  L(((\omega, \dot{v}), X_{n}), u) = \tilde{L}((T_{e}l_{g})(\omega),
  (v, g \cdot \dot{v}), X_{n}),
\end{equation}
for $(((\omega, \dot{v}), X_{n}), u) \in {\frak s} \times T_{n}N
\times {\mathcal O}_{v_{0}}$, with $g \in G$, $u = g^{-1}v_{0}$ and $v
\in V$.

Moreover, we may prove the following result.
\begin{proposition}
  \begin{enumerate}
  \item If $\Phi: TQ \simeq TG \times (V \times V) \times TN \to E =
    ({\frak s} \times TN) \times {\mathcal O}_{v_{0}}$ and $\varphi: G
    \times V \times N \to N \times {\mathcal O}_{v_{0}}$ are the maps
    defined by
    \begin{equation}\label{Defmor}
      \begin{array}{l}
        \Phi(u_{g}, (v, \dot{v}), X_{n}) = ((((T_{g}l_{g^{-1}})(u_{g}),
        g^{-1} \cdot \dot{v}), X_{n}), g^{-1} \cdot v_{0}), \\[5pt]
        \varphi(g, v, n) = (n,
        g^{-1} \cdot v_{0}),
      \end{array}
    \end{equation}
    then $\Phi$ is a fiberwise bijective Lie algebroid morphism over
    $\varphi$.
  \item The nonholonomic Lagrangian systems $(\tilde{L}, \tilde{D})$
    and $(L, D)$ on $TQ$ and $E = ({\frak s} \times TN)\times
    {\mathcal O}_{v_{0}}$ are $\Phi$-related, that is,
    \[
    L \circ \Phi = \tilde{L}, \; \; \; \Phi(\tilde{D}) = D.
    \]
    Here, $D$ is the vector subbundle of the vector bundle $E$ whose
    fiber at the point $(n, v) \in N \times {\mathcal O}_{v_{0}}$ is
    given by (\ref{D}).
  \item If the system $(\tilde{L}, \tilde{D})$ is regular then the
    system $(L, D)$ is also regular. In addition, if $\gamma: I \to
    TQ$ is a solution of the Lagrange-d'Alembert equations for
    $(\tilde{L}, \tilde{D})$ then $\Phi \circ \gamma: I \to ({\frak s}
    \times TN) \times {\mathcal O}_{v_{0}}$ is a solution of the
    Lagrange-d'Alembert equations for $(L, D)$.
  \end{enumerate}
\end{proposition}
\begin{proof}
  $(1)$ Suppose that $\omega_{1}$ and $\omega_{2}$ are elements of
  ${\frak g}$, that $\dot{v}_{1}$ and $\dot{v}_{2}$ are vectors of $V$
  and that $X_{1}$ and $X_{2}$ are vector fields on $N$. Then, we
  consider the vector fields $Z_{1}$ and $Z_{2}$ on $Q$ defined by
  \[
  \begin{array}{l}
    Z_{1}(g, v, n) = (\lvec{\omega}_{1}(g), g \cdot \dot{v}_{1},
    X_{1}(n))
    \in T_{g}G \times V \times T_{n}N, \\[5pt]
    Z_{2}(g, v, n) = (\lvec{\omega}_{2}(g), g \cdot \dot{v}_{2},
    X_{2}(n)) \in T_{g}G \times V \times T_{n}N,
  \end{array}
  \]
  for $(g, v, n) \in G \times V \times N = Q$, where
  $\lvec{\omega}_{1}$ (respectively, $\lvec{\omega}_{2}$) is the
  left-invariant vector field on $G$ such that $\lvec{\omega}_{1}(e)
  = \omega_{1}$ (respectively, $\lvec{\omega}_{2}(e) = \omega_{2}$),
  $e$ being the identity element of $G$.

  A direct computation proves that
  \[
  [Z_{1}, Z_{2}] (g, v, n) = (\lvec{[\omega_{1},
    \omega_{2}]}_{{\frak g}}(g), g(\omega_{1} \cdot \dot{v}_{2} -
  \omega_{2} \cdot \dot{v}_{1}), [X_{1}, X_{2}](n)).
  \]
  Moreover, if $((\omega_{1}, \dot{v}_{1}), X_{1})$ (respectively,
  $((\omega_{2}, \dot{v}_{2}), X_{2})$) is the section of the vector
  bundle $\tau_{E}: E \to M$ induced by $\omega_{1}$, $\dot{v}_{1}$
  and $X_{1}$ (respectively, $\omega_{2}$, $\dot{v}_{2}$ and
  $X_{2}$) then it is clear that
  \[
  \Phi \circ Z_{1} = ((\omega_{1}, \dot{v}_{1}), X_{1}) \circ
  \varphi, \; \; \; \Phi \circ Z_{2} = ((\omega_{2}, \dot{v}_{2}),
  X_{2}) \circ \varphi .
  \]
  Thus, using (\ref{LiealgF}), it follows that
  \begin{equation}\label{Bracket}
    \Phi \circ [Z_{1}, Z_{2}]  = [((\omega_{1}, \dot{v}_{1}), X_{1}),
    ((\omega_{2}, \dot{v}_{2}), X_{2})]_{E} \circ \varphi.
  \end{equation}
  On the other hand, we have that
  \[
  \begin{array}{l}
    (T_{(g, v, n)}\varphi)(u_{g}, \dot{v}, X_{n}) = (X_{n},
    -(T_{g}l_{g^{-1}})(u_{g}) \cdot (g^{-1} \cdot v_{0})) \\[5pt]
    \in T_{n}N \times T_{g^{-1} \cdot v_{0}}{\mathcal O}_{v_{0}}
    \subseteq T_{n}N \times V,
  \end{array}
  \]
  for $(g, v, n) \in Q$ and $(u_{g}, \dot{v}, X_{n}) \in T_{g}G \times
  V \times T_{n}N \simeq T_{(g, v, n)}Q$.

  Therefore, from (\ref{AnclaE}) and (\ref{Defmor}), we deduce that
  \begin{equation}\label{Anchor}
    T\varphi = \rho_{E} \circ \Phi.
  \end{equation}
  Consequently, using (\ref{Bracket}) and (\ref{Anchor}), we
  conclude that the pair $(\Phi, \varphi)$ is a Lie algebroid
  morphism. Note that one may choose a local basis $\{Z_{i}\}$ of
  vector fields on $Q$ such that
  \[
  Z_{i}(g, v, n) = (\lvec{\omega}_{i}(g), g \cdot \dot{v}_{i},
  X_{i}(n)), \; \; \mbox{ for } (g, v, n) \in Q
  \]
  with $\omega_{i} \in {\frak g}$, $\dot{v}_{i} \in V$ and $X_{i}
  \in {\frak X}(N)$.

  Finally, if $(g, v, n) \in Q$, it is clear that
  \[
  \Phi_{|T_{(g,v,n)}Q}: T_{(g,v,n)}Q \simeq T_{g}G \times V \times
  T_{n}N \to E_{(n, g^{-1} \cdot v_{0})} \simeq {\frak g} \times V
  \times T_{n}N
  \]
  is a linear isomorphism.

  $(2)$ It follows from (\ref{D}), (\ref{Dtilv0}), (\ref{Ele}) and
  (\ref{Defmor}).

  $(3)$ It follows using $(1)$, $(2)$ and the results of Section
  \ref{sec:reduction} (see Theorem \ref{t5.6}).
\end{proof}

The above theory may be applied to a particular example of a
mechanical system: \emph{the Chaplygin Gyro} (see \cite{Ma,MT:04}).
This system consists of a Chaplygin sphere (that is, a ball with
nonhomogeneous mass distribution) with a gyro-like mechanism,
consisting of a gimbal and a pendulous mass, installed in it. The
gimbal is a circle-like structure such that its center coincides with
the geometric center of the Chaplygin sphere. It is free to rotate
about the axis connecting the north and south poles of the Chaplygin
sphere. The pendulous mass can move along the smooth track of the
gimbal. For this particular example, the vector space $V$ is $\R^{3}$,
the Lie group $G$ is $SO(3)$ and the manifold $N$ is $\R^{2}$. The
action of $SO(3)$ on $\R^{3}$ is the standard one and $v_{0} = (0, 0,
1)$ is the advected parameter, see~\cite{MT:04} for more details.

\newcommand{\JK}{J\!K}

\subsection{Chaplygin-type systems}
A frequent situation is the following.  Consider a constrained
Lagrangian system $(L,D)$ on a Lie algebroid $\map{\tau}{E}{M}$ such
that the restriction of the anchor to the constraint distribution,
$\map{\rho|D}{D}{TM}$, is an isomorphism of vector bundles.  Let
$\map{h}{TM}{D\subset E}$ be the right-inverse of $\rho|_D$, so that
$\rho\circ h=\id_{TM}$. It follows that $E$ is a transitive Lie
algebroid and $h$ is a splitting of the exact sequence
\[
\xymatrix{0\ar[r]&\Ker(\rho)\ar[r]&E\ar[r]^\rho&TM\ar[r]&0\,.}
\]

Let us define the function $\bar{L}\in\cinfty{TM}$ by $\bar{L}=L\circ
h$.  The dynamics defined by $L$ does not reduce to the dynamics
defined by $\bar{L}$ because, while the map $\Phi=\rho$ is a morphism
of Lie algebroids and $\Phi(D)=TM$, we have $\bar{L}\circ\Phi=L\circ
h\circ\rho\neq L$. Nevertheless, we can use $h$ to express the
dynamics on $TM$, by finding relations between the dynamics defined by
$L$ and $\bar{L}$.

We need some auxiliary properties of the splitting $h$ and its
prolongation. We first notice that $h$ is an admissible map over the
identity in $M$, because $\rho_E\circ h=\id_{TM}$ and
$T\id_M\circ\rho_{TM}=id_{TM}$, but in general $h$ is not a morphism.
We can define the tensor $K$, a $\ker(\rho)$-valued differential
2-form on $M$, by means of
\[
K(X,Y)=[h\circ X,h\circ Y]-h\circ[X,Y]
\]
for every $X,Y\in\mathfrak{X}(M)$. It is easy to see that $h$ is a
morphism if and only if $K=0$. In coordinates $(x^i)$ in $M$,
$(x^i,v^i)$ in $TM$, and linear coordinates $(x^i,y^i,y^A)$ on $E$
corresponding to a local basis $\{e^i,e^A\}$ of sections of $E$
adapted to the splitting $h$, we have that
\[
K=\frac{1}{2}\Omega_{ij}^A\,dx^i\wedge dx^j\otimes e_A,
\]
where $\Omega_{ij}^A$ are defined by $[e_i,e_j]=\Omega_{ij}^Ae_A$.

Since $h$ is admissible, its prolongation $\prol[h]{h}$ is a
well-defined map from $T(TM)$ to $\TEE$. Moreover, it is an
admissible map, which is a morphism if and only if $h$ is a
morphism. In what respect to the energy and the Cartan 1-form, we
have that $(\prol[h]{h})\pb E_L=E_{\bar{L}}$ and
$(\prol[h]{h})\pb\theta_L=\theta_{\bar{L}}$. Indeed, notice that by
definition, $(\prol[h]{h})\pb E_L= E_L\circ h$ and
\begin{align*}
  E_L(h(v))
  &=\frac{d}{dt}L(h(v)+t(h(v))|_{t=0}-L(h(v))
  =\frac{d}{dt}L(h(v+tv))|_{t=0}-L(h(v))\\
  &=\frac{d}{dt}\bar{L}(v+tv)|_{t=0}-\bar{L}(v) =E_{\bar{L}}(v).
\end{align*}
On the other hand, for every $V_v\equiv (v,w,V)\in T(TM)\equiv
\prol[TM]{(TM)}$ where $w=T\tau(V)$, we have
\begin{align*}
  \pai{(\prol[h]{h})\pb\theta_L}{V}
  &=\pai{\theta_L}{\prol[h]{h}(v,w,V)}
  =\pai{\theta_L}{(h(v),h(w),Th(V))}\\
  &=\frac{d}{dt}L(h(v)+t(h(w))|_{t=0}
  =\frac{d}{dt}L(h(v+tw))|_{t=0}\\
  &=\frac{d}{dt}\bar{L}(v+tw)|_{t=0} =\pai{\theta_{\bar{L}}}{V}.
\end{align*}

Nevertheless, since $h$ is not a morphism, and hence
$(\prol[h]{h})\pb\circ d\neq d\circ (\prol[h]{h})\pb$, we have that
$(\prol[h]{h})\pb\omega_L\neq\omega_{\bar{L}}$. Let $\JK$ be the
2-form on $TM$ defined by
\[
\JK_v(V,W)=\pai{J_{h(v)}}{K_{h(v)}(T\tau_M(V),T\tau_M(W))}
\]
where $J$ is the momentum map defined by $L$ and $\Ker{\rho}$ and
$V,W\in T_{h(v)}(TM)$. The notation resembles the contraction of
the momentum map $J$ with the curvature tensor $K$. Instead of
being symplectic, the map $\prol[h]{h}$ satisfies
\[
(\prol[h]{h})\pb\omega_L=\omega_{\bar{L}}+\JK.
\]
Indeed, we have that
\[
(\prol[h]{h})\pb\omega_L-\omega_{\bar{L}}=
[d\circ(\prol[h]{h})\pb-(\prol[h]{h})\pb\circ d]\,\theta_L
\]
and on a pair of projectable vector fields $U,V$ projecting onto
$X,Y$ respectively, one can easily prove that
\[
[d\circ(\prol[h]{h})\pb-(\prol[h]{h})\pb\circ d]\,\theta_L(U,V)
=\pai{\theta_L}{[\prol[h]{h}(U),\prol[h]{h}(V)]-\prol[h]{h}([U,V])}
\]
from where the result follows by noticing that $\prol[h]{h}\circ
U$ is a projectable section and projects to $h\circ X$, and
similarly $\prol[h]{h}\circ V$ projects to $h\circ Y$. Hence
$[\prol[h]{h}(U),\prol[h]{h}(V)]-\prol[h]{h}([U,V]$ is projectable
and projects to $K(X,Y)$.

Let now $\Gamma$ be the solution of the nonholonomic dynamics for $(L,
D)$, so that $\Gamma$ satisfies the equation
$i_\Gamma\omega_L-dE_L\in\tDo$ and the tangency condition
$\Gamma\big|_D\in\TDD$.  From this second condition we deduce the
existence of a vector field $\bar{\Gamma}\in\mathfrak{X}(TM)$ such
that $\prol[h]{h}\circ\bar{\Gamma}=\Gamma\circ h$. Explicitly, the
vector field $\bar{\Gamma}$ is defined by
$\bar{\Gamma}=\prol[\rho]{\rho}\circ\Gamma\circ h$, from where it
immediately follows that $\bar{\Gamma}$ is a \sode\ vector field on
$M$.

Taking the pullback by $\prol[h]{h}$ of the first equation we get
$(\prol[h]{h})\pb\bigl( i_\Gamma\omega_L-dE_L\bigr)=0$ since
$(\prol[h]{h})\pb\tDo=0$. Therefore
\begin{align*}
  0
  &=(\prol[h]{h})\pb i_\Gamma\omega_L-(\prol[h]{h})\pb dE_L\\
  &=i_{\bar{\Gamma}}(\prol[h]{h})\pb\omega_L-d (\prol[h]{h})\pb E_L\\
  &=i_{\bar{\Gamma}}\bigl(\omega_{\bar{L}}+\JK)-dE_{\bar{L}}\\
  &=i_{\bar{\Gamma}}\omega_{\bar{L}}-dE_{\bar{L}}+i_{\bar{\Gamma}}\JK.
\end{align*}
Therefore, the vector field $\bar{\Gamma}$ is determined by the
equations
\[
i_{\bar{\Gamma}}\omega_{\bar{L}}-dE_{\bar{L}}=-\pai{J}{K(\mathbb{T},\,
  \cdot\, )},
\]
where $\mathbb{T}$ is the identity in $TM$ considered as a vector
field along the tangent bundle projection $\tau_M$ (also known as the
total time derivative operator). Equivalently we can write these
equations in the form
\[
d_{\bar{\Gamma}}\theta_{\bar{L}}-d\bar{L}=\pai{J}{K(\mathbb{T},\,
  \cdot\, )}.
\]

Note that if $\bar{a}: I \to TM$ is an integral curve of
$\bar{\Gamma}$ then $a = h \circ \bar{a}: I \to D$ is a solution of
the constrained dynamics for the nonholonomic Lagrangian system $(L,
D)$ on $E$. Conversely, if $a: I \to D$ is a solution of the
constrained dynamics then $\rho \circ a: I \to TM$ is an integral
curve of the vector field $\bar{\Gamma}$.

Finally we mention that extension of the above decomposition for non
transitive Lie algebroids is under development.

\subsection*{Chaplygin systems and Atiyah algebroids}
A particular case of the above theory is that of ordinary Chaplygin
systems(see~\cite{BlKrMaMu, CaCoLeMa,cortes,Ko} and references there
in). In such case we have a principal $G$-bundle
$\map{\pi}{Q}{M=Q/G}$. Then, we may consider the quotient vector
bundle $E = TQ/G \to M=Q/G$ and, it is well-known that, the space of
sections of this vector bundle may be identified with the set of
$G$-invariant vector fields on $Q$. Thus, using that the Lie bracket
of two $G$-invariant vector fields is also $G$-invariant and the fact
that a $G$-invariant vector field is $\pi$-projectable, we may define
a Lie algebroid structure $([\cdot , \cdot], \rho)$ on the vector
bundle $E= TQ/G \to M = Q/G$. The resultant Lie algebroid is called
the {\bf Atiyah (gauge) algebroid} associated with the principal
bundle $\pi: Q \to M = Q/G$ (see \cite{Mackenzie}).  Note that the
canonical projection $\Phi: TQ \to E= TQ/G$ is a fiberwise bijective
Lie algebroid morphism. Now, suppose that $(L_{Q}, D_{Q})$ is an
standard nonholonomic Lagrangian system on $TQ$ such that $L_{Q}$ is
$G$-invariant and $D_{Q}$ is the horizontal distribution of a
principal connection on $\pi: Q \to M= Q/G$.  Then, we have a reduced
nonholonomic Lagrangian system $(L, D)$ on $E$. In fact, $L_{Q} = L
\circ \Phi$ and $\Phi((D_{Q})_{q}) = D_{\pi(q)}$, for all $q \in Q$.
Moreover, $\rho_{|D}: D \to TM = T(Q/G)$ is an isomorphism (over the
identity of M) between the vector bundles $D \to M$ and $TM \to M$.
Therefore, we may apply the above general theory.

Next, we describe the nonholonomic Lagrangian system on the Atiyah
algebroid associated with a particular example of a Chaplygin system:
a two-wheeled planar mobile robot (see \cite{cortes} and the
references there in).  Consider the motion of two-wheeled planar
mobile robot which is able to move in the direction in which it points
and, in addition, can spin about a vertical axis. Let $P$ be the
intersection point of the horizontal symmetry axis of the robot and
the horizontal line connecting the centers of the two wheels. The
position and orientation of the robot is determined, with respect to a
fixed Cartesian reference frame by $(x, y, \theta) \in SE(2)$, where
$\theta \in S^1$ is the heading angle, the coordinates $(x, y) \in
\R^{2}$ locate the point $P$ and $SE(2)$ is the group of Euclidean
motions of the two-dimensional plane $\R^{2}$. Let $\psi_{1}, \psi_{2}
\in S^1$ denote the rotation angles of the wheels which are assumed to
be controlled independently and roll without slipping on the floor.
The configuration space of the system is $Q = \mathbb{T}^{2} \times
SE(2)$, where $\mathbb{T}^{2}$ is the real torus of dimension $2$.

The Lagrangian function $L_{Q}$ is the kinetic energy corresponding to
the metric $g_{Q}$
\[
\begin{array}{rcl}
  g_{Q} &=& m dx \otimes dx + m dy \otimes dy + m_{0}l \cos \theta
  (dy \otimes d\theta + d\theta \otimes dy)\\&& - m_{0}l \sin \theta
  (dx \otimes d\theta + d\theta \otimes dx) + J d\theta \otimes
  d\theta + J_{2} d\psi_{1} \otimes d\psi_{1} + J_{2} d\psi_{2}
  \otimes d\psi_{2},
\end{array}
\]
where $m = m_{0} + 2m_{1}$, $m_{0}$ is the mass of the robot without
the wheels, $J$ its momenta of inertia with respect to the vertical
axis, $m_{1}$ the mass of each wheel, $J_{2}$ the axial moments of
inertia of the wheels, and $l$ the distance between the center of mass
$C$ of the robot and the point $P$. Thus,
\[
\begin{array}{rcl}
  L_{Q} &=& \displaystyle \frac{1}{2} (m \dot{x}^{2} + m \dot{y}^{2}
  + 2 m_{0}l \dot{y}\dot{\theta} \cos \theta - 2 m_{0}l
  \dot{x}\dot{\theta}\sin \theta \\ && + J\dot{\theta}^{2} +
  J_{2}\dot{\psi}_{1}^{2} + J_{2}\dot{\psi}_{2}^{2}).
\end{array}
\]

The constraints, induced by the conditions that there is no lateral
sliding of the robot and that the motion of the wheels also consists
of a rolling without sliding, are
\[
\begin{array}{rcl}
  \dot{x}\sin \theta - \dot{y} \cos \theta &=& 0,\\
  \dot{x}\cos \theta + \dot{y}\sin \theta + c\dot{\theta} +
  R\dot{\psi}_{1} &= & 0,
  \\
  \dot{x}\cos \theta + \dot{y}\sin \theta - c\dot{\theta} +
  R\dot{\psi}_{2} &=& 0,
\end{array}
\]
where $R$ is the radius of the wheels and $2c$ the lateral length
of the robot. The constraint distribution $D$ is then spanned by
\[
\begin{array}{lr}
  \{H_{1} = \displaystyle \frac{\partial}{\partial \psi_{1}} -
  \frac{R}{2}(\cos \theta \frac{\partial}{\partial x} +\sin \theta
  \frac{\partial}{\partial y} + \frac{1}{c} \frac{\partial}{\partial
    \theta}),& \\
  & \kern-38pt H_{2} = \displaystyle \frac{\partial}{\partial \psi_{2}} -
  \frac{R}{2}(\cos \theta \frac{\partial}{\partial x} +\sin \theta
  \frac{\partial}{\partial y} - \frac{1}{c} \frac{\partial}{\partial
    \theta})\},
\end{array}
\]

Note that if $\{\xi_{1}, \xi_{2}, \xi_{3}\}$ is the canonical basis of
${\frak se}(2)$,
\[
[\xi_{1}, \xi_{2}] = 0, \; \; [\xi_{1}, \xi_{3}] = -\xi_{2}, \; \;
[\xi_{2}, \xi_{3}] = \xi_{1},
\]
then
\[
H_{1} = \displaystyle \frac{\partial}{\partial \psi_{1}} -
\frac{R}{2} \lvec{\xi_{1}} - \frac{R}{2c} \lvec{\xi_{3}}, \; \;
H_{2} = \displaystyle \frac{\partial}{\partial \psi_{2}} -
\frac{R}{2} \lvec{\xi_{1}} + \frac{R}{2c} \lvec{\xi_{3}},
\]
where $\lvec{\xi_{i}}$ ($i = 1, 2, 3$) is the left-invariant
vector field of $SE(2)$ such that $\lvec{\xi_{i}}(e) = \xi_{i}$,
$e$ being the identity element of $SE(2)$.

On the other hand, it is clear that $Q = \mathbb{T}^{2} \times SE(2)$
is the total space of a trivial principal $SE(2)$-bundle over $M =
\mathbb{T}^{2}$. Moreover, the metric $g_{Q}$ is $SE(2)$-invariant and
$D_{Q}$ is the horizontal distribution of a principal connection on $Q
= \mathbb{T}^{2} \times SE(2) \to \mathbb{T}^{2}$.

Now, we consider the corresponding Atiyah algebroid
\[
E = TQ/SE(2) \simeq (T\mathbb{T}^{2} \times TSE(2))/SE(2) \to M =
\mathbb{T}^{2}.
\]
Using the left-translations on $SE(2)$, we have that the tangent
bundle to $SE(2)$ may be identified with the product manifold $SE(2)
\times {\frak se}(2)$ and, under this identification, the Atiyah
algebroid is isomorphic to the trivial vector bundle
\[
\tilde{\tau}_{\mathbb{T}^{2}} = \tau_{\mathbb{T}^{2}} \circ
pr_{1}: T\mathbb{T}^{2} \times {\frak se}(2) \to \mathbb{T}^{2},
\]
where $\tau_{\mathbb{T}^{2}}: T\mathbb{T}^{2} \to \mathbb{T}^{2}$
is the canonical projection. In addition, if $([ \cdot , \cdot ],
\rho)$ is the Lie algebroid structure on
$\tilde{\tau}_{\mathbb{T}^{2}}: T\mathbb{T}^{2} \times {\frak
  se}(2) \to \mathbb{T}^{2}$ and $\{\displaystyle
\frac{\partial}{\partial \psi_{1}}, \frac{\partial}{\partial
  \psi_{2}}, \xi_{1}, \xi_{2}, \xi_{3} \}$ is the canonical basis of
sections of $\tilde{\tau}_{\mathbb{T}^{2}}: T\mathbb{T}^{2} \times
{\frak se}(2) \to \mathbb{T}^{2}$ then
\[
\begin{array}{rclrclrclrcl}
  \rho(\displaystyle \frac{\partial}{\partial \psi_{1}}) &=&
  \displaystyle \frac{\partial}{\partial \psi_{1}}, \; \; &
  \rho(\displaystyle \frac{\partial}{\partial \psi_{2}}) & = &
  \displaystyle \frac{\partial}{\partial
    \psi_{2}}, \; \; &\rho(\xi_{i}) & = &0, \; \;& i = 1, 2, 3 \\[8pt]
  [\xi_{1}, \xi_{3}] & = & -\xi_{2}, \; \; & [\xi_{2}, \xi_{3}] & =
  & \xi_{1}, && &&
\end{array}
\]
and the rest of the fundamental Lie brackets are zero.

Denote by $(\psi_{1}, \psi_{2}, \dot{\psi}_{1}, \dot{\psi}_{2},
\omega^{1}, \omega^{2}, \omega^{3})$ the (local) coordinates on
$T\mathbb{T}^{2} \times {\frak se}(2)$ induced by the basis
$\{\displaystyle \frac{\partial}{\partial \psi_{1}},
\frac{\partial}{\partial \psi_{2}}, \xi_{1}, \xi_{2}, \xi_{3} \}$.
Then, the reduced Lagrangian $L: T\mathbb{T}^{2} \times {\frak se}(2)
\to \R$ is given by
\[
L = \displaystyle \frac{1}{2} (m (\omega^{1})^{2} + m
(\omega^{2})^{2} + 2 m_{0}l \omega^{2}\omega^{3} +
J(\omega^{3})^{2} + J_{2}\dot{\psi}_{1}^{2} +
J_{2}\dot{\psi}_{2}^{2})
\]
and the constraint vector subbundle $D$ is generated by the
sections
\[
e_{1} = \displaystyle \frac{\partial}{\partial \psi_{1}} -
\frac{R}{2}\xi_{1} - \frac{R}{2c} \xi_{3}, \; \; e_{2} =
\displaystyle \frac{\partial}{\partial \psi_{2}} - \frac{R}{2}
\xi_{1} + \frac{R}{2c} \xi_{3}.
\]
Since the system $(L_{Q}, D_{Q})$ is regular on the standard Lie
algebroid $\tau_{Q}: TQ \to Q$, we deduce that the nonholonomic
Lagrangian system $(L, D)$ on the Atiyah algebroid
$\tilde{\tau}_{\mathbb{T}^{2}}: T\mathbb{T}^{2} \times {\frak se}(2)
\to \mathbb{T}^{2}$ is also regular.

Now, as in Section \ref{linear}, we consider a basis of sections of
$\tilde{\tau}_{\mathbb{T}^{2}}: T\mathbb{T}^{2} \times {\frak se}(2)
\to \mathbb{T}^{2}$ which is adapted to the constraint subbundle $D$.
This basis is
\[
\{e_{1}, e_{2}, \xi_{1}, \xi_{2}, \xi_{3}\}.
\]
The corresponding (local) coordinates on $T\mathbb{T}^{2} \times
{\frak se}(2)$ are $(\psi_{1}, \psi_{2}, \dot{\psi}_{1},
\dot{\psi}_{2}, \tilde{\omega}^{1}, \tilde{\omega}^{2},
\tilde{\omega}^{3})$, where
\[
\omega^{1} = \tilde{\omega}^{1} - \displaystyle \frac{R}{2}
\dot{\psi}_{1} - \frac{R}{2}\dot{\psi}_{2}, \; \; \omega^{2} =
\tilde{\omega}^{2}, \; \; \omega^{3} = \tilde{\omega}^{3} -
\displaystyle \frac{R}{2c} \dot{\psi}_{1} +
\frac{R}{2c}\dot{\psi}_{2}.
\]
Therefore, using (\ref{LD-edo}), we deduce that the
Lagrange-d'Alembert equations for the system $(L, D)$ are
\[
\begin{array}{rclrcl}
  \ddot{\psi}_{1} & = & \displaystyle \frac{U(\dot{\psi}_{2} -
    \dot{\psi}_{1})}{P^{2} - S^{2}} (P\dot{\psi}_{2} +
  S\dot{\psi}_{1}), \; \; & \ddot{\psi}_{2} & = & \displaystyle -
  \frac{U(\dot{\psi}_{2} - \dot{\psi}_{1})}{P^{2} - S^{2}}
  (P\dot{\psi}_{1} + S\dot{\psi}_{2}),\\ [8pt]
  \tilde{\omega}^{1}
  &=& \tilde{\omega}^{2} = \tilde{\omega}^{3} = 0,&&&
\end{array}
\]
where $P$, $S$ and $U$ are the real numbers
\[
P = \displaystyle \frac{R^{2}}{4} (m + \frac{J}{c^{2}}) + J_{2},
\; \; S = \displaystyle \frac{R^{2}}{4} (m - \frac{J}{c^{2}}), \;
\; U = \displaystyle \frac{R^{3}}{4c^{2}}m_{0}l.
\]
On the other hand, the Lagrangian function $\bar{L}:
T\mathbb{T}^{2} \to \R$ on $T\mathbb{T}^{2}$ is given by
\[
\bar{L}(\psi_{1}, \psi_{2}, \dot{\psi}_{1}, \dot{\psi}_{2}) =
\displaystyle \frac{1}{2} (P \dot{\psi}_{1}^{2} + P
\dot{\psi}_{2}^{2} + 2S \dot{\psi}_{1} \dot{\psi}_{2})
\]
and the $1$-form $\pai{J}{K(\mathbb{T},\, \cdot\, )}$ on
$T\mathbb{T}^{2}$ is
\[
\pai{J}{K(\mathbb{T},\, \cdot\, )} = -U (\dot{\psi}_{2} -
\dot{\psi}_{1}) (\dot{\psi}_{1} d\psi_{2} - \dot{\psi}_{2}
d\psi_{1}).
\]


\section{Nonlinearly constrained Lagrangian systems}
\label{nonlinear}

We show in this section how the main results for linearly constrained
Lagrangian systems can be extended to the case of Lagrangian systems
with nonlinear nonholonomic constraints. This is true under the
assumption that a suitable version of the classical Chetaev's
principle in nonholonomic mechanics is valid (see e.g.,~\cite{LeMa2}
for the study of standard nonholonomic Lagrangian systems subject to
nonlinear constraints).

Let $\tau: E \to M$ be a Lie algebroid and $\cm$ be a submanifold of
$E$ such that $\map{\pi=\tau|_\cm}{\cm}{M}$ is a fibration.  $\cm$ is
the constraint submanifold. Since $\pi$ is a fibration, the
prolongation $\prol[E]{\cm}$ is well-defined. We will denote by $r$
the dimension of the fibers of $\map{\pi}{\cm}{M}$, that is,
$r=\operatorname{dim}\cm-\operatorname{dim} M$.

We define the bundle $\vd\to \cm$ of \emph{virtual displacements} as
the subbundle of $\tau^*E$ of rank $r$ whose fiber at a point
$a\in\cm$ is
\[
\vd[a]=\set{b\in E_{\tau(a)}}{b_a\spV\in T_a\cm}.
\]
In other words, the elements of $\vd$ are pairs of elements
$(a,b)\in E\oplus E$ such that
\[
\frac{d}{dt}\phi(a+tb)\at{t=0}=0,
\]
for every local constraint function $\phi$.

We also define the bundle of \emph{constraint forces} $\cf$ by
$\cf=S^*((\prol[E]{\cm})^\circ)$, in terms of which we set the
Lagrange-d'Alembert equations for a regular Lagrangian function $L \in
C^{\infty}(E)$ as follows:
\begin{equation}\label{8.1}
  \begin{array}{ll}
    &(i_\Gamma\omega_L-dE_L)|_\cm\in\Sec{\cf}, \\[5pt]
    &\Gamma|_\cm\in\Sec{\prol[E]{\cm}},
  \end{array}
\end{equation}
the unknown being the section $\Gamma$. The above equations
reproduce the corresponding ones for standard nonlinear
constrained systems.

From (\ref{2.4'}) and (\ref{8.1}), it follows that
\[
(i_{S\Gamma}\omega_{L} - i_{\Delta}\omega_{L})|_\cm = 0,
\]
which implies that a solution $\Gamma$ of equations (\ref{8.1}) is a
\sode\ section along $\cm$, that is, $(S\Gamma - \Delta)| \cm = 0$.

Note that the rank of the vector bundle $(\prol[E]{\cm})^\circ \to
\cm$ is $s = \rank{E}-r$ and, since $\pi$ is a fibration, the
transformation $S^*: (\prol[E]{\cm})^{\circ} \to \Psi $ defines an
isomorphism between the vector bundles $(\prol[E]{\cm})^{\circ}
\to \cm$ and $\Psi \to \cm$. Therefore, the rank of $\Psi$ is also
$s$. Moreover, if $a \in \cm$ we have
\begin{equation}\label{psia}
  \Psi_a=S^*((\prol[E]{\cm}[a])^\circ)=\set{\zeta\circ
  \prol{\tau}}{\zeta\in {\mathcal V}_a^\circ}.
\end{equation}
In fact, if $\alpha_a\in (\prol[E]{\cm}[a])^\circ$, we may define
$\zeta\in E_{\tau(a)}^*$ by
\[
\zeta(b)=\alpha_a(\xi\spV(a,b)),\mbox{
  for } b\in E_{\tau(a)}.
\]
Then, a direct computation proves that $\zeta\in {\mathcal V}_a^\circ$
and $S^*(\alpha_a)=\zeta\circ \prol{\tau}$. Thus, we obtain
\[
\Psi_a\subseteq \set{\zeta\circ \prol{\tau}}{\zeta\in {\mathcal
V}_a^\circ}
\]
and, using that the dimension of both spaces is $s$, we deduce
(\ref{psia}) holds.  Note that, in the particular case when the
constraints are linear, we have ${\mathcal V}=\tau^*(D)$ and
$\Psi=\widetilde{D^\circ}.$

Next, we consider the vector bundles $F$ and $\prol[\vd]{\cm}$ over
${\cm}$ whose fibers at the point $a\in {\cm}$ are
\[
F_{a} = \omega_{L}^{-1}(\Psi_{a}), \makebox[.4cm]{}
\prol[\vd]{\cm}[a]=\set{(b,v)\in\vd[a]\times
T_a\cm}{T\pi(v)=\rho(b)}.
\]
It follows that
\[
F_{a} = \set{ z \in \prol[E]{E}[a]}{\mbox{ exists } \zeta \in \V_{a}^0
\mbox{ and } i_{z}\omega_{L}(a) = \zeta \circ \prol{\tau}}
\]
and
\begin{equation}\label{8.1'}
  \prol[\vd]{\cm}[a]=\set{z\in\prol[E]{\cm}[a]}{\prol{\pi}(z)\in\vd[a]}=\set{z\in
  \prol[E]{\cm}[a]}{S(z)\in \prol[E]{\cm}[a]}.
\end{equation}
Note that the dimension of $\prol[\vd]{\cm}[a]$ is $2r$ and, when the
constraints are linear, i.e., ${\mathcal M}$ is a vector subbundle $D$
of $E$, we obtain
\[
\prol[\vd]{\cm}[a]=\prol[D]{D}[a], \mbox{ for all } a\in \cm =
D.
\]
Moreover, from (\ref{8.1'}), we deduce that the vertical lift of an
element of $\vd$ is an element of $\prol[\vd]{\cm}$. Thus we can
define for $b,c\in\vd[a]$
\[
\GLV[a](b,c)=\omega_L(a)(\tilde{b},\xi\spV(a,c)),
\]
where $\tilde{b} \in \prol[E]{E}[a]$ and $\prol{\tau}(\tilde{b}) =
b$.

\subsection{Dynamics in local coordinates}

Here we analyze the local nature of equations (\ref{8.1}). We consider
local coordinates $(x^i)$ on an open subset $U$ of $M$ and take a
basis $\{e_{\alpha}\}$ of local sections of $E$. In this way, we have
local coordinates $(x^i, y^{\alpha})$ on $E$.  Suppose that the local
equations defining $\cm$ as a submanifold of $E$ are
\[
\phi^{A}= 0, \makebox[.4cm]{} A = 1, \dots , s ,
\]
where $\phi^{A}$ are independent local constraint functions. Since
$\pi: \cm \to M$ is a fibration, it follows that the matrix
$\displaystyle (\frac{\partial \phi^{A}}{\partial y^{\alpha}})$ is of
rank $s$. Thus, if $d$ is the differential of the Lie algebroid
$\prol[E]{E} \to E$, we deduce that $\{d\phi^{A}|_\cm\}_{A=1, \dots ,
  s}$ is a local basis of sections of the vector bundle
$(\prol[E]{\cm})^{0} \to \cm$. Note that
\[
d\phi^{A} = \displaystyle \rho^j_{\alpha}\frac{\partial
  \phi^{A}}{\partial x^j}\X^{\alpha} + \frac{\partial
  \phi^{A}}{\partial y^{\alpha}}\V^{\alpha}.
\]
Moreover, $\{S^*(d\phi^{A})|_\cm = \displaystyle \frac{\partial
\phi^{A}}{\partial y^{\alpha}} \X^{\alpha}|_\cm\}_{A=1, \dots ,
s}$ is a local basis of sections of the vector bundle $\Psi \to
\cm$.

Next, we introduce the local sections $\{Z_{A}\}_{A=1, \dots , s}$ of
$\prol[E]{E} \to E$ defined by
\[
i_{Z_{A}} \omega_{L} = S^*(d\phi^{A}) = \displaystyle
\frac{\partial \phi^{A}}{\partial y^{\alpha}} \X^{\alpha}.
\]
A direct computation, using (\ref{omegaL}), proves that
\begin{equation}\label{Zeta}
  Z_{A} = \displaystyle -\frac{\partial \phi^{A}}{\partial
    y^{\alpha}} W^{\alpha \beta}\V_{\beta}, \makebox[.4cm]{} \mbox{
    for all } A,
\end{equation}
where $(W^{\alpha \beta})$ is the inverse matrix of $(W_{\alpha
\beta} = \displaystyle \frac{\partial^{2}L}{\partial
y^{\alpha}y^{\beta}})$. Furthermore, it is clear that
$\{Z_{A}|_\cm\}$ is a local basis of sections of the vector bundle
$F \to \cm$.

On the other hand, if $\Gamma_L$ is the Euler-Lagrange section
associated with the regular Lagrangian $L$, then a section $\Gamma$ of
$\prol[E]{\cm}\to \cm$ is a solution of equations (\ref{8.1}) if and
only if
\[
\Gamma=(\Gamma_L + \lambda^A Z_A)|_\cm\
\]
with $\lambda^A$ local real functions on $E$ satisfying
\[
(\lambda^Ad\phi^B(Z_A) + d\phi^B(\Gamma_L))|_\cm\ =0, \mbox{ for
all } B=1,\dots ,s.
\]
Therefore, using (\ref{Zeta}), we conclude that there exists a
unique solution of the Lagrange-d'Alembert equations (\ref{8.1})
if and only if the matrix
\begin{align}\label{eq:matrix}
  \big({\mathcal C}^{AB} =  \frac{\partial \phi^A}{\partial
    y^\alpha}W^{\alpha\beta}\frac{\partial \phi^B}{\partial
    y^\beta}\big)_{A,B=1,\dots ,s}
\end{align}
is regular. We are now ready to prove the following result.

\begin{theorem}\label{regular-nonl}
  The following properties are equivalent:
  \begin{enumerate}
  \item The constrained Lagrangian system $(L,\cm)$ is regular, that
    is, there exists a unique solution of the Lagrange-d'Alembert
    equations,
  \item $\Ker\GLV =\{0\}$,
  \item $\prol[E]{\cm} \cap F=\{0\}$,
  \item $\prol[{\mathcal V}]{\cm} \cap(\prol[{\mathcal
      V}]{\cm})^\perp=\{0\}$.
  \end{enumerate}
\end{theorem}
\begin{proof}
  It is clear that the matrix $({\mathcal C}^{AB})$
  in~\eqref{eq:matrix} is regular if and only if $\prol[E]{\cm}\cap
  F=\{0\}$. Thus, the properties (1) and (3) are equivalent. Moreover,
  proceeding as in the proof of Theorem \ref{regularity}, we deduce
  that the properties (2) and (3) (respectively, (2) and (4)) also are
  equivalent.
\end{proof}

\begin{remark}[Lagrangians of mechanical type]\label{mechty}
  {\rm If $L$ is a Lagrangian function of mechanical type, then, using
    Theorem~\ref{regular-nonl}, we deduce (as in the case of linear
    constraints) that the constrained system $(L, \cm)$ is always
    regular.} \oprocend
\end{remark}

\subsection{Lagrange-d'Alembert solutions and nonholonomic bracket}

Assume that the constrained Lagrangian system $(L, \cm)$ is
regular. Then (3) in Theorem~\ref{regular-nonl} is equivalent to
$(\prol[E]{E})|_\cm\ = \prol[E]{\cm} \oplus F$.  Denote by $P$ and
$Q$ the complementary projectors defined by this decomposition
\[
P_{a}: \prol[E]{E}[a] \to \prol[E]{\cm}[a], \makebox[.3cm]{} Q_{a}:
\prol[E]{E}[a] \to F_a, \; \; \mbox{ for all } a \in \cm.
\]
As in the case of linear constraints, we may prove the following.
\begin{theorem}\label{dym-nonl}
  Let $(L, \cm)$ be a regular constrained Lagrangian system and let
  $\Gamma_{L}$ be the solution of the free dynamics, i.e.,
  $i_{{\Gamma}_{L}}\omega_{L} = dE_{L}$.  Then, the solution of the
  constrained dynamics is the \sode\ $\Gamma_{(L, \cm)}$ obtained as follows
  \[
  \Gamma_{(L, \cm)} = P(\Gamma_{L}|_\cm).
  \]
\end{theorem}
On the other hand, $(4)$ in Theorem~\ref{regular-nonl} is
equivalent to $(\prol[E]{E})|_\cm\ =\prol[\vd]{\cm}\oplus
(\prol[\vd]{\cm})^\perp$ and we will denote by $\bar{P}$ and
$\bar{Q}$ the corresponding projectors induced by this
decomposition, that is,
\[
\bar{P}_a:\prol[E]{E}[a]\to \prol[\vd]{\cm}[a],\;\;\;
\bar{Q}_a:\prol[E]{E}[a]\to (\prol[\vd]{\cm}[a])^\perp, \mbox{ for
all } a\in \cm.
\]

\begin{theorem}\label{t8.3}
  Let $(L,\cm)$ be a regular constrained Lagrangian system, $\Gamma_L$
  (respectively, $\Gamma_{(L, \cm)}$) be the solution of the free
  (respectively, constrained) dynamics and $\Delta$ be the Liouville
  section of $\prol[E]{E}\to E$. Then, $\Gamma_{(L,
    \cm)}=\bar{P}(\Gamma_L|_{\cm})$ if and only if the restriction to
  ${\cm}$ of the vector field $\rho^1(\Delta)$ on $E$ is tangent to
  $\cm$.
\end{theorem}
\begin{proof}
  Proceeding as in the proof of Lemma~\ref{F-TDD}, we obtain that
  \[
  (\prol[\vd]{\cm}[a])^\perp \cap \Ver{\prol[E]{E}[a]}=F_a,\mbox{
    for all }a\in \cm.
  \]
  Thus, it is clear that
  \[
  Q(\Gamma_L(a))\in F_a\subseteq (\prol[\vd]{\cm}[a])^\perp, \mbox{
    for all } a\in \cm.
  \]
  Moreover, from (\ref{8.1'}) and using the fact that the solution of
  the constrained dynamics is a \sode\ along $\cm$, we deduce
  \[
  \Gamma_{(L, \cm)}(a) = P(\Gamma_{L}(a)) \in \prol[\vd]{\cm}[a], \;
  \; \mbox{ for all } a \in \cm,
  \]
  if and only if the restriction to $\cm$ of the vector field
  $\rho^1(\Delta)$ on $E$ is tangent to $\cm$.  This proves the
  result.
\end{proof}

\begin{remark}[Linear constraints]
  {\rm Note that if $\cm$ is a vector subbundle $D$ of $E$, then the
    vector field $\rho^1(\Delta)$ is always tangent to $\cm = D$.} \oprocend
\end{remark}
As in the case of linear constraints, one may develop the
distributional approach in order to obtain the solution of the
constrained dynamics. In fact, if $(L, \cm)$ is regular, then
$\prol[\vd]{\cm} \to \cm$ is a symplectic subbundle of
$(\prol[E]{E}, \omega_{L})$ and, thus, the restriction $\omega^{L,
\cm}$ of $\omega_{L}$ to $\prol[\vd]{\cm}$ is a symplectic section
on that bundle. We may also define $\varepsilon^{L, \cm}$ as the
restriction of $dE_{L}$ to $\prol[\vd]{\cm}$. Then, taking the
restriction of Lagrange-d'Alembert equations to $\prol[\vd]{\cm}$,
we get the following equation
\begin{equation}\label{disapp}
  i_{\bar{\Gamma}}\omega^{L, \cm} = \varepsilon ^{L,
    \cm} ,
\end{equation}
which uniquely determines a section $\bar{\Gamma}$ of $\prol[\vd]{\cm}
\to \cm$.  It is not difficult to prove that $\bar{\Gamma} =
\bar{P}(\Gamma_L|_{\cm})$.  Thus, the unique solution of equation
(\ref{disapp}) is the solution of the constrained dynamics if and only
if the vector field $\rho^{1}(\Delta)$ is tangent to $\cm$.

Let $(L,\cm)$ a regular constrained Lagrangian system. Since $S^*:
(\prol[E]{\cm})^0 \to \Psi$ is a vector bundle isomorphism, it
follows that there exists a unique section $\alpha_{(L, \cm)}$ of
$(\prol[E]{\cm})^0 \to \cm$ such that
\[
i_{Q(\Gamma_{L}|_{\cm})}\omega_{L} = S^*(\alpha_{(L, \cm)}).
\]
Moreover, we have the following result.
\begin{theorem}
  \label{conser-ener}
  If $(L, \cm)$ is a regular constrained Lagrangian system and
  $\Gamma_{(L, \cm)}$ is the solution of the dynamics, then
  $d_{\Gamma_{(L, \cm)}}(E_{L}|_{\cm}) = 0$ if and only if
  $\alpha_{(L, \cm)}(\Delta|_{\cm}) = 0$. In particular, if the vector
  field $\rho^{1}(\Delta)$ is tangent to $\cm$, then $d_{\Gamma_{(L,
      \cm)}}(E_{L}|_{\cm}) = 0$.
\end{theorem}
\begin{proof}
  From Theorem~\ref{dym-nonl}, we deduce
  \[
  (i_{\Gamma_{(L, \cm)}}\omega_{L} - dE_{L})|_{\cm} = -S^*(\alpha_{(L, \cm)}).
  \]
  Therefore, using that $\Gamma_{(L, \cm)}$ is a \sode\ along $\cm$, we obtain
  \[
  d_{\Gamma_{(L, \cm)}}(E_{L}|_{\cm}) = \alpha_{(L, \cm)}(\Delta|_{\cm}).
  \]
\end{proof}
Now, let $(L, \cm)$ be a regular constrained Lagrangian system. In
addition, suppose that $f$ and $g$ are two smooth functions on $\cm$
and take arbitrary extensions to $E$ denoted by the same letters.
Then, as in Section~\ref{NHBracket}, we may define \emph{the
  nonholonomic bracket } of $f$ and $g$ as follows
\[
\{f,g\}_{nh}=\omega_L(\bar{P}(X_f),\bar{P}(X_g))|_{\cm},
\]
where $X_f$ and $X_g$ are the Hamiltonian sections on
$\prol[E]{E}$ associated with $f$ and $g$, respectively.

Moreover, proceeding as in the case of linear constraints, one can
prove that
\[
\dot{f}=\rho^1(R_L)(f)+\{f,E_L\}_{nh}, \;\;\; f\in C^\infty(\cm),
\]
where $R_L$ is the section of $\prol[E]{\cm}\to \cm$ defined by
$R_L=P(\Gamma_L|_\cm)-\bar{P}(\Gamma_L|_{\cm}).$ Thus, in the
particular case when the restriction to $\cm$ of the vector field
$\rho^1(\Delta)$ on $E$ is tangent to ${\cm}$, it follows that
\[
\dot{f}=\{f,E_L\}_{nh},\;\;\; \mbox{ for } f\in C^\infty(\cm).
\]

Alternatively, since $\prol[\vd]{\cm}$ is an anchored vector bundle,
we may consider the differential $\bar{d}f\in
\Sec{(\prol[\vd]{\cm})^*}$ for a function $f\in C^\infty(\cm)$.  Thus,
since the restriction $\omega^{L,\cm}$ of $\omega_L$ to
$\prol[\vd]{\cm}$ is regular, we have a unique section $\bar{X}_f \in
\Sec{\prol[\vd]{\cm}}$ given by $i_{\bar{X}_f}\omega^{L,\cm}=\bar{d}f$
and it follows that
\[
\{f,g\}_{nh}=\omega^{L,\cm}(\bar{X}_f,\bar{X}_g).
\]

\subsection{Morphisms and reduction}

Let $(L,\cm)$ be a regular constrained Lagrangian system on a Lie
algebroid $\tau:E\to M$ and let $(L',\cm')$ be another constrained
Lagrangian system on a second Lie algebroid $\tau':E'\to M'$.
Suppose also that we have a fiberwise surjective morphism of Lie
algebroids $\Phi:E\to E'$ over a surjective submersion $\phi:M\to
M'$ such that:
\begin{itemize}
\item[(i)] $L=L'\circ \Phi$,
\item[(ii)] $\Phi|_\cm :\cm\to \cm'$ is a surjective submersion,
\item[(iii)] $\Phi(\vd[a])=\vd[\Phi(a)]'$, for all $a\in \cm$.
\end{itemize}

Note that condition (ii) implies that $\Phi(\vd[a])\subseteq
\vd[\Phi(a)]'$, for all $a\in \cm$.  Moreover, if $V(\Phi)$ is the
vertical bundle of $\Phi$ and
\[
V_a(\Phi)\subset T_a\cm, \mbox{ for all } a\in \cm ,
\]
then condition (ii) also implies that $\vd[\Phi(a)]'\subseteq
\Phi(\vd[a])$, for all $a\in \cm$.

On the other hand, using condition (iii) and
Proposition~\ref{transformation-omegaL}, it follows that $\ker
G^{L',{\vd[]'}}=\{0\}$ and, thus, the constrained Lagrangian system
$(L',\cm')$ is regular. Moreover, proceeding as in the proof of
Lemma~\ref{l5.5} and Theorem~\ref{t5.6}, we deduce the following
results.

\begin{lemma}\label{l8.4}
  With respect to the decompositions
  \[
  (\prol[E]{E})|_\cm=\prol[E]\cm\oplus F \quad \text{and} \quad
  (\prol[E']E')|_{\cm'}=\prol[E']\cm'\oplus F'
  \]
  we have the following properties
  \begin{enumerate}
  \item[(1)] $\prol[\Phi]{\Phi}(\prol[E]\cm)=\prol[E']{\cm'},$
  \item[(2)] $\prol[\Phi]{\Phi}(F)=F'$,
  \item[(3)] If $P$, $Q$ and $P',Q'$ are the projectors associated
    with $(L,\cm)$ and $(L',\cm')$, respectively, then $P'\circ
    \prol[\Phi]{\Phi}=\prol[\Phi]{\Phi}\circ P$ and $Q'\circ
    \prol[\Phi]{\Phi}=\prol[\Phi]{\Phi}\circ Q$.
  \end{enumerate}

  \noindent With respect to the decompositions
  \[
  (\prol[E]{E})|_{\cm}=\prol[\mathcal V]\cm\oplus (\prol[\mathcal V
  ]\cm)^\perp \mbox{ and } (\prol[E']{E'})|_{\cm'} = \prol[\mathcal
  V']{\cm'}\oplus (\prol[\mathcal V']{\cm'})^\perp
  \]
  we have the following properties
  \begin{enumerate}
  \item[(4)] $(\prol[\Phi]{\Phi})(\prol[\mathcal
    V]{\cm})=\prol[\mathcal V']{\cm'}$,
  \item[(5)] $(\prol[\Phi]{\Phi})((\prol[\mathcal
    V]{\cm})^\perp)=(\prol[\mathcal V']{\cm'})^\perp$,
  \item[(6)] If $\bar{P},\bar{Q}$ and $\bar{P}'$ and $\bar{Q'}$ are
    the projectors associated with $(L,{\cm})$ and $(L',{\cm'})$,
    respectively, then $\bar{P}'\circ \prol[\Phi]{\Phi} =
    \prol[\Phi]{\Phi}\circ \bar{P}$ and $\bar{Q}'\circ
    \prol[\Phi]{\Phi} = \prol[\Phi]{\Phi}\circ \bar{Q}$.
  \end{enumerate}
\end{lemma}

\begin{theorem}[Reduction of the constrained dynamics]\label{t8.5}
  Let $(L, \cm)$ be a regular constrained Lagrangian system on a Lie
  algebroid $E$ and let $(L',\cm')$ be a constrained Lagrangian system
  on a second Lie algebroid $E'$. Assume that we have a fiberwise
  surjective morphism of Lie algebroids $\Phi:E\to E'$ over $\phi:M\to
  M'$ such that conditions (i)-(iii) hold.  If $\Gamma_{(L, \cm)}$ is the
  constrained dynamics for $L$ and $\Gamma_{(L', \cm')}$ is the constrained
  dynamics for $L'$, respectively, then $\prol[\Phi]{\Phi}\circ
  \Gamma_{(L, \cm)}=\Gamma_{(L', \cm')}\circ \Phi$. If $a(t)$ is a solution of
  Lagrange-d'Alembert differential equations for $L$, then
  $\Phi(a(t))$ is a solution of Lagrange-d'Alembert differential
  equations for $L'$.
\end{theorem}

We will say that the constrained dynamics $\Gamma_{(L', \cm')}$ is
\emph{the reduction of the constrained dynamics} $\Gamma_{(L, \cm)}$
by the morphism $\Phi$.  As in the case of linear constraints (see
Theorem~\ref{t5.7}), we also may prove the following result
\begin{theorem}\label{t8.6'}
  Under the same hypotheses as in Theorem~\ref{t8.5}, we have that
  \[
  \{f'\circ \Phi,g'\circ \Phi\}_{nh}=\{f',g'\}_{nh}'\circ \Phi ,
  \]
  for $f',g'\in C^\infty(\cm')$, where $\{\cdot,\cdot\}_{nh}$
  (respectively, $\{\cdot,\cdot\}_{nh}'$) is the nonholonomic bracket
  for the constrained system $(L,\cm)$ (respectively, $(L',\cm')$). In
  other words, $\Phi:\cm\to \cm'$ is an almost-Poisson morphism.
\end{theorem}

Now, let $\phi:Q\to M$ be a principal $G$-bundle and $\tau:E\to Q$ be
a Lie algebroid over $Q$. In addition, assume that we have an action
of $G$ on $E$ such that the quotient vector bundle $E/G$ is defined
and the set $\Sec{E}^G$ of equivariant sections of $E$ is a Lie
subalgebra of $\Sec{E}$. Then, $E'=E/G$ has a canonical Lie algebroid
structure over $M$ such that the canonical projection $\Phi:E\to E'$
is a fiberwise bijective Lie algebroid morphism over $\phi$ (see
Theorem~\ref{quotient-Lie-algebroid}).

Next, suppose that $(L,\cm)$ is a $G$-invariant regular
constrained Lagrangian system, that is, the Lagrangian function
$L$ and the constraint submanifold $\cm$ are $G$-invariant. Then,
one may define a Lagrangian function $L':E'\to \R$ on $E'$ such
that
\[
L=L'\circ \Phi.
\]
Moreover, $G$ acts on $\cm$ and if the set of orbits $\cm'=\cm/G$ of
this action is a quotient manifold, that is, $\cm'$ is a smooth
manifold and the canonical projection $\Phi_{|\cm}:\cm\to \cm'={\cm
}/{G}$ is a submersion, then one may consider the constrained
Lagrangian system $(L',{\cm}')$ on $E'$.

\begin{remark}[Quotient manifold]\label{r8.5'}
  {\rm If $\cm$ is a closed submanifold of $E$, then, using a
    well-known result (see~\cite[Theorem~4.1.20]{AM}), it follows that
    the set of orbits $\cm'=\cm/G$ is a quotient manifold.} \oprocend
\end{remark}

Since the orbits of the action of $G$ on $E$ are the fibers of $\Phi$
and $\cm$ is $G$-invariant, we deduce that
\[
V_a(\Phi)\subseteq T_a\cm, \mbox{ for all } a\in \cm,
\]
which implies that $\Phi_{|\vd[a]}:{\vd[a]}\to \vd[\Phi(a)]'$ is a
linear isomorphism, for all $a\in \cm.$

Thus, from Theorem~\ref{t8.5}, we conclude that the constrained
Lagrangian system $(L',\cm')$ is regular and that
\[
\prol[\Phi]{\Phi}\circ \Gamma_{(L, \cm)}=\Gamma_{(L', \cm')}\circ
\Phi,
\]
where $\Gamma_{(L, \cm)}$ (resp., $\Gamma_{(L', \cm')}$) is the
constrained dynamics for $L$ (resp., $L'$). In addition, using
Theorem~\ref{t8.6'}, we obtain that $\Phi: \cm \to \cm'$ is an
almost-Poisson morphism when on $\cm$ and $\cm'$ we consider the
almost-Poisson structures induced by the corresponding
nonholonomic brackets.

We illustrate the results above in a particular example in the
following subsection.

\subsection{Example: a ball rolling on a rotating table}

The following example is taken from~\cite{BlKrMaMu,CLMM,NF}.  A
(homogeneous) sphere of radius $r>0$, unit mass $m=1$ and inertia
about any axis $k^2,$ rolls without sliding on a horizontal table
which rotates with constant angular velocity $\Omega$ about a vertical
axis through one of its points.  Apart from the constant gravitational
force, no other external forces are assumed to act on the sphere.

Choose a Cartesian reference frame with origin at the center of
rotation of the table and $z$-axis along the rotation axis. Let
$(x,y)$ denote the position of the point of contact of the sphere with
the table. The configuration space for the sphere on the table is
$Q=\R^2\times SO(3)$, where $SO(3)$ may be parameterized by the
Eulerian angles $\theta,\varphi$ and $\psi$. The kinetic energy of the
sphere is then given by
  \[
  T=\frac{1}{2}(\dot{x}^2 + \dot{y}^2 + k^2(\dot\theta^2 + \dot\psi^2
  + 2 \dot\varphi\dot\psi \cos \theta)) ,
  \]
  and with the potential energy being constant, we may put $V=0.$ The
  constraint equations are
  \begin{align*}
    \dot{x}-r\dot{\theta}\sin \psi + r \dot\varphi\sin \theta \cos
    \psi&=-\Omega y,\\
    \dot{y} + r\dot\theta \cos\psi + r \dot{\varphi}\sin \theta \sin
    \psi&=\Omega x.
  \end{align*}
  Since the Lagrangian function is of mechanical type, the constrained
  system is regular. Note that the constraints are affine, and hence
  not linear, and that the restriction to the constraint submanifold
  $\cm$ of the Liouville vector field on $TQ$ is not tangent to $\cm$.
  Indeed, the constraints are linear if and only if $\Omega = 0$.

  Now, we can proceed from here to construct to equations of motion of
  the sphere, following the general theory. However, the use of the
  Eulerian angles as part of the coordinates leads to very complicated
  expressions. Instead, one may choose to exploit the symmetry of the
  problem, and one way to do this is by the use of appropriate
  \emph{quasi-coordinates} (see \cite{NF}).  First of all, observe
  that the kinetic energy may be expressed as
  \[
  T=\frac{1}{2}(\dot{x}^2+\dot{y}^2 + k^2(\omega_x^2 + \omega_y^2 +
  \omega_z^2)),
  \]
  where
  \begin{align*}
    \omega_x&= \dot{\theta}\cos\psi + \dot{\varphi}\sin\theta\sin\psi
    , \\
    \omega_y&= \dot{\theta}\sin\psi - \dot{\varphi}\sin\theta\cos\psi
    , \\
    \omega_z&= \dot{\varphi}\cos\theta + \dot\psi ,
  \end{align*}
  are the components of the angular velocity of the sphere. The
  constraint equations expressing the rolling conditions can be
  rewritten as
  \begin{align*}
    \dot{x}-rw_y & = -\Omega y,\\
    \dot{y} + r\omega_x& = \Omega x.
  \end{align*}
  Next, following~\cite{CLMM}, we consider local coordinates
  $(\bar{x},\bar{y},\bar{\theta}, \bar{\varphi}, \bar{\psi};
  \pi_i)_{i=1,\dots ,5}$ on $TQ=T\R^2 \times T(SO(3))$, where
  \[
  \bar{x}=x,\;\;\; \bar{y}=y,\;\;\; \bar{\theta}=\theta,\;\;\;
  \bar{\varphi}=\varphi,\;\;\; \bar{\psi}=\psi,\]
  \[
  \pi_1=r\dot{x}+k^2\dot{q}_2,\;\;\;
  \pi_2=r\dot{y}-k^2\dot{q}_1,\;\;\; \pi_3=k^2\dot{q}_3,
  \]
  \[
  \pi_4=\frac{k^2}{(k^2+ r^2)}(\dot{x}-r\dot{q}_2 + \Omega y),\;\;\;
  \pi_5=\frac{k^2}{(k^2+ r^2)}(\dot{y}+r\dot{q}_1 - \Omega x),
  \]
  and $(\dot{q}_1,\dot{q}_2,\dot{q}_3)$ are the quasi-coordinates
  defined by
  \[
  \dot{q}_1=\omega_x,\;\;\; \dot{q}_2=\omega_y,\;\;\;
  \dot{q}_3=\omega_z.
  \]
  As is well-known, the coordinates $q_i$ only have a symbolic
  meaning. In fact, $ \displaystyle \{\frac{\partial }{\partial
    q_1}, \frac{\partial }{\partial q_2}, \frac{\partial }{\partial
    q_3}\}$ is the basis of left-invariant vector fields on $SO(3)$
  given by
  \begin{align*}
    \frac{\partial }{\partial q_1} &= (\cos\psi) \frac{\partial
    }{\partial \theta} + \frac{\sin \psi }{\sin \theta}(\frac{\partial
    }{\partial \varphi}-\cos\theta \frac{\partial }{\partial
      \psi}),\\
    \frac{\partial }{\partial q_2} &= (\sin \psi) \frac{\partial
    }{\partial \theta}-\frac{\cos\psi }{\sin \theta}(\frac{\partial
    }{\partial \varphi}-\cos\theta \frac{\partial }{\partial
      \psi}),\\
    \frac{\partial }{\partial q_3} &= \frac{\partial }{\partial \psi} ,
  \end{align*}
  and we have that
  \[
  [\frac{\partial }{\partial q_2}, \frac{\partial }{\partial
    q_1}]=\frac{\partial }{\partial q_3} , \quad [\frac{\partial
  }{\partial q_1}, \frac{\partial }{\partial q_3}]=\frac{\partial
  }{\partial q_2} , \quad [\frac{\partial }{\partial q_3},
  \frac{\partial }{\partial q_2}]=\frac{\partial }{\partial q_1}.
  \]

  Note that in the new coordinates the local equations defining the
  constraint submanifold $\cm$ are $\pi_4=0,$ $\pi_5=0$.  On the other
  hand, if $P:(\prol[TQ]{TQ})|_\cm =T_\cm(TQ)\to \prol[TQ]{\cm}=T\cm$
  and $Q:T_\cm(TQ)\to F$ are the projectors associated with the
  decomposition $T_\cm(TQ)=T\cm\oplus F$, then we have that
  (see~\cite{CLMM})
  \begin{align*}
    Q&= \frac{\partial }{\partial \pi_4}\otimes d\pi_4 +
    \frac{\partial }{\partial \pi_5}\otimes
    d\pi_5,\\
    P&= \text{Id} - \frac{\partial }{\partial \pi_4}\otimes
    d\pi_4-\frac{\partial }{\partial \pi_5}\otimes d\pi_5.
 \end{align*}
 Moreover, using that the unconstrained dynamics $\Gamma_L$ is
 given by
 \begin{align*}
   \Gamma_L &=\dot{x}\displaystyle\frac{\partial }{\partial \bar{x}} +
   \dot{y}\displaystyle\frac{\partial }{\partial \bar{y}}+
   \dot{\theta}\displaystyle\frac{\partial }{\partial \bar{\theta}} +
   \dot{\varphi}\displaystyle\frac{\partial }{\partial \bar{\varphi}}+
   \dot{\psi}\displaystyle\frac{\partial }{\partial \bar{\psi}} +
   \displaystyle\frac{k^2\Omega}{(k^2 +
     r^2)}\dot{y}\displaystyle\frac{\partial }{\partial \pi_4} +
   \displaystyle\frac{k^2\Omega}{(k^2 +
     r^2)}\dot{x}\displaystyle\frac{\partial }{\partial \pi_5}\\
   &= \dot{x}\displaystyle\frac{\partial }{\partial \bar{x}} +
   \dot{y}\displaystyle\frac{\partial }{\partial \bar{y}}+
   \dot{q_1}\displaystyle\frac{\partial }{\partial q_1} +
   \dot{q_2}\displaystyle\frac{\partial }{\partial q_2}+
   \dot{q_3}\displaystyle\frac{\partial }{\partial q_3} +
   \displaystyle\frac{k^2\Omega}{(k^2 +
     r^2)}\dot{y}\displaystyle\frac{\partial }{\partial \pi_4} +
   \displaystyle\frac{k^2\Omega}{(k^2 +
     r^2)}\dot{x}\displaystyle\frac{\partial }{\partial \pi_5},
 \end{align*}
 we deduce that the constrained dynamics is the \sode\ $\Gamma_{(L, \cm)}$ along
 $\cm$ defined by
 \begin{align}
   \label{Gamma}
   \Gamma_{(L, \cm)} = (P\Gamma_L|_{\cm}) &=(\dot{x}\displaystyle\frac{\partial
   }{\partial \bar{x}} + \dot{y}\displaystyle\frac{\partial }{\partial
     \bar{y}}+ \dot{\theta}\displaystyle\frac{\partial }{\partial
     \bar{\theta}} + \dot{\varphi}\displaystyle\frac{\partial
   }{\partial \bar{\varphi}}+ \dot{\psi}\displaystyle\frac{\partial
   }{\partial \bar{\psi}})|_{\cm} \nonumber
   \\&=(\dot{x}\displaystyle\frac{\partial }{\partial \bar{x}} +
   \dot{y}\displaystyle\frac{\partial }{\partial \bar{y}}+
   \dot{q_1}\displaystyle\frac{\partial }{\partial q_1} +
   \dot{q_2}\displaystyle\frac{\partial }{\partial q_2}+
   \dot{q_3}\displaystyle\frac{\partial }{\partial q_3})|_{\cm}.
 \end{align}
 This implies that
 \[
 d_{\Gamma_{(L, \cm)}}(E_{L}|_{\cm}) = d_{\Gamma_{(L, \cm)}}(L|_{\cm}) = \displaystyle
 \frac{\Omega^{2}k^{2}}{(k^{2} + r^{2})}(x \dot{x} + y
 \dot{y})|_{\cm}.
 \]
 Consequently, the Lagrangian energy is a constant of the motion if
 and only if $\Omega = 0$.

 When constructing the nonholonomic bracket on $\cm$, we find that
 the only non-zero fundamental brackets are
 \begin{equation}\label{Cornonh}
   \begin{array}{ll}
     \{x,\pi_1\}_{nh}=r,&\kern-50pt\{y,\pi_2\}_{nh}=r,\\
     \{q_1,\pi_2\}_{nh}=-1,&\kern-50pt\{q_2,\pi_1\}_{nh}=1,\;\;\;\;\;\;\;\;
     \{q_3,\pi_3\}_{nh}=1,\\[5pt]
     \{\pi_1,\pi_2\}_{nh}=\pi_3,&\kern-50pt\{\pi_2,\pi_3\}_{nh} =
     \displaystyle\frac{k^2}{(k^2+r^2)}\pi_1
     + \displaystyle\frac{rk^2\Omega}{(k^2+
       r^2)}y,\\\{\pi_3,\pi_1\}_{nh}=\displaystyle\frac{k^2}{(k^2+r^2)}\pi_2
     - \displaystyle\frac{rk^2\Omega}{(k^2+ r^2)}x,
   \end{array}
 \end{equation}
 in which the ``appropriate operational'' meaning has to be attached
 to the quasi-coordinates $q_i$.  As a result, we have
 \[
 \dot{f}=R_L(f) + \{f,L\}_{nh}, \mbox{ for  $f\in C^\infty(\cm)$}
 \]
 where $R_L$ is the vector field on $\cm$ given by
 \begin{align*}
   R_L&=\displaystyle(\frac{k^2\Omega}{(k^2+
     r^2)}(x\displaystyle\frac{\partial }{\partial
     y}-y\displaystyle\frac{\partial }{\partial x}) +
   \displaystyle\frac{r\Omega}{(k^2+
     r^2)}(x\displaystyle\frac{\partial }{\partial q_1} + y
   \frac{\partial
   }{\partial q_2} \\[5pt]
   & + x(\pi_3-k^2\Omega)\displaystyle\frac{\partial }{\partial \pi_1}
   + y (\pi_3-k^2\Omega)\displaystyle\frac{\partial }{\partial
     \pi_2}-k^2(\pi_1x+\pi_2y)\displaystyle\frac{\partial }{\partial
     \pi_3}))|_{\cm}.
 \end{align*}
 Note that $R_{L} = 0$ if and only if $\Omega = 0$.

 Now, it is clear that $Q=\R^2\times SO(3)$ is the total space of a
 trivial principal $SO(3)$-bundle over $\R^2$ and the bundle
 projection $\phi:Q\to M=\R^2$ is just the canonical projection on the
 first factor.  Therefore, we may consider the corresponding Atiyah
 algebroid $E'=TQ/SO(3)$ over $M=\R^2$.  Next, we describe this Lie
 algebroid.

 Using the left-translations in $SO(3)$, one may define a
 diffeomorphism $\lambda$ between the tangent bundle to $SO(3)$ and
 the product manifold $SO(3)\times \R^3$ (see \cite{AM}). In fact, in
 terms of the Euler angles, the diffeomorphism $\lambda$ is given by
 \begin{equation}\label{lambda}
   \lambda(\theta,\varphi,\psi;\dot\theta,\dot\varphi,\dot\psi) =
   (\theta,\varphi,\psi;
   \omega_x,\omega_y,\omega_z).
 \end{equation}
 Under this identification between $T(SO(3))$ and $SO(3)\times \R^3$,
 the tangent action of $SO(3)$ on $T(SO(3))\cong SO(3)\times \R^3$ is
 the trivial action
 \begin{equation}\label{Action}
   SO(3)\times (SO(3)\times \R^3)\to SO(3)\times \R^3,\;\;\;
   (g,(h,\omega))\mapsto (gh,\omega).
 \end{equation}
 Thus, the Atiyah algebroid $TQ/SO(3)$ is isomorphic to the product
 manifold $T\R^2\times \R^3$, and the vector bundle projection is
 $\tau_{\R^2}\circ pr_1$, where $pr_1:T\R^2\times \R^3\to T\R^2$ and
 $\tau_{\R^2}:T\R^2\to \R^2$ are the canonical projections.

 A section of $E'=TQ/SO(3)\cong T\R^2\times \R^3\to \R^2$ is a pair
 $(X,u)$, where $X$ is a vector field on $\R^2$ and $u:\R^2\to \R^3$
 is a smooth map. Therefore, a global basis of sections of
 $T\R^2\times \R^3\to \R^2$ is
 \[
 \begin{array}{rrr}
   e_1'=(\displaystyle\frac{\partial}{\partial x},0),&
   e_2'=(\displaystyle\frac{\partial}{\partial y},0),&\\[5pt]
   e_3'=(0,u_1),& e_4'=(0,u_2),& e_5'=(0,u_3),
 \end{array}
 \]
 where $u_1,u_2,u_3:\R^2\to \R^3$ are the constant maps
 \[
 u_1(x,y)=(1,0,0),\;\;\; u_2(x,y)=(0,1,0),\;\;\; u_3(x,y)=(0,0,1).
 \]
 In other words, there exists a one-to-one correspondence between the
 space $\Sec{E'=TQ/SO(3)}$ and the $G$-invariant vector fields on $Q$.
 Under this bijection, the sections $e_1'$ and $e_2'$ correspond with
 the vector fields $\displaystyle\frac{\partial }{\partial x}$ and
 $\displaystyle\frac{\partial }{\partial y}$ and the sections $e_3'$,
 $e_4'$ and $e_5'$ correspond with the vertical vector fields
 $\displaystyle\frac{\partial }{\partial q_1}$,
 $\displaystyle\frac{\partial }{\partial q_2}$ and
 $\displaystyle\frac{\partial }{\partial q_3}$, respectively.

 The anchor map $\rho':E'=TQ/SO(3)\cong T\R^2\times \R^3\to T\R^2$ is
 the projection over the first factor and, if $\lcf\cdot, \cdot \rcf'$
 is the Lie bracket on the space $\Sec{E'=TQ/SO(3)}$, then the only
 non-zero fundamental Lie brackets are
 \[
 \lcf e_4',e_3'\rcf'=e_5',\;\;\;\lcf e_5',e_4'\rcf'=e_3',\;\;\; \lcf
 e_3',e_5'\rcf'=e_4'.
 \]
 From (\ref{lambda}) and (\ref{Action}), it follows that the
 Lagrangian function $L=T$ and the constraint submanifold $\cm$ are
 $SO(3)$-invariant. Consequently, $L$ induces a Lagrangian function
 $L'$ on $E'=TQ/SO(3)$ and, since $\cm$ is closed on $TQ$, the set of
 orbits $\cm'=\cm/SO(3)$ is a submanifold of $E'=TQ/SO(3)$ in such a
 way that the canonical projection $\Phi|_\cm:\cm\to \cm'=\cm/SO(3)$
 is a surjective submersion.

 Under the identification between $E'=TQ/SO(3)$ and $T\R^2\times
 \R^3$, $L'$ is given by
 \[
 L'(x,y,\dot{x},\dot{y};\omega_1,\omega_2,\omega_3) =
 \frac{1}{2}(\dot{x}^2 + \dot{y}^2) + \frac{k^2}{2} (\omega_1^2 +
 \omega_2^2 + \omega_3^2) ,
 \]
 where $(x,y,\dot{x},\dot{y})$ and $(\omega_1,\omega_2,\omega_3)$ are
 the standard coordinates on $T\R^2$ and $\R^3$, respectively.
 Moreover, the equations defining $\cm'$ as a submanifold of
 $T\R^2\times \R^3$ are
 \[
 \dot{x}-r\omega_2 + \Omega y=0,\;\;\; \dot{y} + r\omega_1-\Omega x=0.
 \]
 So, we have the constrained Lagrangian system $(L',\cm')$ on the
 Atiyah algebroid $E'=TQ/SO(3)\cong T\R^2\times \R^3$. Note that the
 constraints are not linear, and that, if $\Delta'$ is the Liouville
 section of the prolongation $\prol[E']{E'}$, then the restriction to
 $\cm'$ of the vector field $(\rho')^1(\Delta')$ is not tangent to
 $\cm'$.

 Now, it is clear that the tangent bundle $TQ=T\R^2\times
 T(SO(3))\cong T\R^2\times (SO(3)\times \R^3)$ is the total space of a
 trivial principal $SO(3)$-bundle over $E'=TQ/SO(3)\cong T\R^2\times
 \R^3$ and, in addition (see~\cite[Theorem 9.1]{LeMaMa}), the
 prolongation $\prol[E']{E'}$ is isomorphic to the Atiyah algebroid
 associated with this principal $SO(3)$-bundle. Therefore, the
 sections of the prolongation $\prol[E']{E'}\to E'$ may be identified
 with the $SO(3)$-invariant vector fields on $TQ\cong T\R^2\times
 (SO(3)\times \R^3)$. Under this identification, the constrained
 dynamics $\Gamma_{(L', \cm')}$ for the system $(L',\cm')$ is just the
 $SO(3)$-invariant vector field $\Gamma_{(L, \cm)}=P(\Gamma_L|_{\cm})$. We recall
 that if $\Phi:TQ\to TQ/SO(3)$ is the canonical projection, then
 \begin{equation}\label{Reduc}
   \prol[\Phi]{\Phi}\circ \Gamma_{(L, \cm)}=\Gamma_{(L', \cm')}\circ \Phi.
 \end{equation}

 Next, we give a local description of the vector field
 $(\rho')^1(\Gamma_{(L', \cm')})$ on $E'=TQ/SO(3)\cong T\R^2\times
 \R^3$ and the nonholonomic bracket $\{\cdot,\cdot\}_{nh}'$ for the
 constrained system $(L',\cm')$. For this purpose, we consider a
 suitable system of local coordinates on $TQ/SO(3)\cong T\R^2\times
 \R^3.$ If we set
 \[\begin{array}{lll}
   x'=x,&y'=y,&\\
   \pi_1'=r\dot{x} +
   k^2\omega_2,&\pi_2'=r\dot{y}-k^2\omega_1,&\pi_3'=k^2\omega_3,\\
   \pi_4=\frac{k^2}{(k^2 + r^2)}(\dot{x}-r\omega_2 + \Omega y),
   &\pi_5'=\frac{k^2}{(k^2+r^2)}(\dot{y} + r \omega_1 -\Omega x),
 \end{array}
 \]
 then $(x',y',\pi_1',\pi_2',\pi_3',\pi_4',\pi_5')$ is a system of
 local coordinates on $TQ/SO(3)\cong T\R^2\times \R^3$. In these
 coordinates the equations defining the submanifold $\cm'$ are
 $\pi_4'=0$ and $\pi_5'=0$, and the canonical projection $\Phi:TQ\to
 TQ/SO(3)$ is given by
 \begin{equation}\label{Phi}
   \Phi(\bar{x},\bar{y},\bar{\theta},\bar{\varphi},\bar{\psi};
   \pi_1,\pi_2,\pi_3,\pi_4,\pi_5)=(\bar{x},\bar{y};{\pi}_1, {\pi}_2,
   {\pi}_3, {\pi}_4, {\pi}_5).
 \end{equation}

 Thus, from (\ref{Gamma}) and (\ref{Reduc}), it follows that
 \[
 (\rho')^1(\Gamma_{(L', \cm')})=(\dot{x}'\frac{\partial }{\partial x'} +
 \dot{y}'\frac{\partial }{\partial y'})|_{\cm'} ,
 \]
 or, in the standard coordinates $(x,y,\dot{x},\dot{y};
 \omega_1,\omega_2,\omega_3)$ on $T\R^2\times \R^3,$
 \begin{align*}
   (\rho')^1(\Gamma_{(L', \cm')})&=\{\dot{x}(\displaystyle\frac{\partial
   }{\partial x} + \displaystyle\frac{\Omega k^2}{(k^2 +
     r^2)}\displaystyle\frac{\partial }{\partial \dot{y}} +
   \displaystyle\frac{\Omega r}{(k^2 +
     r^2)}\displaystyle\frac{\partial }{\partial \omega_1})\\
   & \quad + \dot{y}(\displaystyle\frac{\partial }{\partial y} -
   \displaystyle\frac{\Omega k^2}{(k^2 +
     r^2)}\displaystyle\frac{\partial }{\partial \dot{x}} +
   \displaystyle\frac{\Omega r}{(k^2 +
     r^2)}\displaystyle\frac{\partial }{\partial \omega_2})\}|_{\cm'}.
 \end{align*}
 On the other hand, from (\ref{Cornonh}), (\ref{Phi}) and
 Theorem~\ref{t8.6'}, we deduce that the only non-zero fundamental
 nonholonomic brackets for the system $(L',\cm')$ are
 \[
 \begin{array}{lll}
   \{x',\pi_1'\}'_{nh}=r,& \{y',\pi_2'\}'_{nh}=r,&\\
   \{\pi_1',\pi_2'\}_{nh}'=\pi_3',&\{\pi_2',\pi_3'\}_{nh}' =
   \displaystyle\frac{k^2}{(k^2+
     r^2)}\pi_1' + \displaystyle\frac{rk^2\Omega}{(k^2+ r^2)}y',&\\
   \{\pi_3',\pi_1'\}_{nh}'=\displaystyle\frac{k^2}{(k^2 + r^2)}
   \pi_2'- \displaystyle\frac{rk^2\Omega}{(k^2+ r^2)}x'.&&
 \end{array}
 \]
 Therefore, we have that
 \[
 \dot{f}'=(\rho')^1(R_{L'})(f') + \{f',L'\}_{nh}',\mbox{ for } f'\in
 C^\infty(\cm'),
 \]
 where $(\rho')^1(R_{L'})$ is the vector field on $\cm'$ given by
 \begin{align*}
   (\rho')^1(R_{L'}) &= \{ \displaystyle\frac{k^2\Omega}{k^2+
     r^2}(x'\displaystyle\frac{\partial }{\partial
     y'}-y'\displaystyle\frac{\partial }{\partial x'}) +
   \displaystyle\frac{r\Omega}{(k^2 +
     r^2)}(x'(\pi_3'-k^2\Omega)\displaystyle\frac{\partial }{\partial
     \pi_1'} \\[5pt]
   & \quad + y' (\pi_3'-k^2\Omega)\displaystyle\frac{\partial
   }{\partial \pi_2'}-k^2(\pi_1' x' + \pi_2'
   y')\displaystyle\frac{\partial }{\partial \pi_3'})\}|_{\cm'}.
 \end{align*}

\section{Conclusions and outlook}\label{conclusions}

We have developed a geometrical description of nonholonomic mechanical
systems in the context of Lie algebroids. This formalism is the
natural extension of the standard treatment on the tangent bundle of
the configuration space. The proposed approach also allows to deal
with nonholonomic mechanical systems with symmetry, and perform the
reduction procedure in a unified way.  The main results obtained in
the paper are summarized as follows:
\begin{itemize}
\item we have identified the notion of regularity of a nonholonomic
  mechanical system with linear constraints on a Lie algebroid, and we
  have characterized it in geometrical terms;
\item we have obtained the constrained dynamics by projecting the
  unconstrained one using two different decompositions of the
  prolongation of the Lie algebroid along the constraint subbundle;
\item we have developed a reduction procedure by stages and applied it
  to nonholonomic mechanical systems with symmetry.  These results
  have allowed us to get new insights in the technique of
  quasicoordinates;
\item we have defined the operation of nonholonomic bracket to measure
  the evolution of observables along the solutions of the system;
\item we have examined the setup of nonlinearly constrained systems;
\item we have illustrated the main results of the paper in several
  examples.
\end{itemize}
Current and future directions of research include the in-depth study
of the reduction procedure following the steps
of~\cite{BlKrMaMu,CaLeMaMa} for the standard case; the synthesis of
so-called nonholonomic integrators~\cite{cortes,CoSo,LeMaSa} for
systems evolving on Lie algebroids, and the development of a
comprehensive treatment of classical field theories within the Lie
algebroid formalism following the ideas by E.
Mart{\'\i}nez~\cite{CFTLAMF}.

\end{document}